\documentstyle [12pt,epsf,rotate] {article}

\newcommand{\nwc}{\newcommand}
\nwc{\cl}  {\clubsuit}
\nwc{\di}  {\diamondsuit}
\nwc{\sps}  {\spadesuit}

\nwc{\hyp} {\hyphenation}
\nwc{\be}  {\begin{equation}}
\nwc{\ee}  {\end{equation}}
\nwc{\ba}  {\begin{array}}
\nwc{\ea}  {\end{array}}
\nwc{\bdm} {\begin{displaymath}}
\nwc{\edm} {\end{displaymath}}
\nwc{\bea} {\be\ba{rcl}}
\nwc{\eea} {\ea\ee}
\nwc{\ben} {\begin{eqnarray}}
\nwc{\een} {\end{eqnarray}}
\nwc{\bda} {\bdm\ba{lcl}}
\nwc{\eda} {\ea\edm}
\nwc{\bc}  {\begin{center}}
\nwc{\ec}  {\end{center}}
\nwc{\ds}  {\displaystyle}
\nwc{\bmat}{\left(\ba}
\nwc{\emat}{\ea\right)}
\nwc{\non} {\nonumber}
\nwc{\bib} {\bibitem}
\nwc{\lra} {\longrightarrow}
\nwc{\Llra}{\Longleftrightarrow}
\nwc{\ra}  {\rightarrow}
\nwc{\Ra}  {\Rightarrow}
\nwc{\lmt} {\longmapsto}
\nwc{\prl} {\partial}
\nwc{\iy}  {\infty}
\nwc{\ol}  {\overline}
\nwc{\hm}  {\hspace{3mm}}
\nwc{\lf}  {\left}
\nwc{\ri}  {\right}
\nwc{\lm}  {\limits}
\nwc{\lb}  {\lbrack}
\nwc{\rb}  {\rbrack}
\nwc{\ov}  {\over}
\nwc{\pr}  {\prime}
\nwc{\nnn} {\nonumber \vspace{.2cm} \\ }
\nwc{\Sc}  {{\cal S}}
\nwc{\Lc}  {{\cal L}}
\nwc{\Rc}  {{\cal R}}
\nwc{\Dc}  {{\cal D}}
\nwc{\Oc}  {{\cal O}}
\nwc{\Cc}  {{\cal C}}
\nwc{\Pc}  {{\cal P}}
\nwc{\Mc}  {{\cal M}}
\nwc{\Ec}  {{\cal E}}
\nwc{\Fc}  {{\cal F}}
\nwc{\Hc}  {{\cal H}}
\nwc{\Kc}  {{\cal K}}
\nwc{\Xc}  {{\cal X}}
\nwc{\Gc}  {{\cal G}}
\nwc{\Zc}  {{\cal Z}}
\nwc{\Nc}  {{\cal N}}
\nwc{\fca} {{\cal f}}
\nwc{\xc}  {{\cal x}}
\nwc{\Ac}  {{\cal A}}
\nwc{\Bc}  {{\cal B}}
\nwc{\Uc}  {{\cal U}}
\nwc{\Vc}  {{\cal V}}
\nwc{\Th} {\Theta}
\nwc{\th} {\theta}
\nwc{\vth} {\vartheta}
\nwc{\eps}{\epsilon}
\nwc{\si} {\sigma}
\nwc{\Gm} {\Gamma}
\nwc{\gm} {\gamma}
\nwc{\bt} {\beta}
\nwc{\La} {\Lambda}
\nwc{\la} {\lambda}
\nwc{\om} {\omega}
\nwc{\Om} {\Omega}
\nwc{\dt} {\delta}
\nwc{\Si} {\Sigma}
\nwc{\Dt} {\Delta}
\nwc{\al} {\alpha}
\nwc{\vph}{\varphi}
\nwc{\zt} {\zeta}
\def\tr{\mathop{\rm tr}}
\def\Tr{\mathop{\rm Tr}}

\def\VEV#1{\left\langle #1\right\rangle}

\def\abs#1{\left| #1\right|}
\def\pr#1{#1^\prime}
\def\ltap{\raisebox{-.4ex}{\rlap{$\sim$}} \raisebox{.4ex}{$<$}}

\nwc{\Id}  {{\bf 1}}
\nwc{\diag} {{\rm diag}}
\nwc{\inv}  {{\rm inv}}
\nwc{\mod}  {{\rm mod}}
\nwc{\hal} {\frac{1}{2}}
\nwc{\tpi}  {2\pi i}

\def\slash#1{#1\!\!\!/\!\,\,}

\def\pr#1{Phys. Rev. {\bf #1}}

\def\APP#1{Acta Phys.~Pol.~{\bf #1}}

\def\CNPP#1{Comm. Nucl. Part. Phys.~{\bf #1}}

\def\IJMP#1{Int. J. Mod. Phys.~{\bf #1}}

\def\MPL#1{Mod. Phys. Lett.~{\bf #1}}
\def\NP#1{Nucl. Phys.~{\bf #1}}
\def\NPPS#1{Nucl. Phys. Proc. Suppl.~{\bf #1}}
\def\NC#1{Nuovo Cim.~{\bf #1}}

\def\PL#1{Phys. Lett.~{\bf #1}}
\def\PR#1{Phys. Rev.~{\bf #1}}
\def\PRP#1{Phys. Rep.~{\bf #1}}
\def\PRL#1{Phys. Rev. Lett.~{\bf #1}}

\def\PTP#1{Progr. Theor. Phys.~{\bf #1}}
\def\RMP#1{Rev. Mod. Phys.~{\bf #1}}

\def\ZP#1{Z. Phys.~{\bf #1}}

\def\MeV {\,{\rm  MeV}}
\def\GeV {\,{\rm  GeV}}

\def \lta {\mathrel{\vcenter
     {\hbox{$<$}\nointerlineskip\hbox{$\sim$}}}}
\def \gta {\mathrel{\vcenter
     {\hbox{$>$}\nointerlineskip\hbox{$\sim$}}}} 
\newsavebox{\nnin} \sbox{\nnin}{$\hspace{1mm}\in\kern -.8em /
                   \hspace{1mm}$}

\newcommand{\sub}{\subset}
\newsavebox{\nnsub} \sbox{\nnsub}{$\hspace{1mm}\sub\kern -.9em /
            \hspace{1mm}$}

\def\KK{{\rm I\kern -.2em  K}}
\def\NN{{\rm I\kern -.16em N}}
\def\RR{{\rm I\kern -.2em  R}}
\def\ZZ{Z \kern -.43em Z}
\def\QQ{{\rm \kern .25em
             \vrule height1.4ex depth-.12ex width.06em\kern-.31em Q}}
\def\CC{{\rm \kern .25em
             \vrule height1.4ex depth-.12ex width.06em\kern-.31em C}}
\def\ZZZ{Z\kern -0.31em Z}

\newcounter{app}
\def\app{\par
 \addtocounter{app}{1}
 \def\thesection{\Alph{app}}
 \def\ksection{\Alph{app}}}

\def\appendix#1{\app\sect{#1}}

\newcommand{\sect}[1]{ \section{#1} \setcounter{equation}{0} }

\begin{document}

\begin{titlepage}

  \title{\vspace{1cm}Two Flavor Chiral Phase Transition \\ from \\ 
    Nonperturbative Flow Equations\thanks{Supported by the Deutsche
      Forschungsgemeinschaft}}

  \author{ {\sc J.~Berges\thanks{Email:
        J.Berges@thphys.uni-heidelberg.de}}\\[3mm]{\sc
      D.--U.~Jungnickel\thanks{Email:
        D.Jungnickel@thphys.uni-heidelberg.de}} \\[3mm]
    and \\[3mm]
    {\sc C.~Wetterich\thanks{Email: C.Wetterich@thphys.uni-heidelberg.de}}
    \\[6mm]
    {\em Institut f\"ur Theoretische Physik} \\ {\em Universit\"at
      Heidelberg} \\ {\em Philosophenweg 16} \\ {\em 69120 Heidelberg,
      Germany} }

\date{May 30, 1997}
\maketitle

\begin{picture}(5,2.5)(-350,-500)
\put(12,-105){HD--THEP--97--20}
\end{picture}

\thispagestyle{empty}

\begin{abstract}
  We employ nonperturbative flow equations to compute the equation of
  state for two flavor QCD within an effective quark meson model. This
  yields the temperature and quark mass dependence of quantities like
  the chiral condensate or the pion mass. A precision estimate of the
  universal critical equation of state for the three--dimensional
  $O(4)$ Heisenberg model is presented.  We explicitly connect the
  $O(4)$ universal behavior near the critical temperature and zero
  quark mass with the physics at zero temperature and a realistic pion
  mass.  For realistic quark masses the pion correlation length near
  $T_c$ turns out to be smaller than its zero temperature value.
\end{abstract}

\end{titlepage}

\sect{Introduction}
\label{Introduction}

Strong interactions in thermal equilibrium at high temperature $T$ ---
as realized in early stages of the evolution of the Universe ---
differ in important aspects from the well tested vacuum or zero
temperature properties. A phase transition at some critical
temperature $T_c$ or a relatively sharp crossover may separate the
high and low temperature physics~\cite{MO96-1}. Many experimental
activities at heavy ion colliders~\cite{QM96} search for signs of such
a transition.  It was realized early that the transition should be
closely related to a qualitative change in the chiral condensate
according to the general observation that spontaneous symmetry
breaking tends to be absent in a high temperature situation. A series
of stimulating contributions~\cite{PW84-1,RaWi93-1,Raj95-1} pointed
out that for sufficiently small up and down quark masses, $m_u$ and
$m_d$, and for a sufficiently large mass of the strange quark, $m_s$,
the chiral transition is expected to belong to the universality class
of the $O(4)$ Heisenberg model. This means that near the critical
temperature only the pions and the sigma particle play a role for the
behavior of condensates and long distance correlation functions. It
was suggested~\cite{RaWi93-1,Raj95-1} that a large correlation length
may be responsible for important fluctuations or lead to a disoriented
chiral condensate~\cite{Ans88-1}. This was even
related~\cite{RaWi93-1,Raj95-1} to the spectacular ``Centauro
events''~\cite{LFH80-1} observed in cosmic rays. The question how
small $m_u$ and $m_d$ would have to be in order to see a large
correlation length near $T_c$ and if this scenario could be realized
for realistic values of the current quark masses remained, however,
unanswered. The reason was the missing link between the universal
behavior near $T_c$ and zero current quark mass on one hand and the
known physical properties at $T=0$ for realistic quark masses on the
other hand.

It is the purpose of the present paper to provide this link. We
present here the equation of state for two flavor QCD within an
effective quark meson model. The equation of state expresses the
chiral condensate $\VEV{\ol{\psi}\psi}$ as a function of temperature
and the average current quark mass $\hat{m}=(m_u+m_d)/2$.  This
connects explicitly the universal critical behavior for $T\ra T_c$ and
$\hat{m}\ra0$ with the temperature dependence for a realistic value
$\hat{m}_{\rm phys}$. Since our discussion covers the whole
temperature range $0\le T \,\ltap\, 1.7\, T_c$ we can fix
$\hat{m}_{\rm phys}$ such that the (zero temperature) pion mass is
$m_\pi=135\MeV$. The condensate $\VEV{\ol{\psi}\psi}$ plays here the
role of an order parameter. Its precise definition will be given in
section \ref{TheQuarkMesonModelAtT=0}.  Figure \ref{ccc_T} shows our
results for $\VEV{\ol{\psi}\psi}(T,\hat{m})$:
\begin{figure}
\unitlength1.0cm

\begin{center}
\begin{picture}(13.,7.0)

\put(0.0,0.0){
\epsfysize=11.cm
\rotate[r]{\epsffile{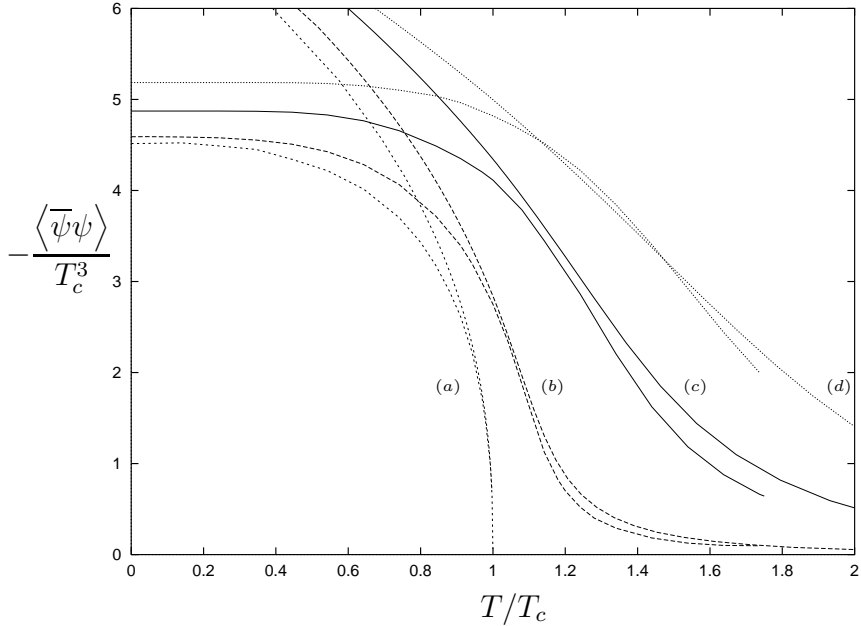}}
}
\put(-0.5,4.2){\bf $\ds{-\frac{\VEV{\ol{\psi}\psi}}{T_{c}^3}}$}
\put(5.8,-0.5){\bf $\ds{T/T_{c}}$}
\put(5.2,2.5){\tiny $(a)$}
\put(6.6,2.5){\tiny $(b)$}
\put(8.5,2.5){\tiny $(c)$}
\put(10.4,2.5){\tiny $(d)$}

\end{picture}
\end{center}
\caption{\footnotesize The plot shows the chiral condensate
  $\VEV{\ol{\psi}\psi}$ as a function of temperature $T$.  Lines
  $(a)$, $(b)$, $(c)$, $(d)$ correspond at zero temperature to
  $m_\pi=0,45\MeV,135\MeV,230\MeV$, respectively. For each pair of
  curves the lower one represents the full $T$--dependence of
  $\VEV{\ol{\psi}\psi}$ whereas the upper one shows for comparison the
  universal scaling form of the equation of state for the $O(4)$
  Heisenberg model. The critical temperature for zero quark mass is
  $T_c=100.7\MeV$. The chiral condensate is normalized at a scale
  $k_{\Phi}\simeq 620\MeV$.}
\label{ccc_T}
\end{figure}
Curve $(a)$ gives the temperature dependence of $\VEV{\ol{\psi}\psi}$
in the chiral limit $\hat{m}=0$. Here the lower curve is the full
result for arbitrary $T$ whereas the upper curve corresponds to the
universal scaling form of the equation of state for the $O(4)$
Heisenberg model.  We see perfect agreement of both curves for $T$
sufficiently close to $T_c=100.7 \MeV$. This demonstrates the
capability of our method to cover the critical behavior and, in
particular, to reproduce the critical exponents of the $O(4)$--model.
We have determined (cf.~section~\ref{CriticalBehavior}) the universal
critical equation of state as well as the non--universal amplitudes.
This provides the full functional dependence of $\VEV{\ol{\psi}\psi}
(T,\hat{m})$ for small $T-T_c$ and $\hat{m}$.  The curves $(b)$, $(c)$
and $(d)$ are for non--vanishing values of the average current quark
mass $\hat{m}$.  Curve $(c)$ corresponds to $\hat{m}_{\rm phys}$ or,
equivalently, $m_\pi(T=0)=135\MeV$. One observes a crossover in the
range $T=(1.2-1.5)T_c$. The $O(4)$ universal equation of state (upper
curve) gives a reasonable approximation in this temperature range. The
transition turns out to be much less dramatic than for $\hat{m}=0$. We
have also plotted in curve $(b)$ the results for comparably small
quark masses $\simeq1\MeV$, i.e.~$\hat{m}=\hat{m}_{\rm phys}/10$, for
which the $T=0$ value of $m_\pi$ equals $45\MeV$. The crossover is
considerably sharper but a substantial deviation from the chiral limit
remains even for such small values of $\hat{m}$. In order to
facilitate comparison with lattice simulations which are typically
performed for larger values of $m_\pi$ we also present results for
$m_\pi(T=0)=230\MeV$ in curve $(d)$. One may define a ``pseudocritical
temperature'' $T_{pc}$ associated to the smooth crossover as the
inflection point of $\VEV{\ol{\psi}\psi}(T)$ as usually done in
lattice simulations. Our results for this definition of $T_{pc}$ are
denoted by $T_{pc}^{(1)}$ and are presented in table \ref{tab11} for
the four different values of $\hat{m}$ or, equivalently, $m_\pi(T=0)$.
\begin{table}
\begin{center}
\begin{tabular}{|c||c|c|c|c|} \hline
  $\stackrel{ }{\frac{m_\pi}{\MeV}}$ &
  $0$ &
  $45$ &
  $135$ &
  $230$
  \\[1.0mm] \hline
  $\stackrel{ }{\frac{T_{pc}^{(1)}}{\MeV}}$ &
  $100.7$ &
  $\simeq110$ &
  $\simeq130$ &
  $\simeq150$
  \\[1mm] \hline
  $\stackrel{ }{\frac{T_{pc}^{(2)}}{\MeV}}$ &
  $100.7$ &
  $$113 &
  $$128 &
  $$---
  \\[1mm] \hline
\end{tabular}
\caption{\footnotesize The table shows the critical and
  ``pseudocritical'' temperatures for various values of the zero
  temperature pion mass. Here $T_{pc}^{(1)}$ is defined as the
  inflection point of $\VEV{\ol{\psi}\psi}(T)$ whereas $T_{pc}^{(2)}$
  is the location of the maximum of the sigma correlation length (see
  section \ref{TheQuarkMesonModelAtTNeq0}).}
\label{tab11}
\end{center}
\end{table}
The value for the pseudocritical temperature for $m_{\pi}=230 \MeV$
compares well with the lattice results for two flavor QCD
(cf.~section~\ref{CriticalBehavior}). One should mention, though, that
a determination of $T_{pc}$ according to this definition is subject to
sizeable numerical uncertainties for large pion masses as the curve in
figure \ref{ccc_T} is almost linear around the inflection point for
quite a large temperature range.  A problematic point in lattice
simulations is the extrapolation to realistic values of $m_\pi$ or
even to the chiral limit. Our results may serve here as an analytic
guide. The overall picture shows the approximate validity of the
$O(4)$ scaling behavior over a large temperature interval in the
vicinity of and above $T_c$ once the (non--universal) amplitudes are
properly computed.

\begin{figure}
\unitlength1.0cm
\begin{center}
\begin{picture}(13.,7.0)

\put(0.0,0.0){
\epsfysize=11.cm
\rotate[r]{\epsffile{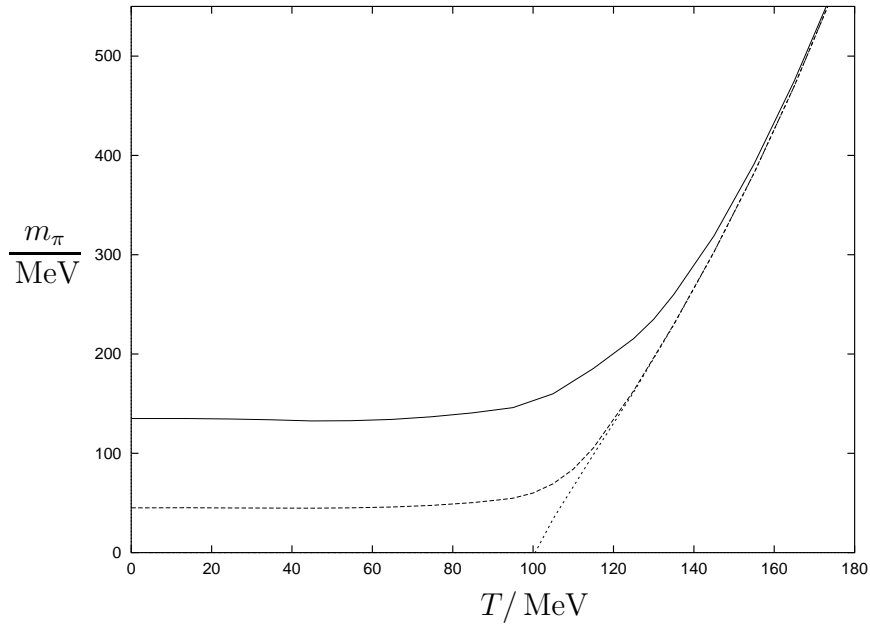}}
}
\put(-0.5,4.2){\bf $\ds{\frac{m_\pi}{\MeV}}$}
\put(5.8,-0.5){\bf $\ds{T/\MeV}$}

\end{picture}
\end{center}
\caption{\footnotesize The plot shows $m_\pi$ as a function of
  temperature $T$ for three different values of the average light
  current quark mass $\hat{m}$. The solid line corresponds to the
  realistic value $\hat{m}=\hat{m}_{\rm phys}$ whereas the dotted line
  represents the situation without explicit chiral symmetry breaking,
  i.e., $\hat{m}=0$. The intermediate, dashed line assumes 
  $\hat{m}=\hat{m}_{\rm phys}/10$.}
\label{mpi_T}
\end{figure}

A second important result of our investigations is the temperature
dependence of the space--like pion correlation length
$m_\pi^{-1}(T)$. (We will often call $m_\pi(T)$ the temperature
dependent pion mass since it coincides with the physical pion mass for
$T=0$.) The plot for $m_\pi(T)$ in figure \ref{mpi_T} again shows the
second order phase transition in the chiral limit $\hat{m}=0$. 
For $T<T_c$ the pions are massless Goldstone bosons whereas for
$T>T_c$ they form with the sigma particle a degenerate vector of
$O(4)$ with mass increasing as a function of temperature.  For
$\hat{m}=0$ the behavior for small positive $T-T_c$ is characterized
by the critical exponent $\nu$, i.e.
$m_\pi(T)=\left(\xi^+\right)^{-1}T_c \left( (T-T_c)/T_c\right)^\nu$
and we obtain $\nu=0.787$, $\xi^+=0.270$. For $\hat{m}>0$ we find that
$m_\pi(T)$ remains almost constant for $T\lta T_c$ with only a very
slight dip for $T$ near $T_c/2$. For $T>T_c$ the correlation length
decreases rapidly and for $T\gg T_c$ the precise value of $\hat{m}$
becomes irrelevant. We see that the universal critical behavior near
$T_c$ is quite smoothly connected to $T=0$.  The full functional
dependence of $m_\pi(T,\hat{m})$ allows us to compute the overall size
of the pion correlation length near the critical temperature and we
find $ m_\pi(T_{pc})\simeq 1.7 m_\pi(0)$ for the realistic value
$\hat{m}_{\rm phys}$. This correlation length is even smaller than the
vacuum ($T=0$) one and gives no indication for strong fluctuations of
pions with long wavelength. It would be interesting to see if a
decrease of the pion correlation length at and above $T_c$ is
experimentally observable.  It should be emphasized, however, that a
tricritical behavior with a massless excitation remains possible for
three flavors. This would not be characterized by the universal
behavior of the $O(4)$--model. We also point out that the present
investigation for the two flavor case does not take into account a
speculative ``effective restoration'' of the axial $U_A(1)$ symmetry
at high temperature \cite{PW84-1,Shu94-1}. We will comment on these
issues in section~\ref{AdditionalDegreesOfFreedom}.

Our method is based on the effective average action
$\Gamma_k$~\cite{Wet91-1} which is the generating functional of the
$1PI$ Green functions in presence of an infrared cutoff $k$. In a
thermal equilibrium context $\Gamma_k$ depends also on temperature and
describes a coarse grained free energy as a functional of appropriate
fields. Here $k^{-1}$ corresponds to the coarse graining length scale.
Varying the infrared cutoff $k$ allows us to consider the relevant
physics in dependence on some momentum--like scale. The results for
the order parameter and correlation functions presented in this paper
are obtained by removing the infrared cutoff ($k\ra0$) in the end. For
scalar fields $\Phi_i$ the $k$--dependence of the effective average
action is given by an exact nonperturbative flow
equation~\cite{Wet93-2}
\begin{equation}
 \frac{\prl}{\prl t} \Gm_k [\Phi] =
 \hal\Tr\left\{\left(
 \Gm_k^{(2)}[\Phi]+R_k\right)^{-1}
 \frac{\prl R_k}{\prl t}\right\}
 \label{ERGE}
\end{equation}
where $t=\ln(k/\Lambda)$ with $\Lambda$ an arbitrary momentum scale.
Here $\Gamma_k^{(2)}$ denotes the matrix of second functional
derivatives of $\Gamma_k$ with respect to the field components:
\begin{equation}
  \label{BBB02}
  \left(\Gamma_k^{(2)}\right)_{i j}[\Phi](q,q^\prime)=
  \frac{\delta^2\Gamma_k[\Phi]}
  {\delta\Phi^i(q)\delta\Phi^j(-q^\prime)}\; 
\end{equation}
and we employ a momentum dependent infrared cutoff
\begin{equation}
 R_k(q)=\frac{Z_{\Phi,k}q^2 e^{-q^2/k^2}}{1- e^{-q^2/k^2}}
 \label{Rk(q)}
\end{equation}
with $Z_{\Phi,k}$ an appropriate wave function renormalization
constant to be specified later. In momentum space the trace contains a
momentum integration, $\Tr=\int\frac{d^d q}{(2\pi)^d}\sum_i$. The flow
equation (\ref{ERGE}) closely resembles a one--loop equation: Indeed,
replacing $\Gamma_k^{(2)}$ by the second functional derivative of the
classical action, $S_{\rm cl}^{(2)}$, it corresponds to the one--loop
result for a theory where an infrared cutoff
$$\frac{1}{2}\int\frac{d^d q}{(2\pi)^d}\vph_i(q)R_k(q)\vph_i(-q)$$ is
added to the classical action $S_{\rm cl}[\vph]$. This cutoff
appears in the inverse ``average propagator''
\begin{equation}
  \label{BBB01}
  P(q)=q^2+Z_{\Phi,k}^{-1}R_k(q)=
  \frac{q^2}{1-\exp\left\{-\frac{q^2}{k^2}\right\}}
\end{equation}
which approaches $k^2$ for $q^2\ll k^2$. Up to exponentially small
corrections the integration of the high momentum modes with $q^2\gg
k^2$ is not affected by the infrared cutoff. The ``renormalization
group improvement'' $S_{\rm cl}^{(2)}\ra\Gamma_k^{(2)}$ contains all
contributions beyond one--loop and makes (\ref{ERGE}) exact. Of
course, it also turns the flow equation into a functional differential
equation which cannot be solved exactly in general. We emphasize that
the flow equation (\ref{ERGE}) is connected to the Wilsonian
renormalization group equation~\cite{Wil71-1}--\cite{Has86-1} (often
also called exact renormalization group equation). Extensions of the
flow equations to fermions~\cite{Wet90-1,CKM97-1} and gauge
fields~\cite{RW93-1}--\cite{Ell97-1} are available.

Since in most cases the flow equation can not be solved exactly the
capacity to devise useful truncations in a nonperturbative context
becomes crucial. This requires first of all an identification of the
degrees of freedom which are most relevant for a given problem. In the
present paper we concentrate on the chiral aspects of QCD\footnote{For
  a study of chiral symmetry breaking in QED using related exact
  renormalization group techniques see~\cite{AMST97-1}.}. Spontaneous
chiral symmetry breaking occurs through the expectation value of a
(complex) scalar field $\Phi_{ab}$ which transforms as $(\ol{\bf
  N},\bf{N})$ under the chiral flavor group $SU_L(N)\times SU_R(N)$
with $N$ the number of light quark flavors. More precisely, the
expectation value
\begin{equation}
  \label{AAA50}
  \VEV{\Phi^{ab}}=\ol{\sigma}_0\dt^{ab} 
\end{equation} 
induces for $\ol{\sigma}_0\neq0$ a spontaneous breaking of the chiral
group to a vector--like subgroup, $SU_L(N)\times
SU_R(N)\longrightarrow SU_{L+R}(N)\equiv SU_V(N)$. In addition,
non--vanishing current quark masses $m_u,m_d,m_s$ break the chiral
group explicitly and also lift the $SU_V(N)$ degeneracy of the
spectrum if they are unequal. The physical degrees of freedom
contained in the field $\Phi_{ab}$ are pseudoscalar and scalar mesons
which can be understood as quark--antiquark bound states. It is
obvious that any analytical description of the chiral transition has
to include at least part of these (pseudo--)scalar fields as the most
relevant degrees of freedom.

In the present work we use for $k$ smaller than a ``compositeness
scale'' $k_\Phi\simeq600\MeV$ a description in terms of $\Phi_{ab}$
and quark degrees of freedom. The quarks acquire a constituent quark mass
$M_q$ through the chiral condensate $\ol{\sigma}_0$ which forms in our
picture for $k_{\chi SB}\simeq400\MeV$.  This effective quark meson
model can be obtained from QCD by ``integrating out'' the gluon
degrees of freedom and introducing fields for composite
operators~\cite{EW94-1,Wet95-2}. This will be explained in more detail
in the first part of section \ref{TheQuarkMesonModelAtT=0}. In this
picture the scale $k_\Phi$ is associated to the scale at which the
formation of mesonic bound states can be observed in the flow of the
effective (momentum dependent) four--quark interaction. We will
restrict our discussion in this paper to two flavor QCD with equal
quark masses $m_u=m_d\equiv\hat{m}$.  Since in this case the scalar
triplet $a_0$ and the pseudoscalar singlet (associated with the
$\eta^\prime$) have typical masses around\footnote{More precisely,
  because of the anomalous $U_A(1)$ breaking in QCD these mesons are
  significantly heavier than the remaining degrees of freedom in the
  range of scales $k$ where the dynamics of the model is strongly
  influenced by mesonic fluctuations. The situation becomes more
  involved if the model is considered at high temperature which is
  discussed in section \ref{AdditionalDegreesOfFreedom}.} $1\GeV$ we
will neglect them for $k<k_\Phi$.  This reduces the scalar degrees of
freedom of our effective model to a four component vector of $O(4)$,
consisting of the three pions and the ``sigma resonance''.

We imagine that all other degrees of freedom besides the quarks $\psi$
and the scalars $\Phi$ are integrated out. This is reflected in the
precise form of the effective average action
$\Gamma_{k_\Phi}[\psi,\Phi]$ at the scale $k_\Phi$ which serves as an
initial value for the solution of the flow equation. The flow of
$\Gamma_{k}[\psi,\Phi]$ for $k<k_\Phi$ is then entirely due to the
quark and meson fluctuations which are not yet included in
$\Gamma_{k_{\Phi}}[\psi,\Phi]$. Obviously, the initial value
$\Gamma_{k_\Phi}$ may be a quite complicated functional of $\psi$ and
$\Phi$ containing, in particular, important non--local behavior. We
will nevertheless use a rather simple truncation in terms of standard
kinetic terms and a most general form of the scalar potential $U_k$,
i.e.\footnote{Our Euclidean conventions ($\ol{h}_k$ is real) are
  specified in refs.~\cite{JW96-1,Wet90-1}.}
\begin{equation}
  \label{AAA60}
  \begin{array}{rcl}
  \ds{\hat{\Gamma}_k} &=& \ds{
    \Gamma_k-\frac{1}{2}\int d^4 x\tr
    \left(\Phi^\dagger\jmath+\jmath^\dagger\Phi\right)}\nnn
    \ds{\Gamma_{k}} &=& \ds{
      \int d^4x\Bigg\{
      Z_{\psi,k}\ol{\psi}_a i\slash{\prl}\psi^a+
      Z_{\Phi,k}\tr\left[\prl_\mu\Phi^\dagger\prl^\mu\Phi\right]+
      U_k(\Phi,\Phi^\dagger)
      }\nnn
    &+& \ds{
      \ol{h}_{k}\ol{\psi}^a\left(\frac{1+\gm_5}{2}\Phi_{ab}-
      \frac{1-\gm_5}{2}(\Phi^\dagger)_{ab}\right)\psi^b
      \Bigg\} }\; .
  \end{array}
\end{equation}
Here $\Gamma_k$ is invariant under the chiral flavor symmetry
$SU_L(2)\times SU_R(2)$ and the only explicit symmetry breaking arises
through the source term $\jmath\sim\hat{m}$. We will consider the flow
of the most general form of $U_k$ consistent with the symmetries
(without any restriction to a polynomial form as typically used in a
perturbative context). On the other hand, our approximations for the
kinetic terms are rather crude and parameterized by only two running
wave function renormalization constants, $Z_{\Phi,k}$ and
$Z_{\psi,k}$. The same holds for the effective Yukawa coupling
$\ol{h}_k$. The main approximations in this work concern
\begin{description}
\item [(i)] the simple form of the derivative terms and the Yukawa
  coupling, in particular, the neglect of higher derivative terms (and
  terms with two derivatives and higher powers of $\Phi$). This is
  partly motivated by the observation that at the scale $k_\Phi$ and
  for small temperatures the possible strong non--localities related
  to confinement affect most likely only the quarks in a momentum
  range $q^2\lta(300\MeV)^2$.  Details of the quark propagator and
  interactions in this momentum range are not very important in our
  context (see section~\ref{TheQuarkMesonModelAtT=0}).
\item[(ii)] the neglect of interactions involving more than two quark
  fields. This is motivated by the fact that the dominant multi--quark
  interactions are already incorporated in the mesonic description.
  Six--quark interactions beyond those contained effectively in $U_k$
  could be related to baryons and play probably only a minor role for
  the meson physics considered here.
\end{description}

We will choose a normalization of $\psi,\Phi$ such that
$Z_{\psi,k_\Phi}=\ol{h}_{k_\Phi}=1$. We therefore need as initial
values at the scale $k_\Phi$ the scalar wave function renormalization
$Z_{\Phi,k_\Phi}$ and the shape of the potential $U_{k_\Phi}$. We will
make here the important assumption that $Z_{\Phi,k}$ is small at the
compositeness scale $k_\Phi$ (similarly to what is usually assumed in
Nambu--Jona-Lasinio--like models). This results in a large value of
the renormalized Yukawa coupling
$h_k=Z_{\Phi,k}^{-1/2}Z_{\psi,k}^{-1}\ol{h}_k$. A large value of
$h_{k_\Phi}$ is phenomenologically suggested by the comparably large
value of the constituent quark mass $M_q$. The latter is related to
the value of the Yukawa coupling for $k\ra0$ and the pion decay
constant $f_\pi=92.4\MeV$ by $M_q=h f_\pi/2$ (with $h=h_{k=0}$), and
$M_q\simeq300\MeV$ implies $h^2/4\pi\simeq3.4$. For increasing $k$ the
value of the Yukawa coupling grows rapidly for $k\gta M_q$.  Our
assumption of a large initial value for $h_{k_\Phi}$ is therefore
equivalent to the assumption that the truncation (\ref{AAA60}) can be
used up to the vicinity of the Landau pole of $h_k$. The existence of
a strong Yukawa coupling enhances the predictive power of our approach
considerably.  It implies a fast approach of the running couplings to
partial infrared fixed points~\cite{JW96-1}. In consequence, the
detailed form of $U_{k_\Phi}$ becomes unimportant, except for the
value of one relevant parameter corresponding to the scalar mass term
$\ol{m}^2_{k_\Phi}$. In this paper we fix $\ol{m}^2_{k_\Phi}$ such
that $f_\pi=92.4\MeV$ for $m_\pi=135\MeV$. The possibility of such a
choice is highly non--trivial since $f_\pi$ can actually be
predicted~\cite{JW96-1} in our setting within a relatively narrow
range. The value$f_\pi=92.4\MeV$ (for $m_\pi=135\MeV$) sets our unit
of mass for two flavor QCD which is, of course, not directly
accessible by observation. Besides $\ol{m}^2_{k_\Phi}$ (or $f_\pi$)
the other input parameter used in this work is the constituent quark
mass $M_q$ which determines the scale $k_\Phi$ at which $h_{k_\Phi}$
becomes very large. We consider a range $300\MeV\lta M_q\lta350\MeV$
and find a rather weak dependence of our results on the precise value
of $M_q$. We also observe that the limit $h_{k_\Phi}\ra\infty$ can be
considered as the lowest order of a systematic expansion in
$h_{k_\Phi}^{-1}$ which is obviously highly nonperturbative.

A generalization of our method to the realistic case of three light
flavors is possible and work in this direction is in progress. For the
time being we expect that many features found for $N=2$ will carry
over to the realistic case, especially the critical behavior for $T\ra
T_c$ and $\hat{m}\ra0$ (for fixed $m_s\neq0$).  Nevertheless, some
quantities like $\VEV{\ol{\psi}\psi}(T=0)$, the difference between
$f_\pi$ for realistic quark masses and $\hat{m}=0$ or the mass of the
sigma resonance at $T=0$ may be modified. This will also affect the
non--universal amplitudes in the critical equation of state and, in
particular, the value of $T_c$.  In the picture of the two flavor
quark meson model these changes occur through an effective temperature
dependence of the initial values of couplings at the scale $k_\Phi$.
This effect, which is due to the temperature dependence of effects
from fluctuations not considered in the present work is discussed
briefly in section \ref{AdditionalDegreesOfFreedom}.  It remains
perfectly conceivable that this additional temperature dependence may
result in a first order phase transition or a tricritical behavior for
realistic values of $\hat{m}$ for the three flavor case. Details will
depend on the strange quark mass.  We observe, however, that the
temperature dependence in the limit $\hat{m}\ra0$ involves for $T\le
T_c$ only information from the running of couplings in the range
$k\lta300\MeV$.  (The running for $k\gta3T$ effectively drops out in
the comparison between the thermal equilibrium results and those for
$T=0$.)  In this range of temperatures our model should be quite
reliable.

Finally, we mention that we have concentrated here only on the
$\Phi$--dependent part of the effective action which is related to
chiral symmetry breaking. The $\Phi$--independent part of the free
energy also depends on $T$ and only part of this temperature
dependence is induced by the scalar and quark fluctuations considered
in the present paper. Most likely, the gluon degrees of freedom cannot
be neglected for this purpose. This is the reason why we do not give
results for ``overall quantities'' like energy density or pressure as
a function of $T$.

This paper is organized as follows: In
section~\ref{TheQuarkMesonModelAtT=0} we review the linear quark meson
model at vanishing temperature. We begin with an overview of the
different scales appearing in strong interaction physics.
Subsequently, the flow equations for the linear quark meson model are
introduced and their approximate partial fixed point behavior is
discussed in detail leading to a ``prediction'' of the chiral
condensate $\VEV{\ol{\psi}\psi}$. In
section~\ref{FiniteTemperatureFormalism} the exact renormalization
group formulation of field theories in thermal equilibrium is given.
It is demonstrated how mass threshold functions in the flow equations
smoothly decouple all massive Matsubara modes as the temperature
increases, therefore leading to a ``dimensional reduction'' of the
model.  Section~\ref{TheQuarkMesonModelAtTNeq0} contains our results
for the linear quark meson model at non--vanishing temperature. Here
we discuss the $T$--dependences of the parameters and physical
observables of the linear quark meson model in detail for a
temperature range $0\le T\lta170\MeV$ including the (pseudo)critical
temperature $T_c$ of the chiral transition. The critical behavior of
the model near $T_c$ and $\hat{m}=0$, where $\hat{m}$ denotes the
light average current quark mass, is carefully analyzed in
section~\ref{CriticalBehavior}.  There we present the universal
scaling form of the equation of state including a semi--analytical fit
for the corresponding scaling function. Also the universal critical
exponents and amplitude ratios are given there.  The effects of
additional degrees of freedom of strong interaction physics not
included in the linear $O(4)$--symmetric quark meson model are
addressed in section~\ref{AdditionalDegreesOfFreedom}. Here we also
comment on differences between the linear quark meson model and chiral
perturbation theory. Some technical details concerning the quark mass
term and the definition of threshold functions at vanishing and
non--vanishing temperature are presented in three appendices.

\sect{The quark meson model at $T=0$}
\label{TheQuarkMesonModelAtT=0}

Before discussing the finite temperature behavior of strong
interaction physics we will review some of its zero temperature
features. This will be done within the framework of a linear quark
meson model as an effective description for QCD for scales below the
mesonic compositeness scale of approximately $k_\Phi\simeq600\MeV$.
Relating this model to QCD in a semi--quantitative way in subsection
\ref{ASemiQuantitativePicture} will allow us to gain some information
on the initial value for the effective average action at the
compositeness scale $k_\Phi$.  We emphasize, however, that the
quantitative aspects of the derivation of the effective quark meson
model from QCD will not be relevant for our practical calculations in
the mesonic sector. This is related to the ``infrared stability'' for
large Yukawa coupling $h_{k_\Phi}$ as discussed in the introduction
and which will be made quantitative in subsection
\ref{FlowEquationsAndInfraredStability}.

\subsection{A short (scale) history of QCD}
\label{ASemiQuantitativePicture}

For an evaluation of the trace on the right hand side 
of the flow equation
(\ref{ERGE}) only a small momentum range $q^2\simeq k^2$ contributes
substantially. One therefore only needs to take into account
those fluctuations which are
important in this momentum interval. Here we are interested
in the description of chiral symmetry breaking.
The relevant fluctuations in relation to this phenomenon may
change with the scale $k$ and we
begin by summarizing the qualitatively different scale intervals
which appear for meson physics in QCD. 
Some of this will be explained
in more detail in the remainder of this section whereas other aspects
are well known. Details of this discussion may also be
found in refs.~\cite{EW94-1,JW96-1,JW96-4}.
We will distinguish five qualitatively different
ranges of scales:
\begin{enumerate}
\item At sufficiently high momentum scales, say,
  \begin{displaymath}
    k\gta k_p\simeq1.5\GeV
  \end{displaymath}
  the relevant degrees of freedom of strong interactions are quarks
  and gluons and their dynamics is well described by perturbative QCD.
\item For decreasing momentum scales in the range
  \begin{displaymath}
    k_\Phi\simeq600\MeV\lta k\lta k_p\simeq1.5\GeV
  \end{displaymath}
  the dynamical degrees of freedom are still quarks and gluons. Yet,
  as $k$ is lowered part of their dynamics becomes dominated by
  effective non--local four quark interactions which cannot be fully
  accessed perturbatively.
\item At still lower scales this situation changes dramatically.
  Quarks and gluons are supplemented by mesonic bound states as
  additional degrees of freedom which are formed at a scale
  $k_\Phi\simeq600\MeV$. We emphasize that $k_\Phi$ is well separated
  from $\Lambda_{\rm QCD}\simeq200\MeV$ where confinement sets in and
  from the constituent masses of the quarks $M_q\simeq(300-350)\MeV$.
  This implies that below the compositeness scale $k_\Phi$ there
  exists a hybrid description in term of quarks {\em and} mesons! It
  is important to note that for scales not too much smaller than
  $k_\Phi$ chiral symmetry remains unbroken. This situation holds down
  to a scale $k_{\chi SB}\simeq400\MeV$ at which the scalar meson
  potential develops a non--trivial minimum thus breaking chiral
  symmetry spontaneously. The meson dynamics within the range
  \begin{displaymath}
    k_{\chi SB}\simeq400\MeV\lta k\lta k_\Phi\simeq600\MeV
  \end{displaymath}
  is dominated by light current quarks with a strong Yukawa coupling
  $h^2_k/(4\pi)\gg\alpha_s(k)$ to mesons. We will thus assume that
  the leading gluon effects are included below $k_\Phi$ already in the
  formation of mesons.  Near $k_{\chi SB}$ also fluctuations of the
  light scalar mesons become important as their initially large
  renormalized mass approaches zero. Other hadronic bound states like
  vector mesons or baryons should have masses larger than those of the
  lightest scalar mesons, in particular near $k_{\chi SB}$, and give
  therefore only subleading contributions to the dynamics. This leads
  us to a simple linear model of quarks and scalar mesons as an
  effective description of QCD for scales below $k_\Phi$.
\item As one evolves to scales below $k_{\chi SB}$ the Yukawa coupling
  decreases whereas $\alpha_s$ increases. Of course, getting closer to
  $\Lambda_{\rm QCD}$ it is no longer justified to neglect in the
  quark sector the QCD effects which go beyond the dynamics of the
  effective quark meson model in our truncation (\ref{AAA60}).  On the
  other hand, the final IR value of the Yukawa coupling $h$ is fixed
  by the typical values of constituent quark masses $M_q\simeq300\MeV$
  to be $h^2/(4\pi)\simeq3.4$. One may therefore speculate that the
  domination of the Yukawa interaction persists even for the interval
  \begin{displaymath}
    M_q\simeq300\MeV\lta k\lta k_{\rm \chi SB}\simeq400\MeV
  \end{displaymath}
  below which the quarks decouple from the evolution of the mesonic
  degrees of freedom altogether. Of course, details of the gluonic
  interactions are expected to be crucial for an understanding of
  quark and gluon confinement. Strong interaction effects may
  dramatically change the momentum dependence of the quark propagator
  for $k$ and $q^2$ around $\Lambda_{\rm QCD}$.  Yet, there is no
  coupling of the gluons to the color neutral mesons. As long as one
  is only interested in the dynamics of the mesons one is led to
  expect that confinement effects are quantitatively not too
  important.
\item Because of the effective decoupling of the quarks and therefore
  of the whole colored sector the details of confinement have only
  little influence on the mesonic dynamics for scales
  \begin{displaymath}
    k\lta M_q\simeq300\MeV\; .
  \end{displaymath}
  Here quarks and gluons disappear effectively from the spectrum and
  one is left with the pions. They are the only particles whose
  propagation is not suppressed by a large mass. For scales below the
  pion mass the flow of the couplings stops.
\end{enumerate}

Let us now try to understand these different ranges of scales in more
detail. We may start at $k_p=1.5\GeV$ where we assume that all gluonic
degrees of freedom have been integrated out while we have kept an
effective infrared cutoff $\sim k_p$ in the quark propagators.
Details of this procedure were outlined in ref.~\cite{Wet95-2}. This
results in a non--trivial momentum dependence of the quark propagator
and effective non--local four and higher quark interactions. Because
of the infrared cutoff the resulting effective action for the quarks
resembles closely the one for heavy quarks (at least for Euclidean
momenta). The dominant effect is the appearance of an effective quark
potential similar to the one for the charm quark which describes the
effective four quark interactions.

\begin{figure}
\unitlength1.0cm
\begin{picture}(13.,8.)
\put(1.7,0.3){$k_p=1.5\GeV$}
\put(9.6,0.3){$k_\Phi\simeq600\MeV$}
\put(10.5,3.4){$\Phi^{ba}$}

\put(0.7,6.2){$\ol{\psi}_a^i(p_1)$}
\put(3.9,6.2){$\psi_a^j(p_3)$}
\put(0.7,1.6){$\psi_b^i(p_2)$}
\put(3.9,1.6){$\ol{\psi}_b^j(p_4)$}

\put(7.2,5.8){$\ol{\psi}_a^i(p_1)$}
\put(13.2,5.8){$\psi_a^j(p_3)$}
\put(7.2,2.0){$\psi_b^i(p_2)$}
\put(13.2,2.0){$\ol{\psi}_b^j(p_4)$}

\put(1.5,1.3){
\epsffile{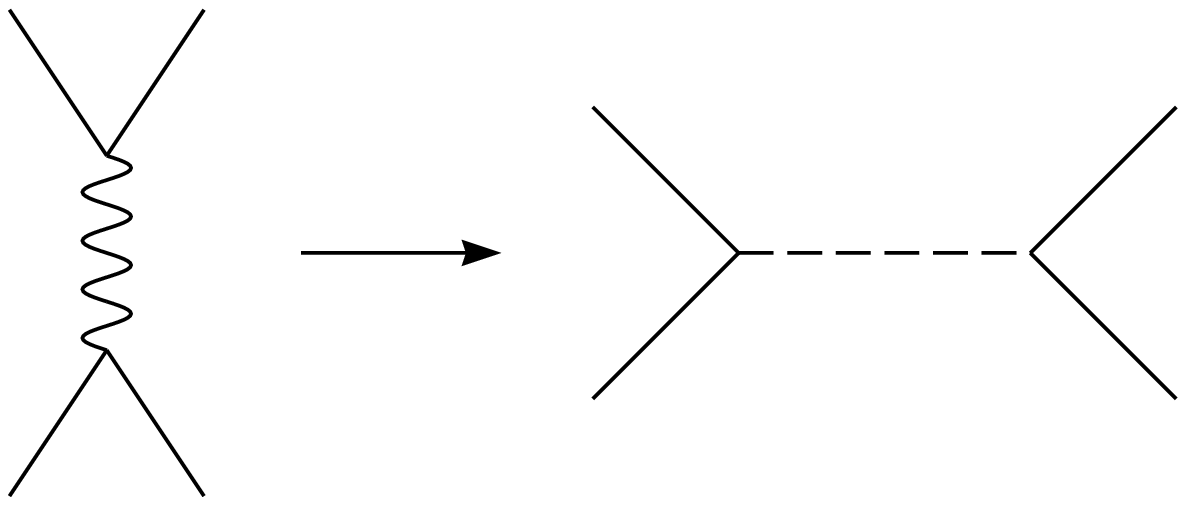}
}
\end{picture}
\caption{\footnotesize The left diagram represents the one gluon
  exchange $t$--channel contribution to the four quark vertex at the
  scale $k_p\simeq1.5\GeV$. It is assumed here that the gluon
  propagator is modified such that it accounts for the linearly rising
  term in the heavy quark potential. The right diagram displays the
  scalar meson $s$--channel exchange found at the compositeness scale
  $k_\Phi\simeq600\MeV$.}
\label{Feyn} 
\end{figure}

We next have to remove the infrared cutoff for the quarks, $k\ra0$.
This task can be carried out by means of the exact flow equation for
quarks only, starting at $k_p$ with an initial value
$\Gamma_{k_p}[\psi]$ as obtained after integrating out the gluons. A
first investigation in this direction~\cite{EW94-1} used a truncation
with a chirally invariant four quark interaction whose most general
momentum dependence was retained. A crucial point is, of course, the
initial value for this momentum dependence at $k_p$. The ansatz used
in ref.~\cite{EW94-1} is obtained by Fierz transforming the heavy
quark potential and keeping, for simplicity, only the scalar meson
channel while neglecting the $\rho$--meson and pomeron channels which
are also present.  The effective heavy quark potential was
approximated there by a one gluon exchange term $\sim\alpha_s(k_p)$
supplemented by a linearly rising string tension term.  This ansatz
corresponds to the four quark interaction generated by the flavor
neutral $t$--channel one gluon exchange depicted in figure \ref{Feyn}
with an appropriately modified gluon propagator and quark gluon vertex
in order to account for the linearly rising part of the potential.

The evolution equation for the four quark interaction can be derived
from the fermionic version of eq.~(\ref{ERGE}). It is by far not clear
that the evolution of the effective four quark vertex will lead at
lower scales to a momentum dependence representing the ($s$--channel)
exchange of colorless mesonic bound states. Yet, at the compositeness
scale
\begin{equation}
  \label{kphi}
  k_\Phi\simeq600\MeV
\end{equation}
one finds \cite{EW94-1} an approximate Bethe--Salpeter factorization
of the four quark amplitude with precisely this property.  This
situation is described by the right Feynman diagram in figure
\ref{Feyn}.  In particular, it was possible to extract the amputated
Bethe--Salpeter wave function as well as the mesonic bound state
propagator displaying a pole--like structure in the $s$--channel if it
is continued to negative $s=(p_1+p_2)^2$. In the limit where the
momentum dependence of the Bethe--Salpeter wave function and the bound
state propagator is neglected the effective action $\Gamma_{k_{\Phi}}$
resembles\footnote{Our solution of the flow equation for $\Gamma_k$
  with $Z_{\Phi,k_\Phi}=0$ (see below) may be considered as a solution
  of the NJL model with a particular form of the ultraviolet cutoff
  dictated by the shape of $R_k(q^2)$ as given in eq.~(\ref{Rk(q)}).}
the Nambu--Jona-Lasinio model~\cite{NJL61-1,Bij95-1}.  It is therefore
not surprising that our description of the dynamics for $k<k_\Phi$
will parallel certain aspects of the investigations of this model,
even though we are not bound to the approximations used typically in
such studies (large--$N_c$ expansion, perturbative renormalization
group, etc.).

It is clear that for scales $k\lta k_\Phi$ a description of strong
interaction physics in terms of quark fields alone would be rather
inefficient. Finding physically reasonable truncations of the
effective average action should be much easier once composite fields
for the mesons are introduced.  The exact renormalization group
equation can indeed be supplemented by an exact formalism for the
introduction of composite field variables or, more generally, a change
of variables~\cite{EW94-1}. In the context of QCD this amounts to the
replacement of the dominant part of the four quark interactions by
scalar meson fields with Yukawa couplings to the quarks.  In turn,
this substitutes the effective quark action at the scale $k_\Phi$ by
the effective quark meson action given in eq.\ (\ref{AAA60}) in the
introduction\footnote{We note that no double counting problem arises
  in this procedure.}.  The term in the effective potential
$U_{k_\Phi}$ which is quadratic in $\Phi$,
$U_{k_\Phi}=\ol{m}^2_{k_\Phi}\tr\Phi^\dagger\Phi+\ldots$, turns out to
be positive as a consequence of the attractiveness of the four quark
interaction inducing it. Its value was found for the simple
truncations used in ref.~\cite{EW94-1} to be
$\ol{m}_{k_\Phi}\simeq120\MeV$.  The higher order terms in
$U_{k_\Phi}$ cannot be determined in the four quark approximation
since they correspond to terms involving six or more quark fields.
(Their values will not be needed for our quantitative investigations
as is discussed in subsection
\ref{FlowEquationsAndInfraredStability}). The initial value of the
(bare) Yukawa coupling corresponds to the amputated Bethe--Salpeter
wave function. Neglecting its momentum dependence it can be normalized
to $\ol{h}_{k_\Phi}=1$.  Moreover, the quark wave function
renormalization $Z_{\psi,k}$ is normalized to one at the scale
$k_\Phi$ for convenience. One may add that we have refrained here for
simplicity from considering four quark operators with vector and
pseudo--vector spin structure.  Their inclusion is straightforward and
would lead to vector and pseudo--vector mesons in the effective
action.

In view of the possible large truncation errors made in
ref.~\cite{EW94-1}
we will take (\ref{kphi}) and the above value of 
$\ol{m}_{k_\Phi}$ only as order of magnitude
estimates. Furthermore, we will assume, as 
motivated in the introduction and usually done in
large--$N_c$ computations within the NJL model, that
\begin{equation}
  \label{AAA32}
  Z_{\Phi,k_\Phi}\ll1\; .
\end{equation}
As a consequence, the initial value of the renormalized Yukawa
coupling $h_{k_\Phi}$ $=$ $Z_{\Phi,k_\Phi}^{-1/2}$
$Z_{\psi,k_\Phi}^{-1}$ $\ol{h}_{k_\Phi}$ is much larger than one and
we will be able to exploit the infrared stable features of the flow
equations. As a typical coupling we take $h_{k_\Phi}=100$ in order to
simulate the limit $h_{k_\Phi}\to \infty$. The effective potential
$U_k(\Phi)$ must be invariant under the chiral $SU_L(N) \times
SU_R(N)$ flavor symmetry. In fact, the axial anomaly of QCD breaks the
Abelian $U_A(1)$ symmetry. The resulting $U_A(1)$ violating
multi--quark interactions\footnote{A first attempt for the computation
  of the anomaly term in the fermionic effective average action can be
  found in ref.~\cite{Paw96-1}.} lead to corresponding $U_A(1)$
violating terms in $U_k(\Phi)$.  Accordingly, the most general
effective potential $U_k$ is a function of the $N+1$ independent $C$
and $P$ conserving $SU_L(N)\times SU_R(N)$ invariants
\begin{equation}
  \begin{array}{rcl}
    \label{Invariants}
    \ds{\rho} &=&
    \ds{\tr\Phi^\dagger\Phi}\; ,\nnn
    \ds{\tau_i} &\sim& \ds{
      \tr\left(\Phi^\dagger\Phi- \frac{1}{N}\rho\right)^i\; ,\;\;\;
      i=2,\ldots,N}\; ,\nnn
    \ds{\xi} &=&
    \ds{\det\Phi+\det\Phi^\dagger}\; .
  \end{array}
\end{equation}
For a given initial form of $U_k$ all quantities in our truncation of
$\Gamma_k$ (\ref{AAA60}) are now fixed and we may follow the flow of
$\Gamma_k$ to $k \to 0$.  In this context it is important that the
formalism for composite fields~\cite{EW94-1} also induces an infrared
cutoff in the meson propagator. The flow equations are therefore
exactly of the form (\ref{ERGE}), with quarks and mesons treated on an
equal footing. At the compositeness scale the quadratic term of
$U_{k_\Phi}=\ol{m}^2_{k_\Phi}\Tr\Phi^\dagger\Phi+\ldots$ is positive
and the minimum of $U_{k_{\Phi}}$ therefore occurs for $\Phi=0$.
Spontaneous chiral symmetry breaking is described by a non--vanishing
expectation value $\VEV{\Phi}$ in absence of quark masses. This
follows from the change of the shape of the effective potential $U_k$
as $k$ flows from $k_\Phi$ to zero.  The large renormalized Yukawa
coupling rapidly drives the scalar mass term to negative values and
leads to a potential minimum away from the origin at some scale
$k_{\rm \chi SB}<k_\Phi$ such that finally
$\VEV{\Phi}=\ol{\sigma}_0\neq0$ for $k \to 0$ \cite{EW94-1,JW96-1}.
This concludes our overview of the general features of chiral symmetry
breaking in the context of flow equations for QCD.

We will concentrate in this work on the two flavor case ($N=2$) and
comment on the effects of including the strange quark in
section~\ref{AdditionalDegreesOfFreedom}. Furthermore we will neglect
isospin violation and therefore consider a singlet source term
$\jmath$ proportional to the average light current quark mass
$\hat{m}\equiv\frac{1}{2}(m_u+m_d)$.  Due to the $U_A(1)$--anomaly
there is a mass split for the mesons described by $\Phi$.  The scalar
triplet $(a_0)$ and the pseudoscalar singlet $(\eta^\prime)$ receive a
large mass whereas the pseudoscalar triplet $(\pi)$ and the scalar
singlet $(\sigma)$ remains light. From the measured values
$m_{\eta^\prime},m_{a_0}\simeq1\GeV$ it is evident that a decoupling
of these mesons is presumably a very realistic limit\footnote{In
  thermal equilibrium at high temperature this decoupling is not
  obvious. We will comment on this point in section
  \ref{AdditionalDegreesOfFreedom}.}.  It can be achieved in a
chirally invariant way and leads to the well known $O(4)$--symmetric
Gell-Mann--Levy linear sigma model~\cite{GML60-1} which is, however,
coupled to quarks now. This is the two flavor linear quark meson model
which we will study in the remainder of this work.  For this model the
effective potential $U_k$ is a function of $\rho$ only.

It remains to determine the source $\jmath$ as a function of the
average current quark mass $\hat{m}$. This is carried out in appendix
\ref{Source} and we obtain in our normalization with
$Z_{\psi,k_{\Phi}}=1$, $\ol{h}_{k_{\Phi}}=1$,
\begin{equation}
  \label{AAA101a}
  \jmath=2 \ol{m}^2_{k_{\Phi}}\hat{m}\; .
\end{equation}
It is remarkable that higher order terms do not influence the relation
between $\jmath$ and $\hat{m}$.  Only the quadratic term
$\ol{m}^2_{k_\Phi}$ enters which is in our scenario the only relevant
coupling. This feature is an important ingredient for the predictive
power of the model as far as the absolute size of the current quark
mass is concerned.

The quantities which are directly connected to chiral symmetry
breaking depend on the $k$--dependent expectation value
$\VEV{\Phi}_k=\ol{\sigma}_{0,k}$ as given by
\begin{equation}
  \label{AAA101}
  \frac{\prl U_k}{\prl\rho}(\rho=2\ol{\sigma}_{0,k}^2)=
  \frac{\jmath}{2\ol{\sigma}_{0,k}}.
\end{equation}
In terms of the renormalized expectation value
\begin{equation}
  \label{AAA100}
  \sigma_{0,k}=Z_{\Phi,k}^{1/2}\ol{\sigma}_{0,k}\; 
\end{equation}
we obtain the following expressions for phenomenological observables
from (\ref{AAA60}) for\footnote{We note that the expressions
  (\ref{AAA65}) obey the well known Gell-Mann--Oakes--Renner relation
  $m_\pi^2
  f_\pi^2=-2\hat{m}\VEV{\ol{\psi}\psi}+\Oc(\hat{m}^2)$~\cite{GMOR68-1}.}
$d=4$
\begin{equation}
  \label{AAA65}
  \begin{array}{rcl}
    \ds{f_{\pi,k}} &=& \ds{2\sigma_{0,k}}\; ,\nnn
    \ds{\VEV{\ol{\psi}\psi}_k} &=& \ds{
      -2\ol{m}^2_{k_\Phi}\left[Z_{\Phi,k}^{-1/2}
      \sigma_{0,k}-\hat{m}\right]}\; ,\nnn
      \ds{M_{q,k}} &=& \ds{
        h_k\sigma_{0,k}}\; ,\nnn
      \ds{m^2_{\pi,k}} &=& \ds{
        Z_{\Phi,k}^{-1/2}
        \frac{\ol{m}^2_{k_\Phi}\hat{m}}{\sigma_{0,k}}=
        Z_{\Phi,k}^{-1/2}\frac{\jmath}{2\sigma_{0,k}}}\; ,\nnn
      \ds{m_{\sigma,k}^2} &=& \ds{
        Z_{\Phi,k}^{-1/2}
        \frac{\ol{m}^2_{k_\Phi}\hat{m}}{\sigma_{0,k}}+
        4\lambda_k\sigma_{0,k}^2}\; .
  \end{array}
\end{equation}
Here we have defined the dimensionless, renormalized couplings
\begin{equation}
  \label{AAA102}
  \begin{array}{rcl}
    \ds{\lambda}_k &=& \ds{Z_{\Phi,k}^{-2}
      \frac{\prl^2U_k}{\prl\rho^2}(\rho=2\ol{\sigma}_{0,k}^2)}\; ,\nnn
    \ds{h_k} &=& \ds{
      Z_{\Phi,k}^{-1/2}Z_{\psi,k}^{-1}\ol{h}_k}\; .
  \end{array}
\end{equation}
We will mainly be interested in the ``physical values'' of the
quantities (\ref{AAA65}) in the limit $k\ra0$ where the infrared
cutoff is removed, i.e.\ $f_{\pi}=f_{\pi,k=0}$,
$m_{\pi}^2=m_{\pi,k=0}^2$, etc.  We point out that the formalism of
composite fields provides the link \cite{EW94-1} to the chiral
condensate $\VEV{\ol{\psi}\psi}$ since the expectation value
$\ol{\sigma}_0$ is related to the expectation value of a composite
quark--antiquark operator.

\subsection{Flow equations and infrared stability}
\label{FlowEquationsAndInfraredStability}

At first sight, a reliable computation of $\Gamma_{k\ra0}$ seems a
very difficult task. Without a truncation $\Gamma_k$ is described by
an infinite number of parameters (couplings, wave function
renormalizations, etc.) as can be seen if $\Gamma_k$ is expanded in
powers of fields and derivatives. For instance, the sigma mass is
obtained as a zero of the exact inverse propagator,
$\lim_{k\ra0}\Gamma_k^{(2)}(q)|_{\Phi=\VEV{\Phi}}$, which formally
receives contributions from terms in $\Gamma_k$ with arbitrarily high
powers of derivatives and the expectation value $\sigma_0$.  Realistic
nonperturbative truncations of $\Gamma_k$ which reduce the problem to
a manageable size are crucial.  We will follow here a twofold
strategy:
\begin{itemize}
\item Physical observables like meson masses, decay constants, etc.,
  can be expanded in powers of (current) quark masses in a similar way
  as in chiral perturbation theory~\cite{GL82-1}. To a given finite
  order of this expansion only a finite number of terms of a
  simultaneous expansion of $\Gamma_k$ in powers of derivatives and
  $\Phi$ are required if the expansion point is chosen properly.
  Details of this procedure and some results can be found
  in~\cite{JW96-2,JW96-3,JW97-1}.
\item Because of an approximate partial IR fixed point behavior of the
  flow equations in the symmetric regime, i.e. for $k_{\chi
    SB}<k<k_\Phi$, the values of many parameters of $\Gamma_k$ for
  $k\ra0$ will be almost independent of their initial values at the
  compositeness scale $k_{\Phi}$. For large enough $h_{k_\Phi}$ only a
  few relevant parameters need to be computed accurately from QCD.
  They can alternatively be determined from phenomenology. Because of
  the present lack of an explicit QCD computation we will pursue the
  latter approach.
\end{itemize}
In combination, these two points open the possibility for a perhaps
unexpected degree of predictive power within the linear quark meson
model.  We wish to stress, however, that a perturbative treatment of
the model at hand, e.g., using perturbative RG techniques, cannot be
expected to yield reliable results. The renormalized Yukawa coupling
is very large at the scale $k_\Phi$. Even the IR value of $h_k$ is
still relatively big
\begin{equation} 
  \label{IRh}
  h_{k=0}=\frac{2M_q}{f_\pi}\simeq6.5
\end{equation} 
and $h_k$ increases with $k$. The dynamics of the linear quark meson
model is therefore clearly nonperturbative for all scales $k\leq
k_\Phi$.

We will now turn to the flow equations for the linear quark meson
model.  We first note that the flow equations for $\Gamma_k$ and
$\Gamma_k-\frac{1}{2}\int d^4
x\tr\left(\jmath^\dagger\Phi+\Phi^\dagger\jmath\right)$ are identical.
The source term therefore does not need to be considered explicitly
and only appears in the condition (\ref{AAA101}) for $\VEV{\Phi}$.  It
is convenient to work with dimensionless and renormalized variables
therefore eliminating all explicit $k$--dependence. With
\begin{equation}
  \label{AAA190}
  u(t,\tilde{\rho})\equiv k^{-d}U_k(\rho)\; ,\;\;\;
  \tilde{\rho}\equiv Z_{\Phi,k} k^{2-d}\rho
\end{equation}
and using (\ref{AAA60}) as a first truncation of the effective average
action $\Gamma_k$ one obtains the flow equation ($t=\ln(k/k_\Phi)$)
\begin{equation}
  \label{AAA68}
  \begin{array}{rcl}
    \ds{\frac{\prl}{\prl t}u} &=& \ds{
      -d u+\left(d-2+\eta_\Phi\right)
      \tilde{\rho}u^\prime}\\[2mm]
    &+& \ds{
      2v_d\left\{
      3l_0^d(u^\prime;\eta_\Phi)+
      l_0^d(u^\prime+2\tilde{\rho}u^{\prime\prime};\eta_\Phi)-
      2^{\frac{d}{2}+1}N_c
      l_0^{(F)d}(\frac{1}{2}\tilde{\rho}h^2;\eta_\psi)
      \right\} }\; .
  \end{array}
\end{equation}
Here $v_d^{-1}\equiv2^{d+1}\pi^{d/2}\Gamma(d/2)$ and primes denote
derivatives with respect to $\tilde{\rho}$. The number of quark colors
is denoted as $N_c$. We will always use in the following $N_c=3$.
Eq.~(\ref{AAA68}) is a partial differential equation for the effective
potential $u(t,\tilde{\rho})$ which has to be supplemented by the flow
equation for the Yukawa coupling and expressions for the anomalous
dimensions $\eta_\Phi$, $\eta_\psi$. The symbols $l_n^d$, $l_n^{(F)d}$
denote bosonic and fermionic mass threshold functions, respectively,
which are defined in appendix \ref{ThresholdFunctions}.  They describe
the decoupling of massive modes and provide an important
nonperturbative ingredient. For instance, the bosonic threshold
functions
\begin{equation}
 \label{AAA85}
 l_n^d(w;\eta_\Phi)=\frac{n+\delta_{n,0}}{4}v_d^{-1}
 k^{2n-d}\int\frac{d^d q}{(2\pi)^d}
 \frac{1}{Z_{\Phi,k}}\frac{\prl R_k}{\prl t}
 \frac{1}{\left[ P(q^2)+k^2w\right]^{n+1}}
\end{equation}
involve the inverse average propagator
$P(q^2)=q^2+Z_{\Phi,k}^{-1}R_k(q^2)$ where the infrared cutoff is
manifest. These functions decrease $\sim w^{-(n+1)}$ for $w\gg1$.
Since typically $w=M^2/k^2$ with $M$ a mass of the model, the main
effect of the threshold functions is to cut off fluctuations of
particles with masses $M^2\gg k^2$. Once the scale $k$ is changed
below a certain mass threshold, the corresponding particle no longer
contributes to the evolution and decouples smoothly.

The dimensionless renormalized expectation value
$\kappa\equiv2k^{2-d}Z_{\Phi,k}\ol{\sigma}_{0,k}^2$, with
$\ol{\sigma}_{0,k}$ the $k$--dependent VEV of $\Phi$, may be computed
for each $k$ directly from the condition (\ref{AAA101})
\begin{equation}
  \label{AAA90}
  u^\prime(t,\kappa)=\frac{\jmath}{\sqrt{2\kappa}}
  k^{-\frac{d+2}{2}}Z_{\Phi,k}^{-1/2}\equiv
  \epsilon_g \; .
\end{equation}
Note that $\kappa\equiv0$ in the symmetric regime for vanishing source
term. Equation (\ref{AAA90}) allows us to follow the flow of $\kappa$
according to
\begin{eqnarray}
  \label{AAA91}
  \ds{\frac{d}{d t}\kappa} &=& \ds{
    \frac{\kappa}{\epsilon_g+2\kappa\lambda}
    \Bigg\{\left[\eta_\Phi-d-2\right]\epsilon_g-
    2\frac{\prl}{\prl t}u^\prime(t,\kappa)\Bigg\} }
\end{eqnarray}
with $\lambda\equiv u^{\prime\prime}(t,\kappa)$.  We define the Yukawa
coupling for $\tilde{\rho}=\kappa$ and its flow equation
reads~\cite{JW96-1}
\begin{equation}
  \label{AAA70}
  \begin{array}{rcl}
  \ds{\frac{d}{d t}h^2} &=& \ds{
  \left(d-4+2\eta_\psi+\eta_\Phi\right)h^2-
    2v_d h^4\Bigg\{
    3l_{1,1}^{(F B)d}(\frac{1}{2}h^2\kappa,
    \epsilon_g;\eta_\psi,\eta_\Phi) }\\[2mm]
  &-& \ds{
    l_{1,1}^{(F B)d}(\frac{1}{2}h^2\kappa,
    \epsilon_g+2\lambda\kappa;
    \eta_\psi,\eta_\Phi)
  \Bigg\} }\; .
  \end{array}
\end{equation}
Similarly, the the scalar and quark anomalous
dimensions are infered from
\begin{equation}
  \label{AAA69}
  \begin{array}{rcl}
    \ds{\eta_\Phi} &\equiv& \ds{
      -\frac{d}{d t}\ln Z_{\Phi,k}=
      4\frac{v_d}{d}\Bigg\{
      4\kappa\lambda^2
      m_{2,2}^d(\epsilon_g,\epsilon_g+2\lambda\kappa;
      \eta_\Phi) }\nnn
    &+& \ds{
      2^{\frac{d}{2}}N_c h^2
      m_4^{(F)d}(\frac{1}{2}h^2\kappa;
      \eta_\psi)
      \Bigg\}\; , }\nnn
    \ds{\eta_\psi} &\equiv& \ds{
      -\frac{d}{d t}\ln Z_{\psi,k}=
      2\frac{v_d}{d}h^2\Bigg\{
      3m_{1,2}^{(F B)d}(\frac{1}{2}h^2\kappa,
      \epsilon_g;\eta_\psi,\eta_\Phi) }\\[2mm]
    &+& \ds{
      m_{1,2}^{(F B)d}(\frac{1}{2}h^2\kappa,
      \epsilon_g+2\lambda\kappa;
      \eta_\psi,\eta_\Phi)
      \Bigg\}\; , }
  \end{array}
\end{equation}
which is a linear set of equations for the anomalous dimensions.  The
threshold functions $l_{n_1,n_2}^{(FB)d}$, $m_{n_1,n_2}^d$,
$m_4^{(F)d}$ and $m_{n_1,n_2}^{(FB)d}$ are also specified in appendix
\ref{ThresholdFunctions}.

The flow equations (\ref{AAA68}), (\ref{AAA91})---(\ref{AAA69}),
constitute a coupled system of ordinary and partial differential
equations which can be integrated numerically.  Here we take the
effective current quark mass dependence of $h_k$, $Z_{\Phi,k}$ and
$Z_{\psi,k}$ into account by stopping the evolution according to
eqs.~(\ref{AAA70}), (\ref{AAA69}), evaluated for the chiral limit,
below the pion mass $m_\pi$.  (For details of the algorithm used here
see refs.~\cite{ABBFTW95-1,BW97-1}.)  One finds for $d=4$ that chiral
symmetry breaking indeed occurs for a wide range of initial values of
the parameters including the presumably realistic case of large
renormalized Yukawa coupling and a bare mass $\ol{m}_{k_\Phi}$ of
order $100\MeV$. Driven by the strong Yukawa coupling, the
renormalized mass term $u^\prime(t,\tilde{\rho}=0)$ decreases rapidly
and goes through zero at a scale $k_{\chi{\rm SB}}$ not far below
$k_\Phi$.  Here the system enters the spontaneously broken regime and
the effective average potential develops an absolute minimum away from
the origin.  The evolution of of the potential minimum
$\sigma_{0,k}^2=k^2\kappa/2$ turns out to be reasonably stable already
before $k\simeq m_\pi$ where it stops. We take this result as an
indication that our truncation of the effective action $\Gm_k$ leads
at least qualitatively to a satisfactory description of chiral
symmetry breaking. The reason for the relative stability of the IR
behavior of the VEV (and all other couplings) is that the quarks
acquire a constituent mass $M_q=h\sigma_0\simeq300\MeV$ in the
spontaneously broken regime. As a consequence they decouple once $k$
becomes smaller than $M_q$ and the evolution is then dominantly driven
by the light Goldstone bosons.  This is also important for our
approximation of neglecting the residual gluonic interactions in the
quark sector of the model as outlined in
subsection~\ref{ASemiQuantitativePicture}.

Most importantly, one finds that the system of flow equations exhibits
an approximate IR fixed point behavior in the symmetric
regime~\cite{JW96-1}. To see this explicitly we study the flow
equations (\ref{AAA68}), (\ref{AAA91})---(\ref{AAA69}) subject to the
condition (\ref{AAA32}). For the relevant range of $\tilde{\rho}$ both
$u^\prime(t,\tilde{\rho})$ and $u^\prime(t,\tilde{\rho})+
2\tilde{\rho}u^{\prime\prime}(t,\tilde{\rho})$ are then much larger
than $\tilde{\rho}h^2(t)$ and we may therefore neglect in the flow
equations all scalar contributions with threshold functions involving
these large masses.  This yields the simplified equations
($d=4,v_4^{-1}=32\pi^2$)
\begin{equation}
  \label{AAA110}
  \begin{array}{rcl}
    \ds{\frac{\prl}{\prl t}u} &=& \ds{
      -4u+\left(2+\eta_\Phi\right)
      \tilde{\rho}u^\prime
      -\frac{N_c}{2\pi^2}
      l_0^{(F)4}(\frac{1}{2}\tilde{\rho}h^2)\; ,
      }\nnn
    \ds{\frac{d}{d t}h^2} &=& \ds{
      \eta_\Phi h^2 m\; ,
      }\nnn
      \ds{\eta_\Phi} &=& \ds{
        \frac{N_c}{8\pi^2}
        m_4^{(F)4}(0)h^2\; ,
        }\nnn
      \ds{\eta_\psi} &=& \ds{0}\; .
    \end{array}
\end{equation}
Of course, it should be clear that this approximation is only valid
for the initial range of running below $k_\Phi$ before the
(dimensionless) renormalized scalar mass squared
$u^\prime(t,\tilde{\rho}=0)$ approaches zero near the chiral symmetry
breaking scale.  The system (\ref{AAA110}) is exactly soluble. Using
$m_4^{(F)4}(0)=1$ which holds independently of the choice of the IR
cutoff we find
\begin{equation}
  \label{AAA113}
  \begin{array}{rcl}
    \ds{h^2(t)} &=& \ds{
      Z_\Phi^{-1}(t)=
      \frac{h_I^2}{1-\frac{N_c}{8\pi^2}h_I^2 t}\; ,
      }\nnn
    \ds{u(t,\tilde{\rho})} &=& \ds{
      e^{-4t}u_I(e^{2t}\tilde{\rho}\frac{h^2(t)}{h_I^2})-
      \frac{N_c}{2\pi^2}\int_0^t d r e^{-4r}
      l_0^{(F)4}(\frac{1}{2}h^2(t)\tilde{\rho}e^{2r}) }\; .
  \end{array}
\end{equation}
(The integration over $r$ on the right hand side of the solution for
$u$ can be carried out by first exchanging it with the one over
momentum implicit in the definition of the threshold function
$l_0^{(F)4}$ (see appendix \ref{ThresholdFunctions}).) Here
$u_I(\tilde{\rho})\equiv u(0,\tilde{\rho})$ denotes the effective
average potential at the compositeness scale and $h_I^2$ is the
initial value of $h^2$ at $k_\Phi$, i.e. for $t=0$.  For simplicity we
will use an expansion of the initial value effective potential
$u_I(\tilde{\rho})$ in powers of $\tilde{\rho}$ around
$\tilde{\rho}=0$
\begin{equation}
  \label{AAA140}
  u_I(\tilde{\rho})=
  \sum_{n=0}^\infty
  \frac{u_I^{(n)}(0)}{n!}\tilde{\rho}^n
\end{equation}
even though this is not essential for the forthcoming reasoning.
Expanding also $l_0^{(F)4}$ in eq.~(\ref{AAA113}) in powers of its
argument one finds for $n>2$
\begin{equation}
  \label{LLL00}
  \ds{\frac{u^{(n)}(t,0)}{h^{2n}(t)}} = \ds{
    e^{2(n-2)t}\frac{u_I^{(n)}(0)}{h_I^{2n}}+
    \frac{N_c}{\pi^2}
    \frac{(-1)^n (n-1)!}{2^{n+2}(n-2)}
    l_n^{(F)4}(0)
    \left[1-e^{2(n-2)t}\right]}\; .
\end{equation}
For decreasing $t\ra-\infty$ the initial values $u_I^{(n)}$ become
rapidly unimportant and $u^{(n)}/h^{2n}$ approaches a fixed point.
For $n=2$, i.e., for the quartic coupling, one finds
\begin{equation}
  \label{LLL01}
  \frac{u^{(2)}(t,0)}{h^2(t)}=
  1-\frac{1-\frac{u_I^{(2)}(0)}{h_I^2}}
  {1-\frac{N_c}{8\pi^2}h_I^2 t}
\end{equation}
leading to a fixed point value $(u^{(2)}/h^2)_*=1$. As a consequence
of this fixed point behavior the system looses all its ``memory'' on
the initial values $u_I^{(n\ge2)}$ at the compositeness scale
$k_\Phi$! This typically happens before the approximation
$u^\prime(t,\tilde{\rho}),u^\prime(t,\tilde{\rho})+
2\tilde{\rho}u^{\prime\prime}(t,\tilde{\rho})\gg\tilde{\rho}h^2(t)$
breaks down and the solution (\ref{AAA113}) becomes invalid.
Furthermore, the attraction to partial infrared fixed points continues
also for the range of $k$ where the scalar fluctuations cannot be
neglected anymore.  The initial value for the bare dimensionless mass
parameter
\begin{equation}
  \label{AAA142}
  \frac{u_I^\prime(0)}{h_I^2}=
  \frac{\ol{m}^2_{k_\Phi}}{k_\Phi^2}
\end{equation}
is never negligible. (In fact, using the values for
$\ol{m}^2_{k_\Phi}$ and $k_\Phi$ computed in ref.~\cite{EW94-1} one
obtains $\ol{m}^2_{k_\Phi}/k_\Phi^2\simeq0.036$.)  For large $h_I$
(and dropping the constant piece $u_I(0)$) the solution (\ref{AAA113})
therefore behaves with growing $\abs{t}$ as
\begin{equation}
  \label{AAA150}
  \begin{array}{rcl}
    \ds{Z_\Phi(t)} &\simeq& \ds{
      -\frac{N_c}{8\pi^2}t\; ,
      }\nnn
    \ds{h^2(t)} &\simeq& \ds{
      -\frac{8\pi^2}{N_c t}\; ,
      }\nnn
    \ds{u(t,\tilde{\rho})} &\simeq& \ds{
      \frac{u_I^\prime(0)}{h_I^2} e^{-2t} h^2(t)\tilde{\rho}
      -\frac{N_c}{2\pi^2}\int_0^t d r e^{-4r}
      l_0^{(F)4}(\frac{1}{2}h^2(t)\tilde{\rho}e^{2r}) }\; .  
  \end{array}
\end{equation}
In other words, for $h_I\ra\infty$ the IR behavior of the linear quark
meson model will depend (in addition to the value of the compositeness
scale $k_\Phi$ and the quark mass $\hat{m}$) only on one parameter,
$\ol{m}^2_{k_\Phi}$.  We have numerically verified this feature by
starting with different values for $u_I^{(2)}(0)$.  Indeed, the
differences in the physical observables were found to be small.  This
IR stability of the flow equations leads to a perhaps surprising
degree of predictive power!  For definiteness we will perform our
numerical analysis of the full system of flow equations (\ref{AAA68}),
(\ref{AAA91})---(\ref{AAA69}) with the idealized initial value
$u_I(\tilde{\rho})=u_I^\prime(0)\tilde{\rho}$ in the limit
$h_I^2\ra\infty$. It should be stressed, though, that deviations from
this idealization will lead only to small numerical deviations in the
IR behavior of the linear quark meson model as long as the condition
(\ref{AAA32}) holds, say for $h_I\gta15$~\cite{JW96-1}.

With this knowledge at hand we may now fix the remaining three
parameters of our model, $k_\Phi$, $\ol{m}^2_{k_\Phi}$ and $\hat{m}$ by
using $f_\pi=92.4\MeV$, the pion mass $M_\pi=135\MeV$ and the
constituent quark mass $M_q$ as phenomenological input.  Because of
the uncertainty regarding the precise value of $M_q$ we give in table
\ref{tab1} the results for several values of $M_q$.
\begin{table}
\begin{center}
\begin{tabular}{|c|c||c|c|c||c|c|c|c|} \hline
  $\frac{M_q}{\MeV}$ &
  $\frac{\lambda_I}{h_I^2}$ &
  $\frac{k_\Phi}{\MeV}$ &
  $\frac{\ol{m}^2_{k_\Phi}}{k_\Phi^2}$ &
  $\frac{\jmath^{1/3}}{\MeV}$ &
  $\frac{\hat{m}(k_\Phi)}{\MeV}$ &
  $\frac{\hat{m}(1\GeV)}{\MeV}$ &
  $\frac{\VEV{\ol{\psi}\psi}(1\GeV)}{\MeV^3}$ &
  $\frac{f_\pi^{(0)}}{\MeV}$
  \\[0.5mm] \hline\hline
  $303$ &
  $1$ &
  $618$ &
  $0.0265$ &
  $66.8$ &
  $14.7$ &
  $11.4$ &
  $-(186)^3$ &
  $80.8$
  \\ \hline
  $300$ &
  $0$ &
  $602$ &
  $0.026$ &
  $66.8$ &
  $15.8$ &
  $12.0$ &
  $-(183)^3$ &
  $80.2$
  \\ \hline
  $310$ &
  $0$ &
  $585$ &
  $0.025$ &
  $66.1$ &
  $16.9$ &
  $12.5$ &
  $-(180)^3$ &
  $80.5$
  \\ \hline
  $339$ &
  $0$ &
  $552$ &
  $0.0225$ &
  $64.4$ &
  $19.5$ &
  $13.7$ &
  $-(174)^3$ &
  $81.4$
  \\ \hline
\end{tabular}
\caption{\footnotesize The table shows the dependence on the
  constituent quark mass $M_q$ of the input parameters $k_\Phi$,
  $\ol{m}^2_{k_\Phi}/k_\Phi^2$ and $\jmath$ as well as some of our
  ``predictions''. The phenomenological input used here besides $M_q$
  is $f_\pi=92.4\MeV$, $m_\pi=135\MeV$.The first line corresponds to
  the values for $M_q$ and $\lambda_I$ used in the remainder of this
  work. The other three lines demonstrate the insensitivity of our
  results with respect to the precise values of these
  parameters.}
\label{tab1}
\end{center}
\end{table}
The first line of table~\ref{tab1} corresponds to the choice of $M_q$
and $\lambda_I\equiv u_I^{\prime\prime}(\kappa)$ which we will use for
the forthcoming analysis of the model at finite temperature.  As
argued analytically above the dependence on the value of $\lambda_I$
is weak for large enough $h_I$ as demonstrated numerically by the
second line. Moreover, we notice that our results, and in particular
the value of $\jmath$, are rather insensitive with respect to the
precise value of $M_q$. It is remarkable that the values for $k_\Phi$
and $\ol{m}_{k_\Phi}$ are not very different from those computed in
ref.~\cite{EW94-1}.  As compared to the analysis of ref.~\cite{JW96-1}
the present truncation of $\Gamma_k$ is of a higher level of accuracy:
We now consider an arbitrary form of the effective average potential
instead of a polynomial approximation and we have included the pieces
in the threshold functions which are proportional to the anomalous
dimensions. It is encouraging that the results are rather robust with
respect to these improvements of the truncation.

Once the parameters $k_\Phi$, $\ol{m}^2_{k_\Phi}$ and $\hat{m}$ are
fixed there are a number of ``predictions'' of the linear meson model
which can be compared with the results obtained by other methods or
direct experimental observation. First of all one may compute the
value of $\hat{m}$ at a scale of $1\GeV$ which is suitable for
comparison with results obtained from chiral perturbation
theory~\cite{Leu96-1} and sum rules~\cite{JM95-1}.  For this purpose
one has to account for the running of this quantity with the
normalization scale from $k_\Phi$ as given in table \ref{tab1} to the
commonly used value of $1\GeV$:
$\hat{m}(1\GeV)=A^{-1}\hat{m}(k_\Phi)$. A reasonable estimate of the
factor $A$ is obtained from the three loop running of $\hat{m}$ in the
$\ol{MS}$ scheme~\cite{JM95-1}. For $M_q\simeq300\MeV$ corresponding
to the first two lines in table~\ref{tab1} its value is $A\simeq1.3$.
The results for $\hat{m}(1\GeV)$ are in acceptable agreement with
recent results from other methods~\cite{Leu96-1,JM95-1} even though
they tend to be somewhat larger. Closely related to this is the value
of the chiral condensate
\begin{equation}
  \label{LLL04}
  \VEV{\ol{\psi}\psi}(1\GeV)\equiv
  -A\ol{m}^2_{k_\Phi}
  \left[f_\pi Z_{\Phi,k=0}^{-1/2}-
    2\hat{m}\right]
    \; .
\end{equation}
These results are quite non--trivial since not only $f_\pi$ and
$\ol{m}^2_{k_\Phi}$ enter but also the computed IR value
$Z_{\Phi,k=0}$.  We emphasize in this context that there may be
substantial corrections both in the extrapolation from $k_\Phi$ to
$1\GeV$ and in the factor $a_q$ (see~(\ref{PPF00})). The latter is due
to the neglected influence of the strange quark which may be important
near $k_\Phi$. These uncertainties have only little effect on the
physics at lower scales as relevant for our analysis of the
temperature effects. Only the value of $\jmath$ which is fixed by
$m_\pi$ enters here.

A further more qualitative test concerns the mass of the sigma
resonance or radial mode whose renormalized mass squared may be
computed according to (\ref{AAA65}) in the limit $k\ra0$. From our
numerical analysis we obtain $\lambda_{k=0}\simeq20$ which translates
into $m_\sigma\simeq 430\MeV$.  One should note, though, that this
result is presumably not very accurate as we have employed in this
work the approximation of using the Goldstone boson wave function
renormalization constant also for the radial mode. Furthermore, the
explicit chiral symmetry breaking contribution to $m_\sigma^2$ is
certainly underestimated as long as the strange quark is neglected.
In any case, we observe that the sigma meson is significantly heavier
than the pions. This is a crucial consistency check for the linear
quark meson model. A low sigma mass would be in conflict with the
numerous successes of chiral perturbation theory~\cite{GL82-1} which
requires the decoupling of all modes other than the Goldstone bosons
in the IR--limit of QCD. The decoupling of the sigma meson is, of
course, equivalent to the limit $\lambda\ra\infty$ which formally
describes the transition from the linear to the non--linear sigma
model and which appears to be reasonably well realized by the large
IR--values of $\lambda$ obtained in our analysis. We also note that
the issue of the sigma mass is closely connected to the value of
$f_\pi^{(0)}$, the value of $f_\pi$ in the chiral limit $\hat{m}=0$
also given in table~\ref{tab1}.  To lowest order in
$(f_\pi-f_\pi^{(0)})/f_\pi$ or, equivalently, in $\hat{m}$ one
has
\begin{equation}
  \label{ABC00}
  f_\pi-f_\pi^{(0)}=\frac{\jmath}{Z_\Phi^{1/2}m_\sigma^2}=
  \frac{f_\pi m_\pi^2}{m_\sigma^2}\; .
\end{equation}
A larger value of $m_\sigma$ would therefore reduce the difference
between $f_\pi^{(0)}$ and $f_\pi$.

\begin{figure}
\unitlength1.0cm
\begin{center}
\begin{picture}(13.,7.0)
\put(0.0,0.0){
\epsfysize=11.cm
\rotate[r]{\epsffile{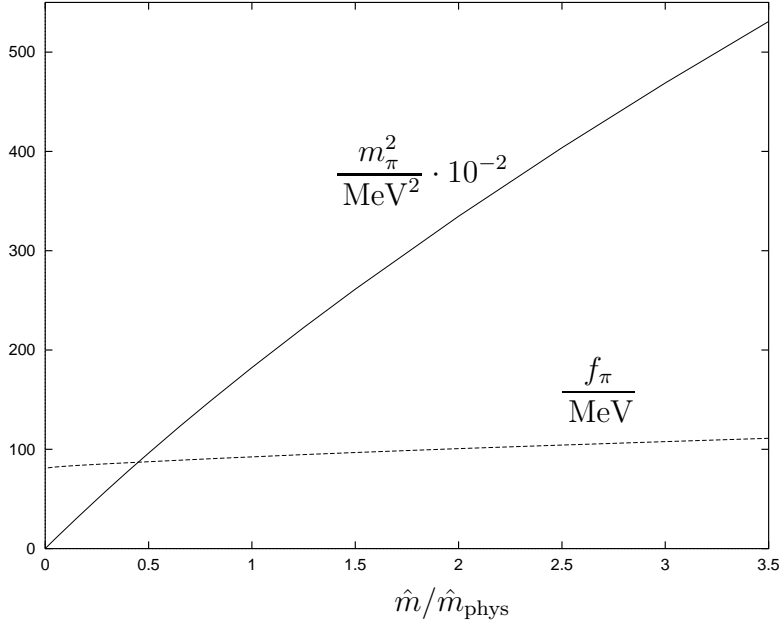}}
}
\put(8.0,2.3){\bf $\ds{\frac{f_\pi}{\MeV}}$}
\put(5.0,5.2){\bf $\ds{\frac{m^2_\pi}{\MeV^2}\cdot10^{-2}}$}
\put(5.8,-0.5){\bf $\ds{\hat{m}/\hat{m}_{\rm phys}}$}
\end{picture}
\end{center}
\caption{\footnotesize The plot shows $m_\pi^2$ (solid line) and
  $f_\pi$ (dashed line) as functions of the current quark mass
  $\hat{m}$ in units of the physical value $\hat{m}_{\rm phys}$.}
\label{mm}
\end{figure}

In figure~\ref{mm} we show the dependence of the pion mass and decay
constant on the average current quark mass $\hat{m}$.
These curves depend very little on the values of the initial
parameters as demonstrated in table~\ref{tab1} by $f_\pi^{(0)}$. We
observe a relatively large difference of $12\MeV$ between the pion
decay constants at $\hat{m}=\hat{m}_{\rm phys}$ and $\hat{m}=0$.
According to (\ref{ABC00}) this difference is related to the mass of
the sigma particle and will be modified in the three flavor case. We
will later find that the critical temperature $T_c$ for the second
order phase transition in the chiral limit is almost independent of
the initial conditions. The values of $f_\pi^{(0)}$ and $T_c$
essentially determine the non--universal amplitudes in the critical
scaling region (cf.~section~\ref{CriticalBehavior}). In summary, we
find that the behavior of our model for small $k$ is quite robust as
far as uncertainties in the initial conditions at the scale $k_\Phi$
are concerned. We will see that the difference of observables between
non--vanishing and vanishing temperature is entirely determined by the
flow of couplings in the range $0<k\lta3T$.

\sect{Thermal equilibrium and dimensional reduction}
\label{FiniteTemperatureFormalism}

The extension of flow equations to thermal equilibrium situations at
non--vanishing temperature $T$ is straightforward~\cite{TW93-1}. In
the Euclidean formalism non--zero temperature results in
(anti--)periodic boundary conditions for (fermionic) bosonic fields in
the Euclidean time direction with periodicity $1/T$~\cite{Kap89-1}.
This leads to the replacement
\begin{equation}
  \label{AAA120}
  \int\frac{d^d q}{(2\pi)^d}f(q^2)\ra
  T\sum_{l\in\ZZZ}\int\frac{d^{d-1}\vec{q}}{(2\pi)^{d-1}}
  f(q_0^2(l)+\vec{q}^{\,2})
\end{equation}
in the trace in (\ref{ERGE}) when represented as a momentum
integration, with a discrete spectrum for the zero component
\begin{equation}
  \label{AAA121}
  q_0(l)=\left\{
  \begin{array}{lll}
    2l\pi T &{\rm for}& {\rm bosons}\\
    (2l+1)\pi T &{\rm for}& {\rm fermions}\; .
  \end{array}\right.
\end{equation}
Hence, for $T>0$ a four--dimensional QFT can be interpreted as a
three--dimensional model with each bosonic or fermionic degree of
freedom now coming in an infinite number of copies labeled by
$l\in\ZZ$ (Matsubara modes). Each mode acquires an additional
temperature dependent effective mass term $q_0^2(l)$. In a high
temperature situation where all massive Matsubara modes decouple from
the dynamics of the system one therefore expects to observe an
effective three--dimensional theory with the bosonic zero modes as the
only relevant degrees of freedom. In other words, if the
characteristic length scale associated with the physical system is
much larger than the inverse temperature the compactified Euclidean
``time'' dimension cannot be resolved anymore. This phenomenon is
known as ``dimensional reduction''~\cite{Gin80-1}.

\begin{figure}
\unitlength1.0cm
\begin{center}
\begin{picture}(13.,18.0)

\put(0.0,9.5){
\epsfysize=11.cm
\rotate[r]{\epsffile{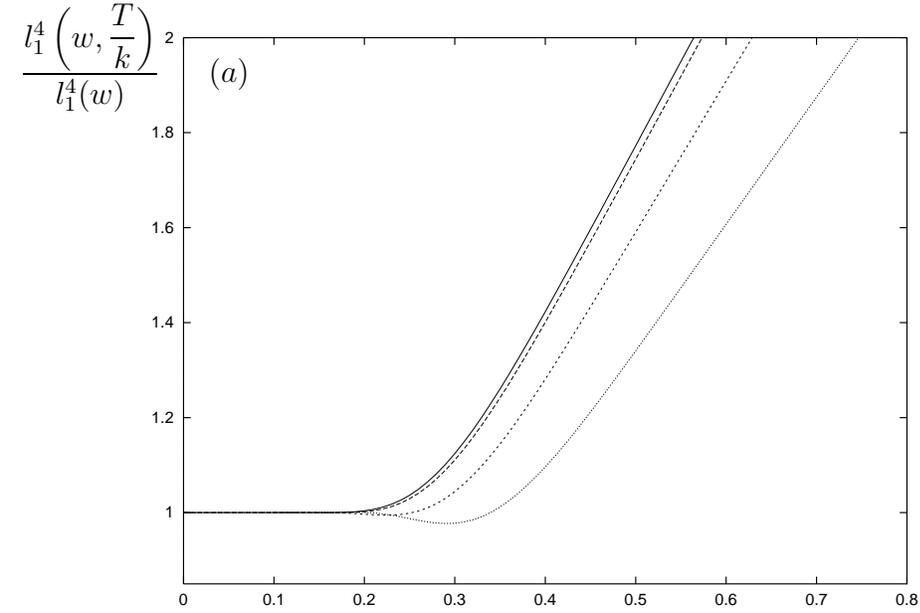}}
}
\put(-1.0,16.5){\bf $\ds{\frac{l_1^4\left(w,\ds{\frac{T}{k}}\right)}
    {l_1^4(w)}}$}
\put(1.5,16.5){\bf $\ds{(a)}$}

\put(0.0,0.5){
\epsfysize=11.cm
\rotate[r]{\epsffile{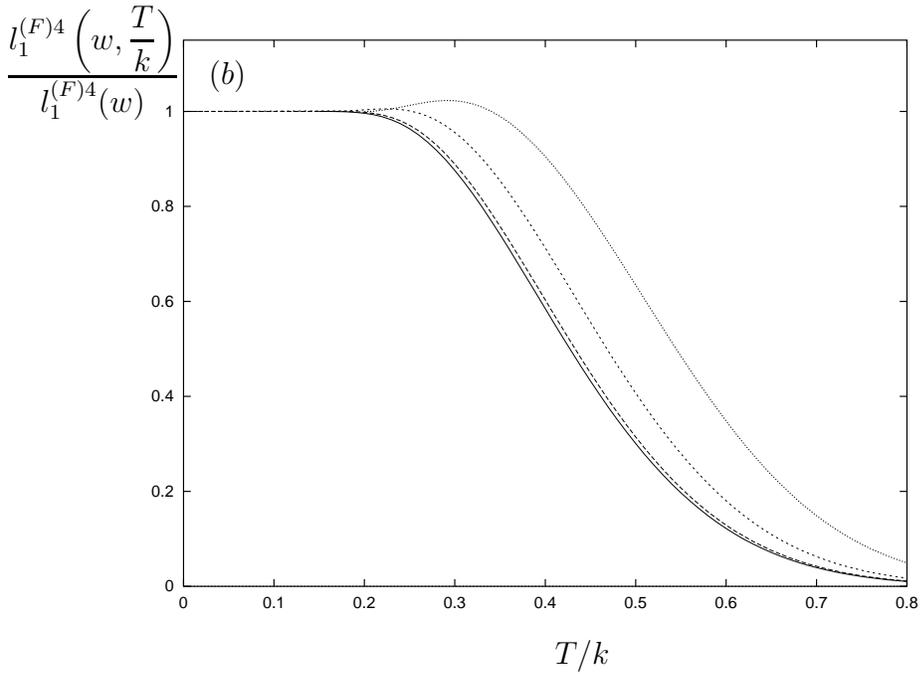}}
}
\put(-1.2,7.5){\bf
  $\ds{\frac{l_1^{(F)4}\left(w,\ds{\frac{T}{k}}\right)}
    {l_1^{(F)4}(w)}}$}
\put(6.1,-0.2){\bf $\ds{T/k}$}
\put(1.5,7.5){\bf $\ds{(b)}$}
\end{picture}
\end{center}
\caption{\footnotesize The plot shows the temperature 
  dependence of the bosonic (a) and the fermionic (b) threshold
  functions $l_1^4(w,T/k)$ and $l_1^{(F)4}(w,T/k)$, respectively, for
  different values of the dimensionless mass term $w$.  The solid line
  corresponds to $w=0$ whereas the dotted ones correspond to $w=0.1$,
  $w=1$ and $w=10$ with decreasing size of the dots.  For $T \gg k$
  the bosonic threshold function becomes proportional to $T/k$ whereas
  the fermionic one tends to zero.  In this range the theory with
  properly rescaled variables behaves as a classical
  three--dimensional theory.}
\label{Thresh}
\end{figure}

The formalism of the effective average action automatically provides
the tools for a smooth decoupling of the massive Matsubara modes as
the scale $k$ is lowered from $k\gg T$ to $k\ll T$.  It therefore
allows us to directly link the low--$T$, four--dimensional QFT to the
effective three--dimensional high--$T$ theory. The replacement
(\ref{AAA120}) in (\ref{ERGE}) manifests itself in the flow equations
(\ref{AAA68}), (\ref{AAA91})---(\ref{AAA69}) only through a change to
$T$--dependent threshold functions.  For instance, the dimensionless
functions $l_n^d(w;\eta_\Phi)$ defined in eq.~(\ref{AAA85}) are
replaced by
\begin{equation}
  \label{AAA200}
  l_n^d(w,\frac{T}{k};\eta_\Phi)\equiv
  \frac{n+\delta_{n,0}}{4}v_d^{-1}k^{2n-d}
  T\sum_{l\in\ZZZ}\int
  \frac{d^{d-1}\vec{q}}{(2\pi)^{d-1}}
  \left(\frac{1}{Z_{\Phi,k}}\frac{\prl R_k(q^2)}{\prl t}\right)
  \frac{1}{\left[P(q^2)+k^2 w\right]^{n+1}}
\end{equation}
where $q^2=q_0^2+\vec{q}^{\,2}$ and $q_0=2\pi l T$. A list of the
various temperature dependent threshold functions appearing in the
flow equations can be found in appendix~\ref{AnomalousDimensions}.
There we also discuss some subtleties regarding the definition of the
Yukawa coupling and the anomalous dimensions for $T\neq0$. In the
limit $k\gg T$ the sum over Matsubara modes approaches the integration
over a continuous range of $q_0$ and we recover the zero temperature
threshold function $l_n^d(w;\eta_\Phi)$.  In the opposite limit $k\ll
T$ the massive Matsubara modes ($l\neq0$) are suppressed and we expect
to find a $d-1$ dimensional behavior of $l_n^d$. In fact, one obtains
from~(\ref{AAA200})
\begin{equation}
  \label{AAA201}
  \begin{array}{rclcrcl}
    \ds{l_n^d(w,T/k;\eta_\Phi)} &\simeq& \ds{
      l_n^{d}(w;\eta_\Phi)}
    &{\rm for}& \ds{T\ll k}\; ,\nnn
    \ds{l_n^d(w,T/k;\eta_\Phi)} &\simeq& \ds{
      \frac{T}{k}\frac{v_{d-1}}{v_d}
      l_n^{d-1}(w;\eta_\Phi)}
    &{\rm for}& \ds{T\gg k}\; .
  \end{array}
\end{equation}
For our choice of the infrared cutoff function $R_k$,
eq.~(\ref{Rk(q)}), the temperature dependent Matsubara modes in
$l_n^d(w,T/k;\eta_\Phi)$ are exponentially suppressed for $T\ll k$
whereas the behavior is more complicated for other threshold functions
appearing in the flow equations (\ref{AAA68}),
(\ref{AAA91})---(\ref{AAA69}).  Nevertheless, all bosonic threshold
functions are proportional to $T/k$ for $T\gg k$ whereas those with
fermionic contributions vanish in this limit\footnote{For the present
  choice of $R_k$ the temperature dependence of the threshold
  functions is considerably smoother than in
  ref.~\cite{TW93-1}.}. This behavior is demonstrated in figure
\ref{Thresh} where we have plotted the quotients
$l_1^4(w,T/k)/l_1^4(w)$ and $l_1^{(F)4}(w,T/k)/l_1^{(F)4}(w)$ of
bosonic and fermionic threshold functions, respectively.
One observes that for $k\gg T$ both threshold functions essentially
behave as for zero temperature. For growing $T$ or decreasing $k$ this
changes as more and more Matsubara modes decouple until finally all
massive modes are suppressed. The bosonic threshold function $l^4_1$
shows for $k \ll T$ the linear dependence on $T/k$ derived in
eq.~(\ref{AAA201}).  In particular, for the bosonic excitations the
threshold function for $w\ll1$ can be approximated with reasonable
accuracy by $l_n^4(w;\eta_\Phi)$ for $T/k<0.25$ and by
$(4T/k)l_n^3(w;\eta_\Phi)$ for $T/k>0.25$. The fermionic threshold
function $l_1^{(F)4}$ tends to zero for $k\ll T$ since there is no
massless fermionic zero mode, i.e.~in this limit all fermionic
contributions to the flow equations are suppressed.  On the other
hand, the fermions remain quantitatively relevant up to $T/k\simeq0.6$
because of the relatively long tail in figure~\ref{Thresh}b.  The
transition from four to three--dimensional threshold functions leads
to a {\em smooth dimensional reduction} as $k$ is lowered from $k\gg
T$ to $k\ll T$!  Whereas for $k\gg T$ the model is most efficiently
described in terms of standard four--dimensional fields $\Phi$ a
choice of rescaled three--dimensional variables
$\Phi_{3}=\Phi/\sqrt{T}$ becomes better adapted for $k\ll T$.
Accordingly, for high temperatures one will use a potential
\begin{equation}
  \label{CCC01}
  u_{3}(t,\tilde{\rho}_{3})=\frac{k}{T}
  u(t,\tilde{\rho})\; ;\;\;\;
  \tilde{\rho}_{3}=\frac{k}{T}\tilde{\rho}\; .
\end{equation}
In this regime $\Gamma_{k\ra0}$ corresponds to the free energy of
classical statistics and $\Gamma_{k>0}$ is a classical coarse grained
free energy.

For our numerical calculations at non--vanishing temperature we
exploit the discussed behavior of the threshold functions by using the
zero temperature flow equations in the range $k\ge10T$. For smaller
values of $k$ we approximate the infinite Matsubara sums
(cf.~eq.~(\ref{AAA200})) by a finite series such that the numerical
uncertainty at $k=10T$ is better than $10^{-4}$. This approximation
becomes exact in the limit $k\ll10T$.

\sect{The quark meson model at $T\neq0$}
\label{TheQuarkMesonModelAtTNeq0}

In section \ref{ASemiQuantitativePicture} we have considered the
relevant fluctuations that contribute to the flow of $\Gamma_k$ in
dependence on the scale $k$. In a thermal equilibrium situation
$\Gamma_k$ also depends on the temperature $T$ and one may ask for the
relevance of thermal fluctuations at a given scale $k$.  In
particular, for not too high values of $T$
(cf.~sect.~\ref{AdditionalDegreesOfFreedom}) the ``initial condition''
$\Gamma_{k_\Phi}$ for the solution of the flow equations should
essentially be independent of temperature.  This will allow us to fix
$\Gamma_{k_\Phi}$ from phenomenological input at $T=0$ and to compute
the temperature dependent quantities in the infrared ($k \to 0$).  We
note that the thermal fluctuations which contribute to the r.h.s.\ of
the flow equation for the meson potential (\ref{AAA68}) are
effectively suppressed for $T \lta k/4$ (cf.~section
\ref{FiniteTemperatureFormalism}).  Clearly for $T \gta k_{\Phi}/3$
temperature effects become important at the compositeness scale. We
expect the linear quark meson model with a compositeness scale
$k_{\Phi} \simeq 600 \MeV$ to be a valid description for two flavor
QCD below a temperature of about\footnote{There will be an effective
  temperature dependence of $\Gamma_{k_{\Phi}}$ induced by the
  fluctuations of other degrees of freedom besides the quarks, the
  pions and the sigma which are taken into account here.  We will
  comment on this issue in section \ref{AdditionalDegreesOfFreedom}.
  For realistic three flavor QCD the thermal kaon fluctuations will
  become important for $T\gta170\MeV$.} $170 \MeV$.

We compute the quantities of interest for temperatures $T\lta170\MeV$
by solving numerically the $T$--dependent version of the flow
equations (\ref{AAA68}), (\ref{AAA91})---(\ref{AAA69})
(cf.~section~\ref{FiniteTemperatureFormalism} and
appendix~\ref{AnomalousDimensions}) by lowering $k$ from $k_\Phi$ to
zero. For this range of temperatures we use the initial values as
given in the first line of table \ref{tab1}.  This corresponds to
choosing the zero temperature pion mass and the pion decay constant
($f_{\pi}=92.4 \MeV$ for $m_{\pi}=135 \MeV$) as phenomenological
input. The only further input is the constituent quark mass $M_q$
which we vary in the range $M_q \simeq 300 - 350 \MeV$. We observe
only a minor dependence of our results on $M_q$ for the considered
range of values. In particular, the value for the critical temperature
$T_c$ of the model remains almost unaffected by this variation.

\begin{figure}
\unitlength1.0cm

\begin{center}
\begin{picture}(13.,7.0)

\put(0.0,0.0){
\epsfysize=11.cm
\rotate[r]{\epsffile{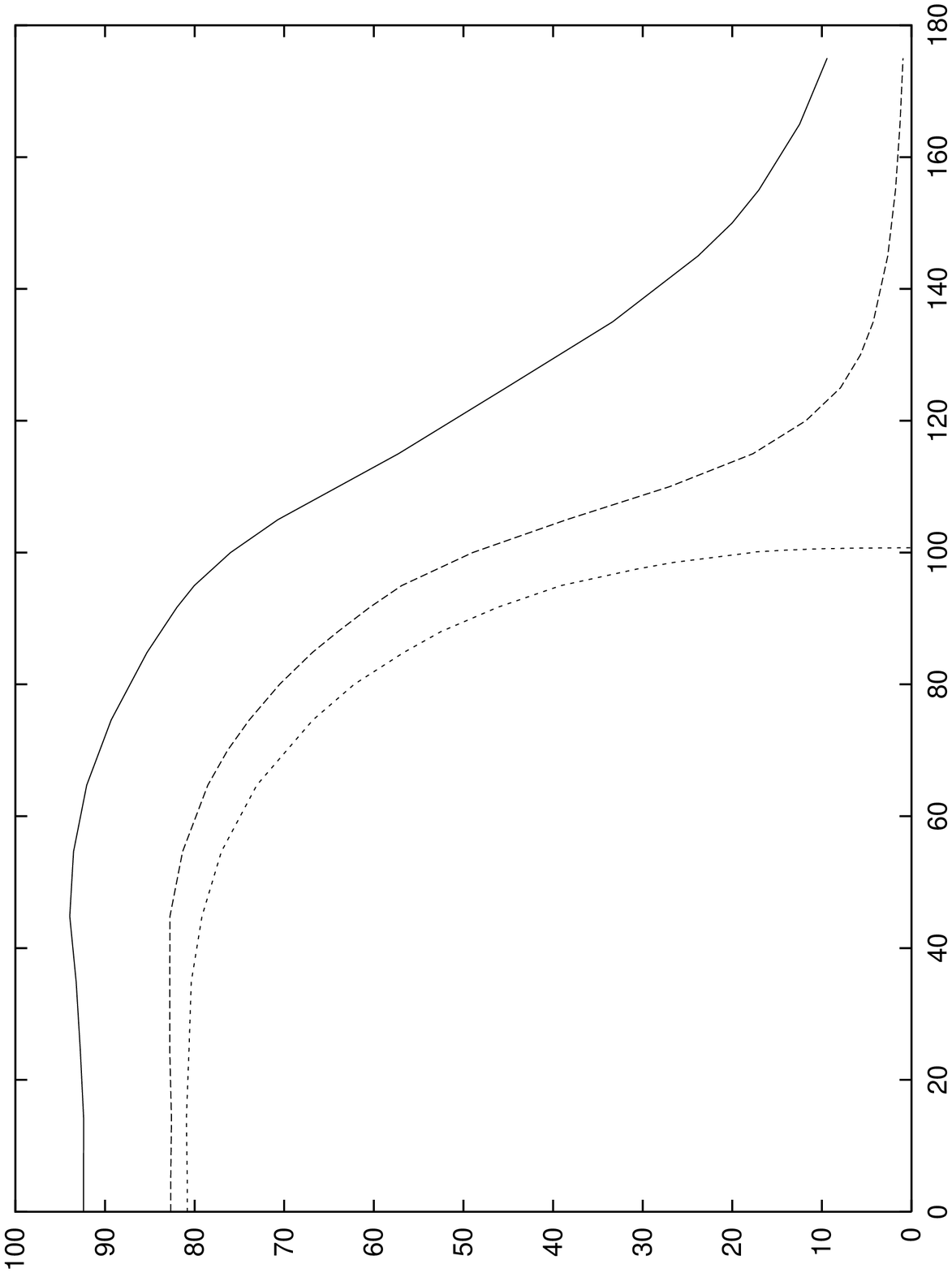}}
}
\put(-0.5,4.2){\bf $\ds{\frac{2\sigma_0(T)}{\MeV}}$}
\put(5.8,-0.5){\bf $\ds{T/\MeV}$}
\put(5.3,0.8){\footnotesize\bf $m_\pi=0$}
\put(8.2,0.8){\footnotesize\bf $m_\pi=45\MeV$}
\put(6.7,5.8){\footnotesize\bf $m_\pi=135\MeV$}

\end{picture}
\end{center}
\caption{\footnotesize The expectation value $2\sigma_0$ is shown as a
  function of temperature $T$ for three different values of the
  zero temperature pion mass.}
\label{fpi_T}
\end{figure}

We have plotted in figure \ref{fpi_T} the renormalized expectation
value $2\sigma_0$ of the scalar field as a function of temperature for
three different values of the average light current quark mass
$\hat{m}$. (We remind that $2\sigma_0(T=0)=f_{\pi}$.)
For $\hat{m}=0$ the order parameter $\sigma_0$ of chiral symmetry
breaking continuously goes to zero for $T\ra T_c = 100.7\MeV$
characterizing the phase transition to be of second order.  The
universal behavior of the model for small $T-T_c$ and small $\hat{m}$
is discussed in more detail in section \ref{CriticalBehavior}.  We
point out that the value of $T_c$ corresponds to
$f_\pi^{(0)}=80.8\MeV$, i.e.~the value of the pion decay constant for
$\hat{m}=0$, which is significantly lower than $f_\pi=92.4\MeV$
obtained for the realistic value $\hat{m}_{\rm phys}$.  If we would
fix the value of the pion decay constant to be $92.4\MeV$ also in the
chiral limit ($\hat{m}=0$), the value for the critical temperature
would raise to $115\MeV$.  The nature of the transition changes
qualitatively for $\hat{m}\neq0$ where the second order transition is
replaced by a smooth crossover.  The crossover for a realistic
$\hat{m}_{\rm phys}$ or $m_{\pi}(T=0)=135 \MeV$ takes place in a
temperature range $T \simeq(120-150)\MeV$.  The middle curve in figure
\ref{fpi_T} corresponds to a value of $\hat{m}$ which is only a tenth
of the physical value, leading to a zero temperature pion mass
$m_\pi=45\MeV$. Here the crossover becomes considerably sharper but
there remain substantial deviations from the chiral limit even for
such small quark masses $\hat{m}\simeq 1 \MeV$. The temperature
dependence of $m_\pi$ has already been mentioned in the introduction
(see fig.~\ref{mpi_T}) for the same three values of $\hat{m}$.  As
expected, the pions behave like true Goldstone bosons for $\hat{m}=0$,
i.e.~their mass vanishes for $T\le T_c$. Interestingly, $m_\pi$
remains almost constant as a function of $T$ for $T<T_c$ before it
starts to increase monotonically. We therefore find for two flavors no
indication for a substantial decrease of $m_\pi$ around the critical
temperature.

\begin{figure}
\unitlength1.0cm
\begin{center}
\begin{picture}(13.,7.0)

\put(0.0,0.0){
\epsfysize=11.cm
\rotate[r]{\epsffile{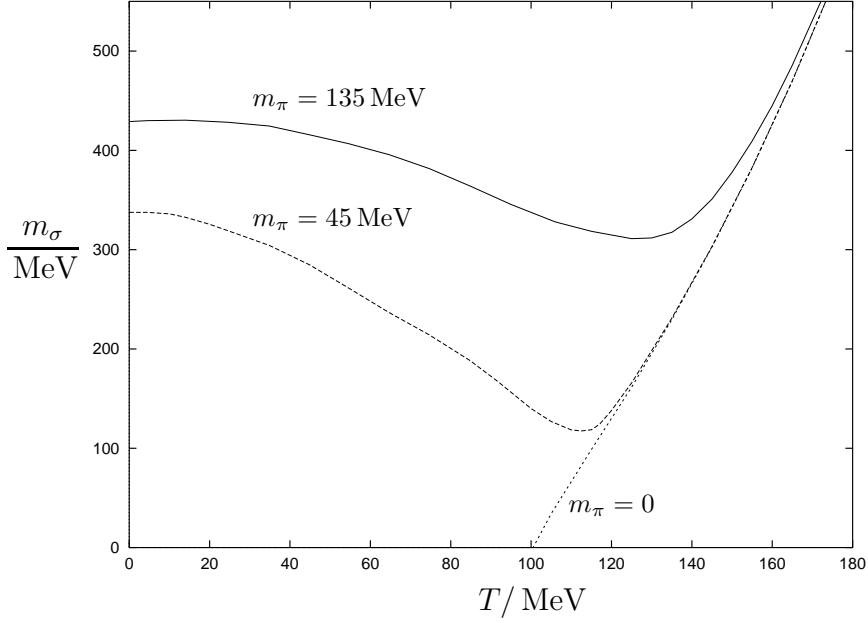}}
}
\put(-0.5,4.2){\bf $\ds{\frac{m_\sigma}{\MeV}}$}
\put(5.8,-0.5){\bf $\ds{T/\MeV}$}
\put(7.0,0.8){\footnotesize\bf $m_\pi=0$}
\put(2.8,4.6){\footnotesize\bf $m_\pi=45\MeV$}
\put(2.8,6.2){\footnotesize\bf $m_\pi=135\MeV$}

\end{picture}
\end{center}
\caption{\footnotesize The plot shows the $m_\sigma$ as a function of
  temperature $T$ for three different values of the 
  zero temperature pion mass.}
\label{ms_T}
\end{figure}

\begin{figure}
\unitlength1.0cm
\begin{center}
\begin{picture}(13.,7.0)
\put(0.0,0.0){
\epsfysize=11.cm
\rotate[r]{\epsffile{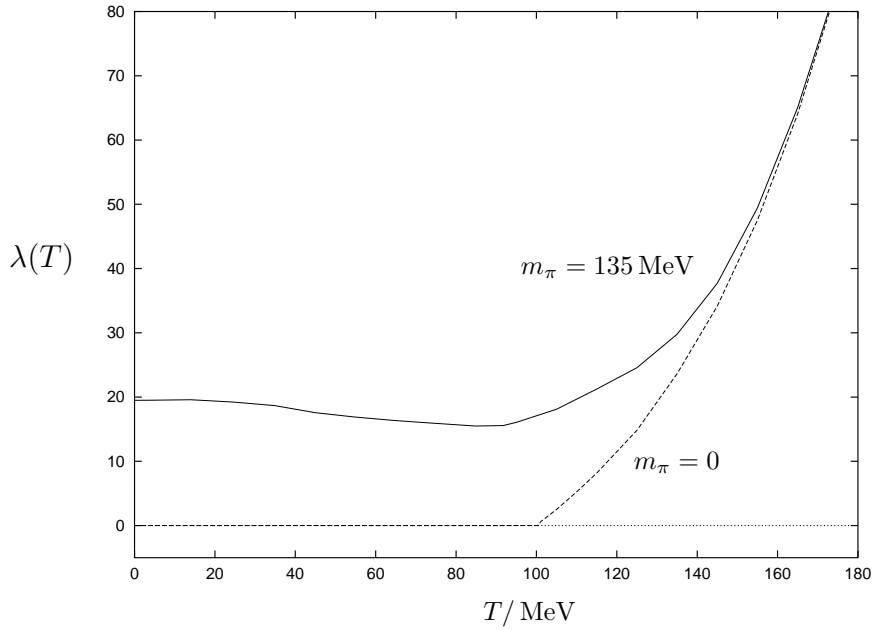}}
}
\put(-0.5,4.2){\bf $\ds{\lambda(T)}$}
\put(5.8,-0.5){\footnotesize\bf $\ds{T/\MeV}$}
\put(7.8,1.5){\footnotesize\bf $\ds{m_\pi=0}$}
\put(6.3,4.1){\footnotesize\bf $\ds{m_\pi=135\MeV}$}
\end{picture}
\end{center}
\caption{\footnotesize The plot shows the renormalized quartic scalar
  self coupling $\lambda$ as a function of temperature $T$ for the
  physical value of $\hat{m}$ (solid line) as well as for $\hat{m}=0$
  (dashed line).}
\label{lambda_T}
\end{figure}

\begin{figure}
\unitlength1.0cm
\begin{center}
\begin{picture}(13.,7.0)
\put(0.0,0.0){
\epsfysize=11.cm
\rotate[r]{\epsffile{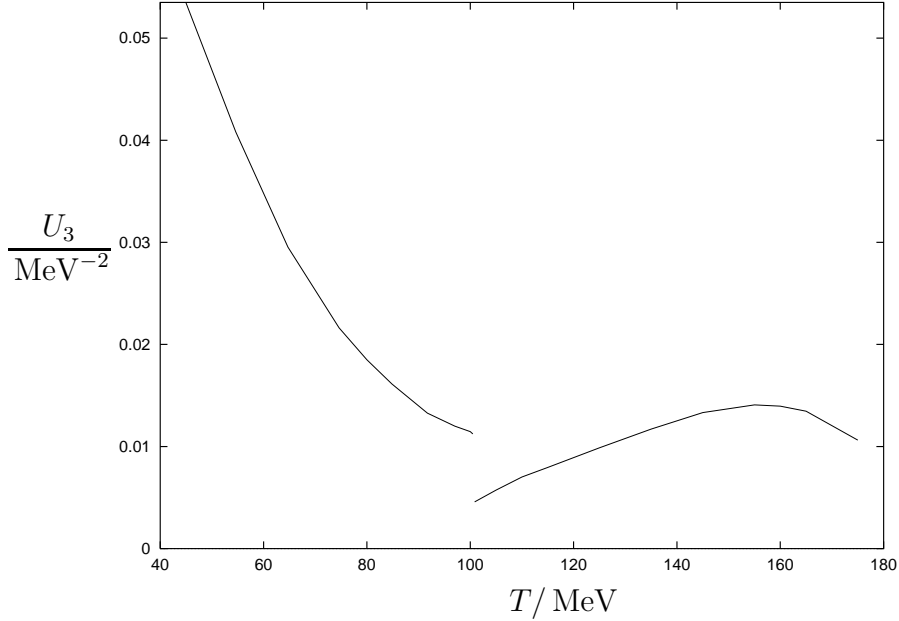}}
}
\put(-0.9,4.2){\bf $\ds{\frac{U_3}{\MeV^{-2}}}$}
\put(5.8,-0.5){\bf $\ds{T/\MeV}$}
\end{picture}
\end{center}
\caption{\footnotesize The plot shows the renormalized $\Phi^6$ scalar
  self coupling $U_3$ as a function of temperature $T$ in the chiral
  limit.}
\label{U3_T}
\end{figure}

The dependence of the mass of the sigma resonance $m_\sigma$ on the
temperature is displayed in figure~\ref{ms_T} for the above three
values of $\hat{m}$.
In the absence of explicit chiral symmetry breaking, $\hat{m}=0$, the
sigma mass vanishes for $T\le T_c$. For $T<T_c$ this is a consequence
of the presence of massless Goldstone bosons in the Higgs phase which
drive the renormalized quartic coupling $\lambda$ to zero.  In fact,
$\lambda$ runs linearly with $k$ for $T \gta k/4$ and only evolves
logarithmically for $T \lta k/4$.  Once $\hat{m}\neq0$ the pions
acquire a mass even in the spontaneously broken phase and the
evolution of $\lambda$ with $k$ is effectively stopped at $k\sim
m_\pi$. Because of the temperature dependence of $\sigma_{0,k=0}$
(cf.~figure~\ref{fpi_T}) this leads to a monotonically decreasing
behavior of $m_\sigma$ with $T$ for $T\lta T_c$. This changes into the
expected monotonic growth once the system reaches the symmetric phase
for\footnote{See section~\ref{FlowEquationsAndInfraredStability} for a
  discussion of the zero temperature sigma mass.} $T>T_c$. For low
enough $\hat{m}$ one may use the minimum of $m_{\sigma}(T)$ for an
alternative definition of the (pseudo-)critical temperature denoted as
$T_{pc}^{(2)}$. Table \ref{tab11} in the introduction shows our
results for the pseudocritical temperature for different values of
$\hat{m}$ or, equivalently, $m_{\pi}(T=0)$. For a zero temperature
pion mass $m_{\pi}=135 \MeV$ we find $T_{pc}^{(2)}=128 \MeV$. At
larger pion masses of about $230 \MeV$ we observe no longer a
characteristic minimum for $m_{\sigma}$ apart from a very broad,
slight dip at $T \simeq 90 \MeV$.  A comparison of our results with
lattice data is given in section \ref{CriticalBehavior}.  In
fig.~\ref{lambda_T} we show the renormalized quartic coupling
$\lambda$ as a function of temperature for two fixed values of the
average current quark mass.
The upper curve corresponds to
the physical value of $\hat{m}$ or, equivalently,  
$m_\pi(T=0)=135 \MeV$ whereas the lower
curve shows $\lambda$ for $\hat{m}=0$. One observes the 
vanishing of the renormalized quartic coupling in the chiral limit 
for $T \leq T_c$ as discussed above. The renormalized scalar $\Phi^6$
self interaction
\begin{equation}
  \label{ABC20}
  U_3(T)=Z_\Phi^{-3}
  \frac{\prl^3 U(\rho,T)}{\prl\rho^3}
  (\rho=2\ol{\sigma}_0^2(T))
\end{equation}
assumes a small negative value for realistic quark masses in the
temperature range $T\lta35\MeV$ with
$2U_3(T)\sigma_0^2(T)\simeq-0.5\ll\lambda(T)$ and
$2U_3(T)\sigma_0^2(T)\simeq8.0,8.5,1.5$ for $T=80,120,160\MeV$. We
display $U_3(T)$ in figure~\ref{U3_T} for the chiral limit where one
observes a discontinuity of $U_3(T)$ at the critical temperature
$T_c$.

\begin{figure}
\unitlength1.0cm
\begin{center}
\begin{picture}(13.,7.0)

\put(0.0,0.0){
\epsfysize=11.cm
\rotate[r]{\epsffile{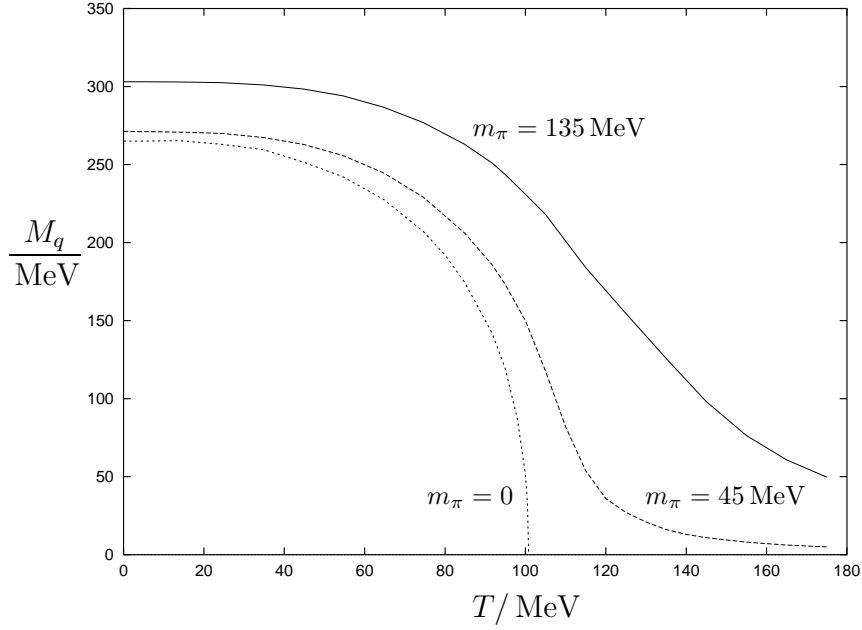}}
}
\put(-0.4,4.2){\bf $\ds{\frac{M_q}{\MeV}}$}
\put(5.8,-0.5){\bf $\ds{T/\MeV}$}
\put(5.2,1.0){\footnotesize\bf $m_\pi=0$}
\put(8.1,1.0){\footnotesize\bf $m_\pi=45\MeV$}
\put(5.8,5.9){\footnotesize\bf $m_\pi=135\MeV$}

\end{picture}
\end{center}
\caption{\footnotesize The plot shows the constituent quark mass $M_q$ 
  as a function of $T$ for three different values of the average light
  current quark mass $\hat{m}$. The solid line corresponds to the
  realistic value $\hat{m}=\hat{m}_{\rm phys}$ whereas the dotted line
  represents the situation without explicit chiral symmetry breaking,
  i.e., $\hat{m}=0$. The intermediate, dashed line assumes
  $\hat{m}=\hat{m}_{\rm phys}/10$.}
\label{mq_T}
\end{figure}

\begin{figure}
\unitlength1.0cm
\begin{center}
\begin{picture}(13.,7.0)
\put(0.0,0.0){
\epsfysize=11.cm
\rotate[r]{\epsffile{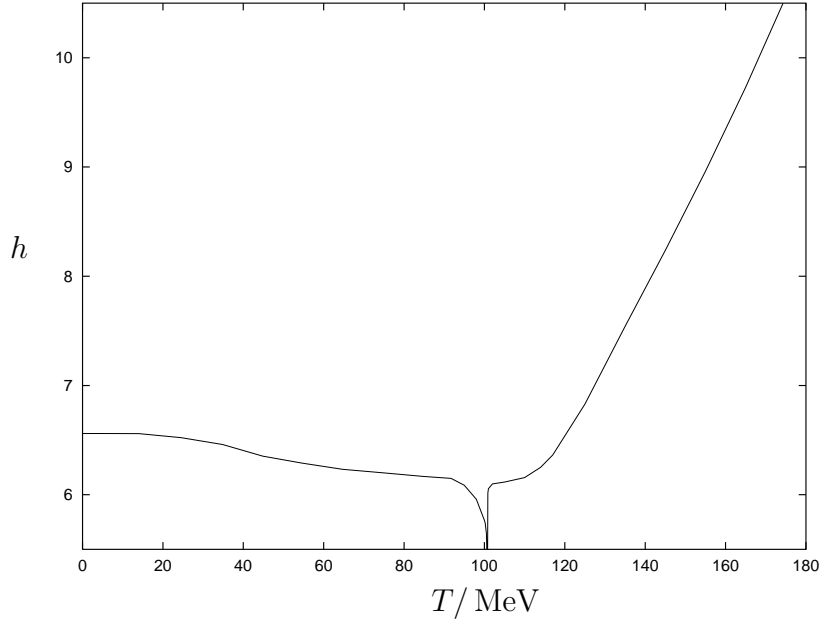}}
}
\put(0.2,4.2){\bf $\ds{h}$}
\put(5.8,-0.5){\bf $\ds{T/\MeV}$}
\end{picture}
\end{center}
\caption{\footnotesize The plot shows the Yukawa coupling, $h$, as a
  function of temperature $T$ in the chiral limit.}
\label{h2_T}
\end{figure}

\begin{figure}
\unitlength1.0cm
\begin{center}
\begin{picture}(13.,7.0)

\put(0.0,0.0){
\epsfysize=11.cm
\rotate[r]{\epsffile{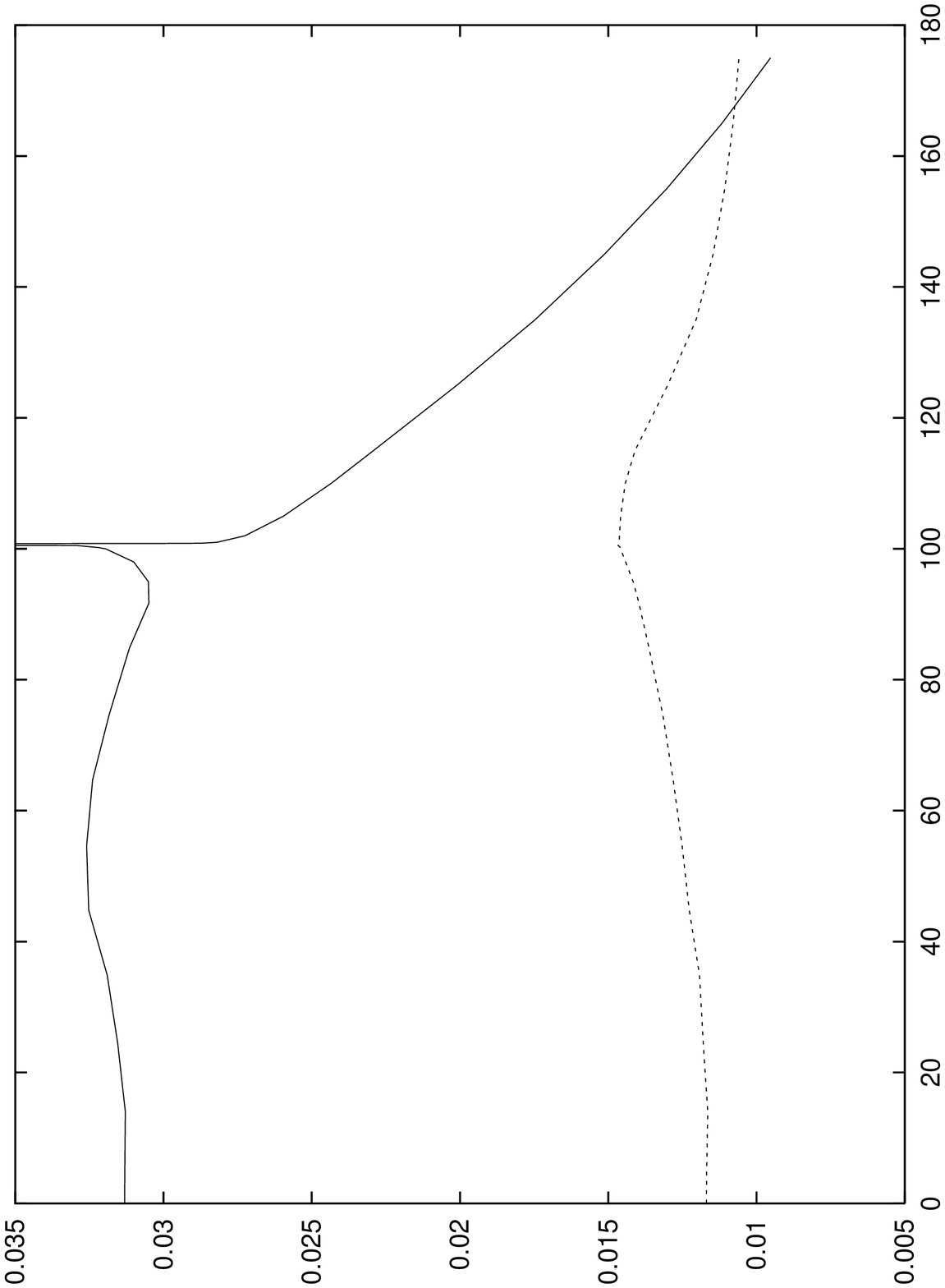}}
}
\put(5.8,-0.5){\bf $\ds{T/\MeV}$}
\put(4.8,2.7){\bf $\ds{Z_\psi\cdot10^{-2}}$}
\put(7.0,5.5){\bf $\ds{Z_\Phi}$}
\end{picture}
\end{center}
\caption{\footnotesize The plot shows the scalar (solid line) and
  quark (dashed line) wave function renormalization constants,
  $Z_\Phi(T)$ and $Z_\psi(T)\cdot10^{-2}$, respectively, as functions
  of temperature $T$ for $\hat{m}=0$.}
\label{Z_T}
\end{figure}

Our results for the chiral condensate $\VEV{\ol{\psi}\psi}$ as a
function of temperature for different values of the average current
quark mass are presented in figure~\ref{ccc_T} in the introduction. We
will compare $\VEV{\ol{\psi}\psi}(T,\hat{m})$ with its universal
scaling form for the $O(4)$ Heisenberg model in
section~\ref{CriticalBehavior}.  Another interesting quantity is the
temperature dependence of the constituent quark mass. Figure
\ref{mq_T} shows $M_q(T)$ for $\hat{m}=0$, $\hat{m}=\hat{m}_{\rm
  phys}/10$ and $\hat{m}=\hat{m}_{\rm phys}$, respectively.
Its behavior is related to the temperature dependence of the
renormalized order parameter $\sigma_{0}(T)\equiv\sigma_{0,k=0}(T)$
and the renormalized Yukawa coupling $h(T)\equiv h_{k=0}(T)$.  The
temperature dependence of $h$ in the chiral limit can be found in
fig.~\ref{h2_T}.
Near the critical temperature one notices a characteristic dip.  This
results from the long wavelength pion fluctuations through a
non--analytic behavior of the mesonic wave function renormalization
$Z_\Phi(T)\equiv Z_{\Phi,k=0}(T)$ which is displayed in
figure~\ref{Z_T}.
There we also present the temperature dependence of the fermionic wave
function renormalization $Z_\psi(T)\equiv Z_{\psi,k=0}(T)$.  Away from
the chiral limit we take the effective quark mass dependence of
$h_k(T)$, $Z_{\Phi,k}(T)$ and $Z_{\psi,k}(T)$ into account by stopping
their evolution when $k$ reaches the temperature dependent pion mass.
In this way we observe no substantial quark mass dependence of these
quantities except for $Z_\Phi(T)$, and consequently for $h(T)$, in the
vicinity of the critical temperature.  A more complete truncation
would incorporate field dependent wave function renormalization
constants and a field dependent Yukawa coupling.

\begin{figure}
\unitlength1.0cm
\begin{center}
\begin{picture}(13.,7.0)
\put(0.0,0.0){
\epsfysize=11.cm
\rotate[r]{\epsffile{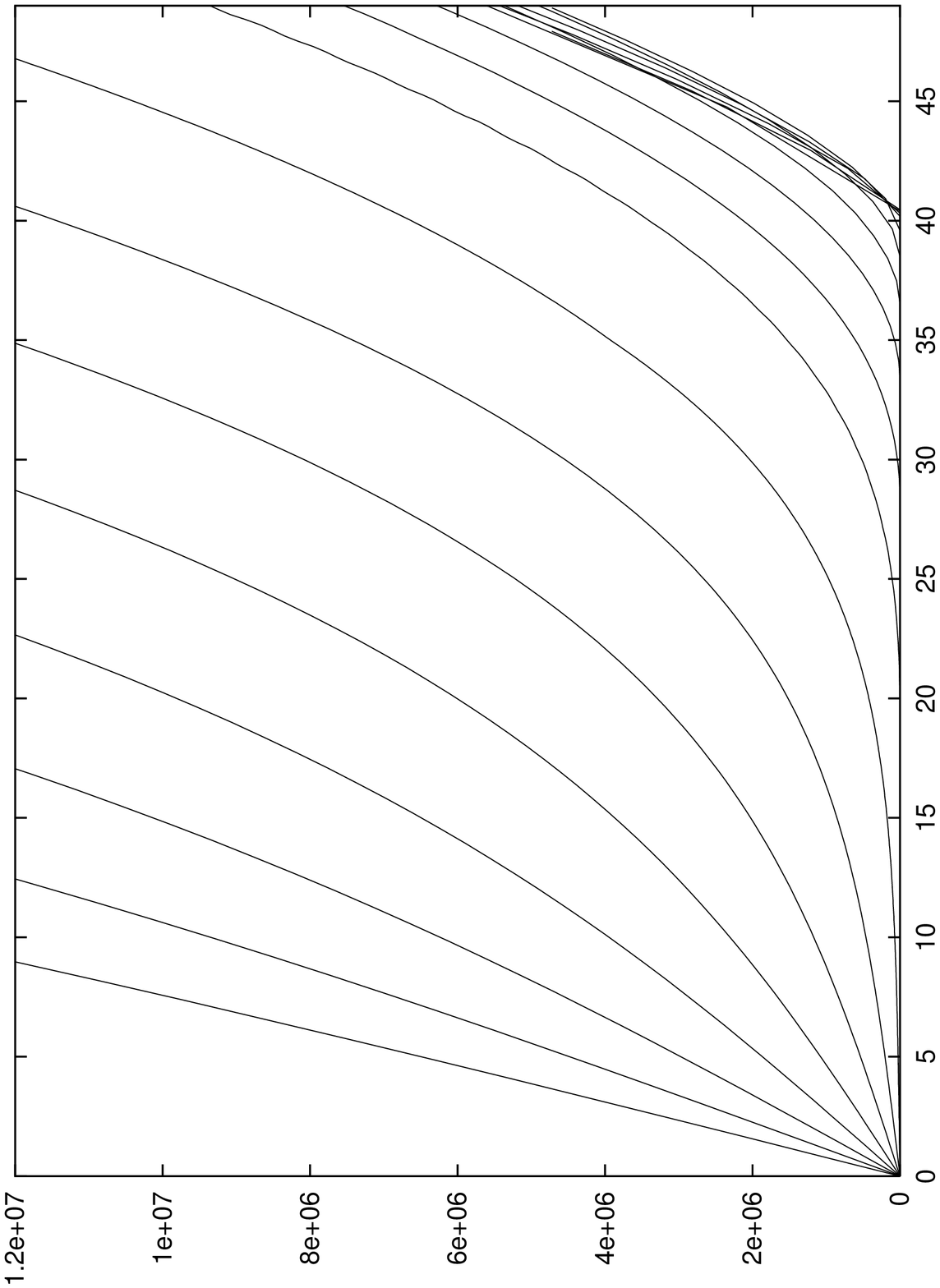}}
}
\put(-1.3,4.2){\bf $\ds{\frac{\partial U(T)/\partial \phi_R}
    {\MeV^{3}}}$}
\put(5.8,-0.5){\bf $\ds{\phi_R/\MeV}$}
\end{picture}
\end{center}
\caption{\footnotesize The plot shows the derivative of the
  meson potential $U(T)$ with respect to the renormalized field
  $\phi_R=(Z_\Phi\rho/2)^{1/2}$ for different values of $T$.  The
  first curve on the left corresponds to $T=175 \MeV$. The successive
  curves to the right differ in temperature by $\Delta T=10 \MeV$ down
  to $T=5 \MeV$. }
\label{Usig}
\end{figure}

Our ability to compute the complete temperature dependent effective
meson potential $U$ is demonstrated in fig.~\ref{Usig} where we
display the derivative of the potential with respect to the
renormalized field $\phi_R=(Z_\Phi\rho/2)^{1/2}$, for different values
of $T$.
The curves cover a temperature range $T = (5 - 175) \MeV$.  The first
one to the left corresponds to $T=175 \MeV$ and neighboring curves
differ in temperature by $\Delta T = 10 \MeV$. One observes only a
weak dependence of $\partial U(T)/\partial\phi_R$ on the temperature
for $T\lta60\MeV$.  Evaluated for $\phi_R=\sigma_{0}$ this
function connects the renormalized field expectation value with
$m_{\pi}(T)$, the source $\jmath$ and the mesonic wave function
renormalization $Z_{\Phi}(T)$ according to
\begin{equation}
  \label{Usigeq}
  \ds{\frac{\partial U(T)}{\partial\phi_R}}
  (\phi_R=\sigma_{0})=
  \ds{\frac{2\jmath}{Z_{\Phi}^{1/2}(T)}}=4 \sigma_{0}(T) 
  m_{\pi}^2(T) \; .
\end{equation} 

We close this section with a short assessment of the validity of our
effective quark meson model as an effective description of two flavor
QCD at non--vanishing temperature.  The identification of
qualitatively different scale intervals which appear in the context of
chiral symmetry breaking, as presented in section
\ref{ASemiQuantitativePicture} for the zero temperature case, can be
generalized to $T \neq 0$: For scales below $k_{\Phi}$ there exists a
hybrid description in terms of quarks and mesons. For $k_{\chi SB}
\leq k \lta 600 \MeV$ chiral symmetry remains unbroken where the
symmetry breaking scale $k_{\chi SB}(T)$ decreases with increasing
temperature. Also the constituent quark mass decreases with $T$
(cf.~figure~\ref{mq_T}). The running Yukawa coupling depends only
mildly on temperature for $T\lta120\MeV$ (see fig.~\ref{h2_T}).  (Only
near the critical temperature and for $\hat{m}=0$ the running is
extended because of massless pion fluctuations.) On the other hand,
for $k\lta4T$ the effective three--dimensional gauge coupling
increases faster than at $T=0$ leading to an increase of $\Lambda_{\rm
  QCD}(T)$ with $T$~\cite{RW93-1}. As $k$ gets closer to the scale
$\Lambda_{\rm QCD}(T)$ it is no longer justified to neglect in the
quark sector confinement effects which go beyond the dynamics of our
present quark meson model.  Here it is important to note that the
quarks remain quantitatively relevant for the evolution of the meson
degrees of freedom only for scales $k \gta T/0.6$
(cf.~fig.~\ref{Thresh}, section~\ref{FiniteTemperatureFormalism}).  In
the limit $k \ll T/0.6$ all fermionic Matsubara modes decouple from
the evolution of the meson potential according to the temperature
dependent version of eq.\ (\ref{AAA68}). Possible sizeable confinement
corrections to the meson physics may occur if $\Lambda_{\rm QCD}(T)$
becomes larger than the maximum of $M_q(T)$ and $T/0.6$. From
fig.~\ref{mq_T} we infer that this is particularly dangerous for small
$\hat{m}$ in a temperature interval around $T_c$. Nevertheless, the
situation is not dramatically different from the zero temperature case
since only a relatively small range of $k$ is concerned. We do not
expect that the neglected QCD non--localities lead to qualitative
changes.  Quantitative modifications, especially for small $\hat{m}$
and $\abs{T-T_c}$ remain possible. This would only effect the
non--universal amplitudes (see sect.~\ref{CriticalBehavior}). The size
of these corrections depends on the strength of (non--local)
deviations of the quark propagator and the Yukawa coupling from the
values computed in the quark meson model.

\sect{Universal critical behavior}
\label{CriticalBehavior}

In this section we study the linear quark meson model in the vicinity
of the critical temperature $T_c$ close to the chiral limit
$\hat{m}=0$. In this region we find that the sigma mass
$m_\sigma^{-1}$ is much larger than the inverse temperature $T^{-1}$,
and one observes an effectively three--dimensional behavior of the
high temperature quantum field theory.  We also note that the fermions
are no longer present in the dimensionally reduced system as has been
discussed in section \ref{FiniteTemperatureFormalism}. We therefore
have to deal with a purely bosonic $O(4)$--symmetric linear sigma
model.  At the phase transition the correlation length becomes
infinite and the effective three--dimensional theory is dominated by
classical statistical fluctuations. In particular, the critical
exponents which describe the singular behavior of various quantities
near the second order phase transition are those of the corresponding
classical system.

Many properties of this system are universal, i.e.~they only depend
on its symmetry ($O(4)$), the dimensionality of space (three) and its
degrees of freedom (four real scalar components). Universality means
that the long--range properties of the system do not depend on the
details of the specific model like its short distance
interactions. Nevertheless, important properties as the value of the
critical temperature are non--universal. We emphasize that although we
have to deal with an effectively three--dimensional bosonic theory,
the non--universal properties of the system crucially depend on the
details of the four--dimensional theory and, in particular, on the
fermions. 

Our aim is a computation of the critical equation of state which
relates for arbitrary $T$ near $T_c$ 
the derivative of the free energy or effective potential $U$
to the average current quark mass $\hat{m}$. The equation of state
then permits to study the temperature and quark mass dependence of
properties of the chiral phase transition.

At the critical temperature and in the chiral limit there is no scale
present in the theory. In the vicinity of $T_c$ and for small enough
$\hat{m}$ one therefore expects a scaling behavior of the effective
average potential $u_k$~\cite{TW94-1} and accordingly a universal
scaling form of the equation of state. There are only two independent
scales close to the transition point which can be related to the
deviation from the critical temperature, $T-T_c$, and to the explicit
symmetry breaking through the quark mass $\hat{m}$.  As a consequence,
the properly rescaled potential can only depend on one scaling
variable.  A possible choice for the parameterization of the rescaled
``unrenormalized'' potential is the use of the Widom scaling
variable~\cite{Wid65-1}
\begin{equation}
  \label{XXX20}
  x=\frac{\left( T-T_c\right)/T_c}
  {\left(2\ol{\sigma}_0/T_c\right)^{1/\beta}}\; .
\end{equation}
Here $\beta$ is the critical exponent of the order parameter
$\ol{\sigma}_0$ in the chiral limit $\hat{m}=0$ (see equation
(\ref{NNN21})).  With
$U^\prime(\rho=2\ol{\sigma}_0^2)=\jmath/(2\ol{\sigma}_0)$ the Widom
scaling form of the equation of state reads~\cite{Wid65-1}
\begin{equation}
  \label{XXX21}
  \frac{\jmath}{T_c^3}=
  \left(\frac{2\ol{\sigma}_0}{T_c}\right)^\delta f(x)
\end{equation}
where the exponent $\delta$ is related to the behavior of the order
parameter according to (\ref{NNN21b}).  The equation of state
(\ref{XXX21}) is written for convenience directly in terms of
four--dimensional quantities.  They are related to the corresponding
effective variables of the three--dimensional theory by appropriate
powers of $T_c$.  The source $\jmath$ is determined by the average
current quark mass $\hat{m}$ as $\jmath=2\ol{m}^2_{k_\Phi}\hat{m}$.
The mass term at the compositeness scale, $\ol{m}^2_{k_\Phi}$, also
relates the chiral condensate to the order parameter according to
$\VEV{\ol{\psi}\psi}=-2\ol{m}^2_{k_\Phi}(\ol{\sigma}_0-\hat{m})$.  The
critical temperature of the linear quark meson model was found in
section \ref{TheQuarkMesonModelAtTNeq0} to be $T_c=100.7\MeV$.

\begin{figure}
\unitlength1.0cm
\begin{center}
\begin{picture}(17.,12.)
\put(0.3,5.5){$\ds{\frac{2\ol{\sigma}_0/T_c}
{(\jmath/T_c^3 D)^{1/\delta}}}$}
\put(8.5,-0.2){$\ds{\frac{(T-T_c)/T_c}
{(\jmath/T_c^3 B^{\delta} D)^{1/\beta \delta}}}$}
\put(8.,2.5){\footnotesize $\mbox{average action}$}
\put(4.19,11.2){\footnotesize $\mbox{average action}$}
\put(13.8,2.2){\footnotesize $\epsilon$}
\put(3.53,11.19){\footnotesize $\epsilon$}
\put(11.3,2.05){\footnotesize $\mbox{MC}$}
\put(3.9,10.){\footnotesize $\mbox{MC}$}
\put(13.2,1.8){\footnotesize $\mbox{mf}$}
\put(6.5,9.2){\footnotesize $\mbox{mf}$}
\put(-1.,-5.9){
\epsfysize=21.cm
\epsfxsize=18.cm
\epsffile{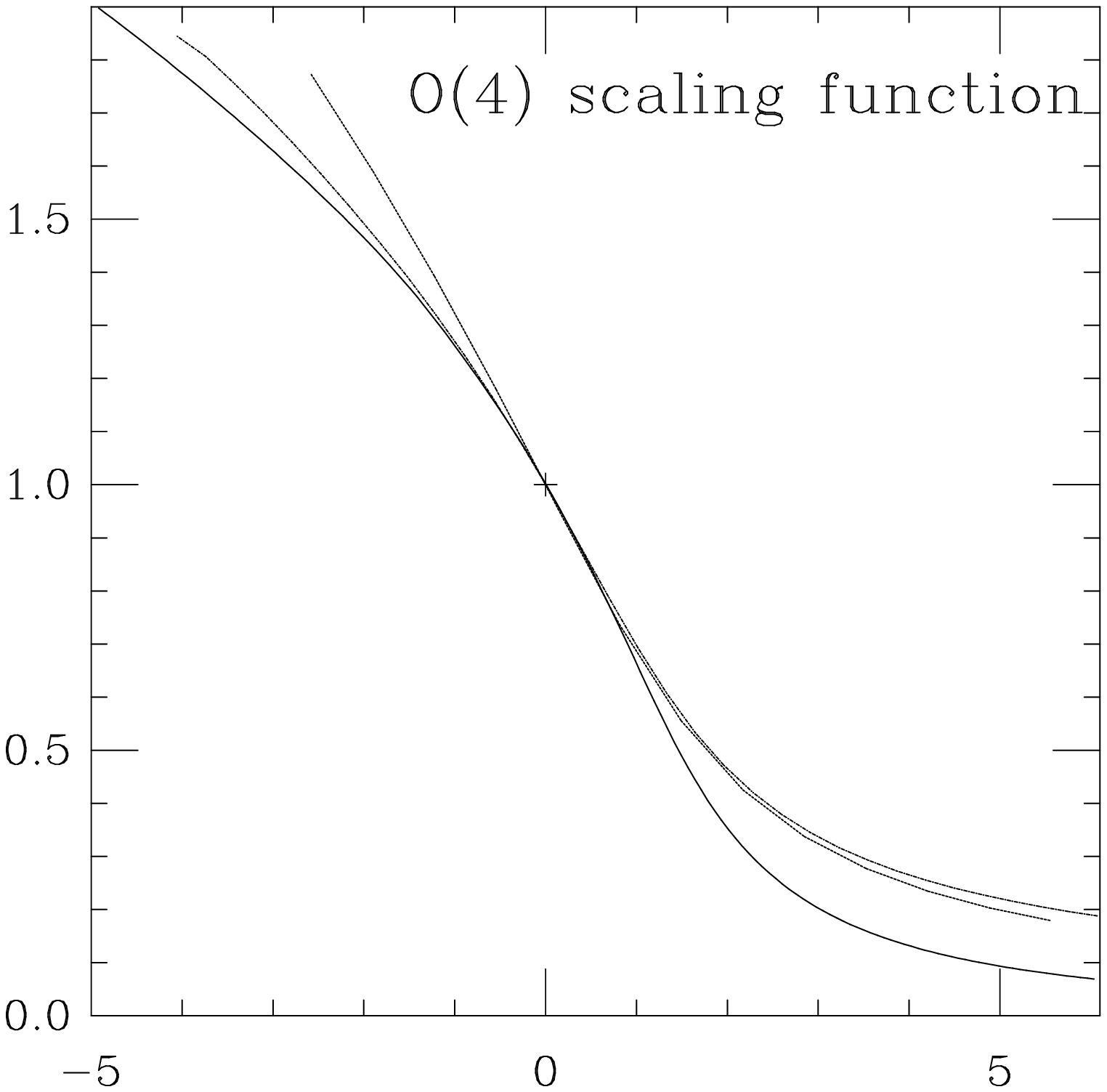}}
\put(1.25,0.483){
\epsfysize=13.45cm
\epsfxsize=11.22cm
\rotate[r]{\epsffile{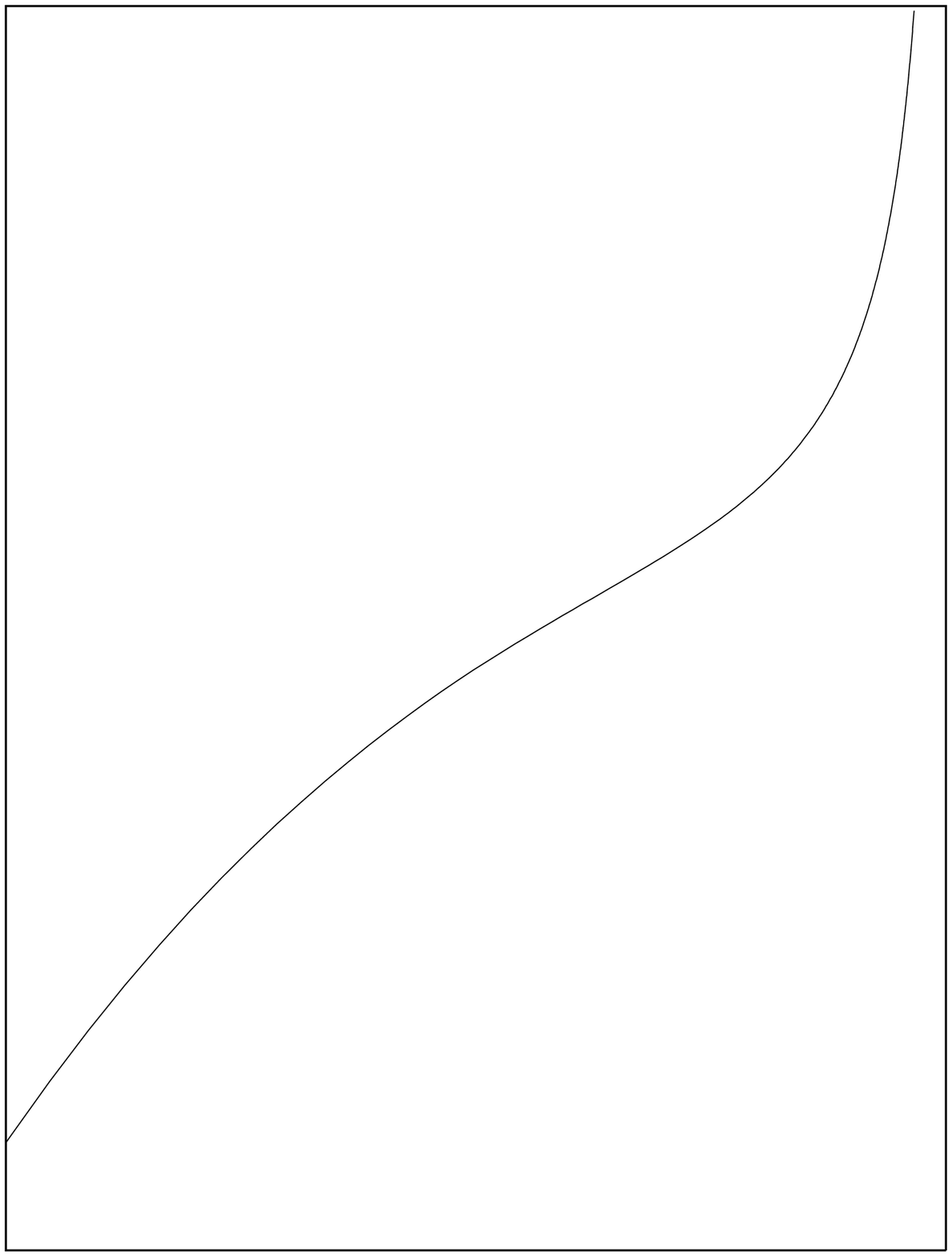}}}
\end{picture}
\end{center}
\caption[]{\footnotesize
  The figure shows a comparison of our results, denoted by ``average
  action'', with results of other methods for the scaling function of
  the three--dimensional $O(4)$ Heisenberg model. We have labeled the
  axes for convenience in terms of the expectation value
  $\ol{\sigma}_0$ and the source $\jmath$ of the corresponding
  four--dimensional theory.  The constants $B$ and $D$ specify the
  non--universal amplitudes of the model (cf.~eq.~\ref{norm}).  The
  curve labeled by ``MC'' represents a fit to lattice Monte Carlo
  data. The second order epsilon expansion \cite{BWW73-1} and mean
  field results are denoted by ``$\epsilon$'' and ``mf'',
  respectively.  Apart from our results the curves are taken
  from~\cite{Tou97-1}.
  \label{scalfunc}
  }
\end{figure}

The scaling function $f$ is universal up to the model specific
normalization of $x$ and itself. Accordingly, all models in the same
universality class can be related by a rescaling of $\ol{\sigma}_0$
and $T-T_c$. The non--universal normalizations for the quark meson
model discussed here are defined according to
\begin{equation}
  \label{norm}
  f(0)=D\quad, \qquad f(-B^{-1/\beta})=0\; .
\end{equation}
We find $D=1.82\cdot10^{-4}$, $B=7.41$ and our result for $\beta$ is
given in table~\ref{tab2}. Apart from the immediate vicinity of the
zero of $f(x)$ we find the following two parameter fit
(cf.~ref.~\cite{BTW96-1}) for the scaling function,
\begin{equation}
  \label{ffit}
  \begin{array}{rcl}
    \ds{f_{\rm fit}(x)}&=&\ds{1.816 \cdot 10^{-4} (1+136.1\, x)^2 \,
      (1+160.9\, \theta\,
      x)^{\Delta}}\nnn
    &&
    \ds{(1+160.9\, (0.9446\, \theta^{\Delta})^{-1/(\gamma-2-\Delta)} 
      \, x)^{\gamma-2-\Delta}}
  \end{array}
\end{equation}
to reproduce the numerical results for $f$ and $df/dx$ at the $1-2\%$
level with $\theta=0.625$ $(0.656)$, $\Delta=-0.490$ $(-0.550)$ for $x
> 0$ $(x < 0)$ and $\gamma$ as given in table \ref{tab2}.  The
universal properties of the scaling function can be compared with
results obtained by other methods for the three--dimensional $O(4)$
Heisenberg model.  In figure \ref{scalfunc} we display our results
along with those obtained from lattice Monte Carlo simulation
\cite{Tou97-1}, second order epsilon expansion \cite{BWW73-1} and mean
field theory.
We observe a good agreement of average action, lattice and epsilon
expansion results within a few per cent for $T < T_c$. Above $T_c$ the
average action and the lattice curve go quite close to each other with
a substantial deviation from the epsilon expansion and mean field
scaling function\footnote{ We note that the question of a better
  agreement of the curves for $T < T_c$ or $T > T_c$ depends on the
  chosen non--universal normalization conditions for $x$ and $f$ (cf.
  eq.\ (\ref{norm})).}.

Before we use the scaling function $f(x)$ to discuss the general
temperature and quark mass dependent case, we consider the limits
$T=T_c$ and $\hat{m}=0$, respectively.  In these limits the behavior
of the various quantities is determined solely by critical amplitudes
and exponents. In the spontaneously broken phase ($T<T_c$) and in the
chiral limit we observe that the renormalized and unrenormalized order
parameters scale according to
\begin{equation}
  \label{NNN21}
  \begin{array}{rcl}
    \ds{\frac{2\sigma_0(T)}{T_c}} &=& \ds{
      \left(2E\right)^{1/2}
        \left(\frac{T_c-T}{T_c}\right)^{\nu/2}
      }\; ,\nnn
    \ds{\frac{2\ol{\sigma}_0(T)}{T_c}} &=& \ds{
      B \left(\frac{T_c-T}{T_c}\right)^{\beta}
      }\; ,
  \end{array}
\end{equation}
respectively, with $E=0.814$ and the value of $B$ given above.  In the
symmetric phase the renormalized mass $m=m_\pi=m_\si$ and the
unrenormalized mass $\ol{m}=Z_\Phi^{1/2}m$ behave as
\begin{equation}
  \label{NNN21a}
  \begin{array}{rcl}
    \ds{\frac{m(T)}{T_c}} &=& \ds{
      \left(\xi^+\right)^{-1}
      \left(\frac{T-T_c}{T_c}\right)^\nu
      }\; ,\nnn
    \ds{\frac{\ol{m}(T)}{T_c}} &=& \ds{
      \left( C^+\right)^{-1/2}
      \left(\frac{T-T_c}{T_c}\right)^{\gamma/2}
       \; , }
  \end{array}
\end{equation}
where $\xi^+=0.270$, $C^+=2.79$. For $T=T_c$ and
non--vanishing current quark mass we have
\begin{equation}
  \label{NNN21b}
  \begin{array}{rcl}
    \ds{\frac{2\ol{\sigma}_0}{T_c}} &=& \ds{
      D^{-1/\delta}
        \left(\frac{\jmath}{T_c^3}\right)^{1/\delta}
      }
  \end{array}
\end{equation}
with the value of $D$ given above. 

Though the five amplitudes $E$, $B$, $\xi^+$, $C^+$ and $D$ are not
universal there are ratios of amplitudes which are invariant under a
rescaling of $\ol{\sigma}_0$ and $T-T_c$. Our results for the
universal amplitude ratios are
\begin{equation}
  \label{ABC01}
  \begin{array}{rcl}
    \ds{R_\chi} &=& \ds{C^+ D B^{\delta-1}=1.02}\; ,\nnn
    \ds{\tilde{R}_\xi} &=& \ds{
      (\xi^+)^{\beta/\nu}D^{1/(\delta+1)}B=0.852}\; ,\nnn
    \ds{\xi^+ E} &=& \ds{0.220}\; .
  \end{array}
\end{equation}
Those for the critical exponents are given in table \ref{tab2}.
\begin{table}
\begin{center}
\begin{tabular}{|c||l|l|l|l|l|} \hline
   &
  $\nu$ &
  $\gamma$ &
  $\delta$ &
  $\beta$ &
  $\eta$
  \\[0.5mm] \hline\hline
  average action &
  $0.787$ &
  $1.548$ &
  $4.80$ &
  $0.407$ &
  $0.0344$
  \\ \hline
  $\epsilon$ &
  $0.73(2)$ &
  $1.44(4)$ &
  $4.82(5)$ &
  $0.38(1)$ &
  $0.03(1)$
  \\ \hline
  MC &
  $0.7479(90)$ &
  $1.477(18)$ &
  $4.851(22)$ &
  $0.3836(46)$ &
  $0.0254(38)$
  \\ \hline
\end{tabular}
\caption[]{\footnotesize The table shows the critical exponents 
  corresponding to the three--dimensional $O(4)$--Heisenberg model.
  Our results are denoted by ``average action'' whereas ``$\epsilon$''
  labels the exponents obtained from the $4-\epsilon$ expansion to
  seven loops. The bottom line contains lattice Monte Carlo
  results.\cite{KK95-1}
  \label{tab2}}
\end{center}
\end{table}
Here the value of $\eta$ is obtained from the temperature dependent
version of eq.~(\ref{AAA69}) (cf.~appendix~\ref{AnomalousDimensions})
at the critical temperature. For comparison table~\ref{tab2} also
gives the results of the $4-\epsilon$ expansion to seven
loops~\cite{BMN78-1,Zin93-1}\footnote{The values are quoted e.g.~in
  ref.~\cite{KK95-1}.} as well as lattice Monte Carlo
results~\cite{KK95-1} which have been used for the lattice form of the
scaling function in fig.~\ref{scalfunc}.~\footnote{See also
  ref.~\cite{MT97-1} and references therein for a recent calculation
  of critical exponents using similar methods as in this work. For
  high precision estimates of the critical exponents see also
  ref.~\cite{BC95-1,Rei95-1}.} There are only two independent
amplitudes and critical exponents, respectively. They are related by
the usual scaling relations of the three--dimensional scalar
$O(N)$--model~\cite{Zin93-1} which we have explicitly verified by the
independent calculation of our exponents.

We turn to the discussion of the scaling behavior of the chiral
condensate $\VEV{\ol{\psi}\psi}$ for the general case of a temperature
and quark mass dependence.  In figure~\ref{ccc_T} in the introduction
we have displayed our results for the scaling equation of state in
terms of the chiral condensate\footnote{In the literature also a
  different definition of the chiral condensate is used, corresponding
  to $\VEV{\ol{\psi}\psi}= -\ol{m}^2_{k_\Phi}T_c[\jmath/(T_c^3
  f(x))]^{1/\delta}$.}
\begin{equation}
  \label{XXX30}
  \VEV{\ol{\psi}\psi}=
  -\ol{m}^2_{k_\Phi}T_c
  \left(\frac{\jmath/T_c^3}{f(x)}\right)^{1/\delta}+
    \jmath
\end{equation}
as a function of $T/T_c=1+x(\jmath/T_c^3 f(x))^{1/\beta\delta}$ for
different quark masses or, equivalently, different values of $\jmath$.
The curves shown in figure \ref{ccc_T} correspond to quark masses
$\hat{m}=0$, $\hat{m}=\hat{m}_{\rm phys}/10$, $\hat{m}=\hat{m}_{\rm
  phys}$ and $\hat{m}=3.5\hat{m}_{\rm phys}$ or, equivalently, to zero
temperature pion masses $m_\pi=0$, $m_\pi=45\MeV$, $m_\pi=135\MeV$ and
$m_\pi=230\MeV$, respectively (cf.~figure~\ref{mm}). One observes that
the second order phase transition with a vanishing order parameter at
$T_c$ for $\hat{m}=0$ is turned into a smooth crossover in the
presence of non--zero quark masses.

The scaling form (\ref{XXX30}) for the chiral condensate is exact only
in the limit $T\to T_c$, $\jmath\ra0$.  It is interesting to find the
range of temperatures and quark masses for which $\VEV{\ol{\psi}\psi}$
approximately shows the scaling behavior (\ref{XXX30}).  This can be
infered from a comparison (see fig.\ \ref{ccc_T}) with our full
non--universal solution for the $T$ and $\jmath$ dependence of
$\VEV{\ol{\psi}\psi}$ as described in
section~\ref{TheQuarkMesonModelAtTNeq0}. For $m_\pi=0$ one observes
approximate scaling behavior for temperatures $T\gta90\MeV$. This
situation persists up to a pion mass of $m_\pi=45\MeV$. Even for the
realistic case, $m_\pi=135\MeV$, and to a somewhat lesser extent for
$m_\pi=230\MeV$ the scaling curve reasonably reflects the physical
behavior for $T\gta T_c$. For temperatures below $T_c$, however, the
zero temperature mass scales become important and the scaling
arguments leading to universality break down.

The above comparison may help to shed some light on the use of
universality arguments away from the critical temperature and the
chiral limit. One observes that for temperatures above $T_c$ the
scaling assumption leads to quantitatively reasonable results even for
a pion mass almost twice as large as the physical value. This in turn
has been used for two flavor lattice QCD as theoretical input to guide
extrapolation of results to light current quark masses.  From
simulations based on a range of pion masses $0.3\lta
m_\pi/m_\rho\lta0.7$ and temperatures $0<T\lta250\MeV$ a
``pseudocritical temperature'' of approximately $140\MeV$ with a weak
quark mass dependence is reported~\cite{MILC97-1}. Here the
``pseudocritical temperature'' $T_{pc}$ is defined as the inflection
point of $\VEV{\ol{\psi}\psi}$ as a function of temperature.  The
values of the lattice action parameters used in~\cite{MILC97-1} with
$N_t=6$ were $a\hat{m}=0.0125$, $6/g^2=5.415$ and $a\hat{m}=0.025$,
$6/g^2=5.445$. For comparison with lattice data we have displayed in
figure \ref{ccc_T} the temperature dependence of the chiral condensate
for a pion mass $m_\pi=230\MeV$.  From the free energy of the linear
quark meson model we obtain in this case a pseudocritical temperature
of about $150\MeV$ in reasonable agreement with the results of
ref.~\cite{MILC97-1}.  In contrast, for the critical temperature in
the chiral limit we obtain $T_c=100.7\MeV$.  This value is
considerably smaller than the lattice results of about $(140 - 150)
\MeV$ obtained by extrapolating to zero quark mass in
ref.~\cite{MILC97-1}.  We point out that for pion masses as large as
$230\MeV$ the condensate $\VEV{\ol{\psi}\psi}(T)$ is almost linear
around the inflection point for quite a large range of temperature.
This makes a precise determination of $T_c$ somewhat difficult.
Furthermore, figure \ref{ccc_T} shows that the scaling form of
$\VEV{\ol{\psi}\psi}(T)$ underestimates the slope of the physical
curve. Used as a fit with $T_c$ as a parameter this can lead to an
overestimate of the pseudocritical temperature in the chiral limit.
We also mention here the results of ref.~\cite{Got97-1}.  There two
values of the pseudocritical temperature, $T_{pc}=150(9)\MeV$ and
$T_{pc}=140(8)$, corresponding to $a\hat{m}=0.0125$, $6/g^2=5.54(2)$
and $a\hat{m}=0.00625$, $6/g^2=5.49(2)$, respectively, (both for
$N_t=8$) were computed.  These values show a somewhat stronger quark
mass dependence of $T_{pc}$ and were used for a linear extrapolation
to the chiral limit yielding $T_c=128(9)\MeV$.

The linear quark meson model exhibits a second order phase transition
for two quark flavors in the chiral limit. As a consequence the model
predicts a scaling behavior near the critical temperature and the
chiral limit which can, in principle, be tested in lattice
simulations. For the quark masses used in the present lattice studies
the order and universality class of the transition in two flavor QCD
remain a partially open question. Though there are results from the
lattice giving support for critical scaling~\cite{Kar94-1,IKKY97-1}
there are also recent simulations with two flavors that reveal
significant finite size effects and problems with
$O(4)$ scaling~\cite{BKLO96-1,Uka97-1}.

\sect{Additional degrees of freedom}
\label{AdditionalDegreesOfFreedom}

So far we have investigated the chiral phase transition of QCD as
described by the linear $O(4)$--model containing the three pions and
the sigma resonance as well as the up and down quarks as degrees of
freedom. Of course, it is clear that the spectrum of QCD is much
richer than the states incorporated in our model. It is therefore
important to ask to what extent the neglected degrees of freedom like
the strange quark, strange (pseudo)scalar mesons, (axial)vector
mesons, baryons, etc., might be important for the chiral dynamics of
QCD.  Before doing so it is perhaps instructive to first look into the
opposite direction and investigate the difference between the linear
quark meson model described here and chiral perturbation theory based
on the non--linear sigma model~\cite{GL82-1}.  In some sense, chiral
perturbation theory is the minimal model of chiral symmetry breaking
containing only the Goldstone degrees of freedom. By construction it
is therefore only valid in the spontaneously broken phase and can not
be expected to yield realistic results for temperatures close to $T_c$
or for the symmetric phase.  However, for small temperatures (and
momentum scales) the non--linear model is expected to describe the
low--energy and low--temperature limit of QCD reliably as an expansion
in powers of the light quark masses. For vanishing temperature it has
been demonstrated recently~\cite{JW96-3,JW97-1} that the results of
chiral perturbation theory can be reproduced within the linear meson
model once certain higher dimensional operators in its effective
action are taken into account for the three flavor case.  Moreover,
some of the parameters of chiral perturbation theory
($L_4,\ldots,L_8$) can be expressed and therefore also numerically
computed in terms of those of the linear model. For non--vanishing
temperature one expects agreement only for low $T$ whereas deviations
from chiral perturbation theory should become large close to $T_c$.
Yet, even for $T\ll T_c$ small quantitative deviations should exist
because of the contributions of (constituent) quark and sigma meson
fluctuations in the linear model which are not taken into account in
chiral perturbation theory.

\begin{figure}
\unitlength1.0cm
\begin{center}
\begin{picture}(13.,7.0)

\put(0.0,0.0){
\epsfysize=11.cm
\rotate[r]{\epsffile{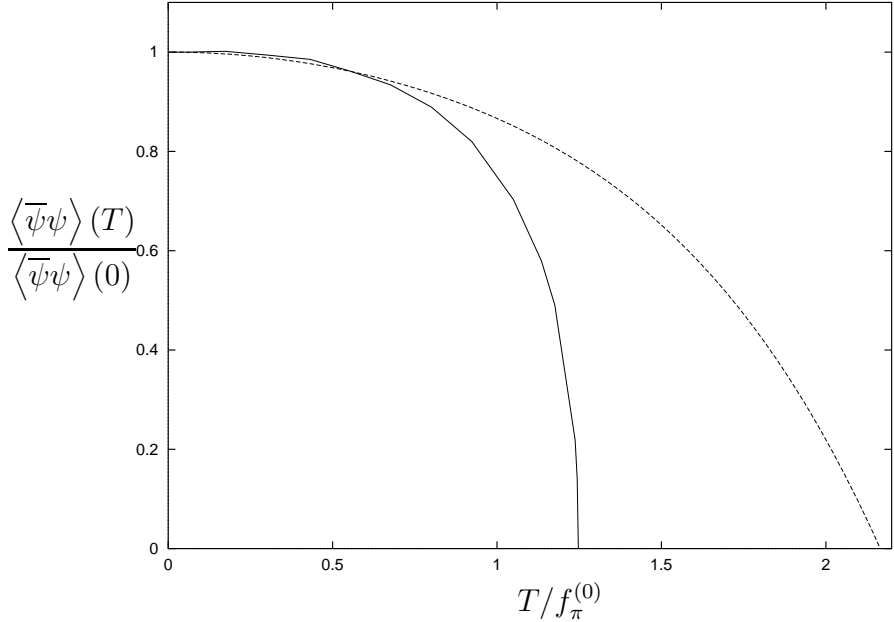}}
}
\put(-1.0,4.2){\bf 
  $\ds{\frac{\VEV{\ol{\psi}\psi}(T)}{\VEV{\ol{\psi}\psi}(0)} }$}
\put(5.8,-0.5){\bf $\ds{T/f_\pi^{(0)}}$}

\end{picture}
\end{center}
\caption{\footnotesize The plot displays the chiral condensate
  $\VEV{\ol{\psi}\psi}$ as a function of $T/f_\pi^{(0)}$. The solid
  line corresponds to our results for vanishing average current quark
  mass $\hat{m}=0$ whereas the dashed line shows the corresponding
  three--loop chiral perturbation theory result for
  $\Gamma_1=470\MeV$.}
\label{cc_T}
\end{figure}

From~\cite{GL87-1} we infer the three--loop result for the
temperature dependence of the chiral condensate in the chiral limit
for $N$ light flavors
\begin{equation}
  \label{BBB100}
  \begin{array}{rcl}
    \ds{\VEV{\ol{\psi}\psi}(T)_{\chi PT}} &=& \ds{
      \VEV{\ol{\psi}\psi}_{\chi PT}(0)
      \Bigg\{1-\frac{N^2-1}{N}\frac{T^2}{12F_0^2}-
        \frac{N^2-1}{2N^2}
        \left(\frac{T^2}{12F_0^2}\right)^2}\nnn
    &+& \ds{
      N(N^2-1)\left(\frac{T^2}{12F_0^2}\right)^3
      \ln\frac{T}{\Gamma_1}
      \Bigg\} +\Oc(T^8)}\; .
  \end{array}
\end{equation}
The scale $\Gamma_1$ can be determined from the $D$--wave isospin zero
$\pi\pi$ scattering length and is given by $\Gamma_1=(470\pm100)\MeV$.
The constant $F_0$ is (in the chiral limit) identical to the pion
decay constant $F_0=f_\pi^{(0)}=80.8\MeV$. In figure \ref{cc_T} we
have plotted the chiral condensate as a function of $T/F_0$ for both,
chiral perturbation theory according to (\ref{BBB100}) and for the
linear quark meson model.
As expected the agreement for small $T$ is very good. Nevertheless,
the anticipated small numerical deviations present even for $T\ll T_c$
due to quark and sigma meson loop contributions are manifest.  For
larger values of $T$, say for $T\gta0.8f_\pi^{(0)}$ the deviations
become significant because of the intrinsic inability of chiral
perturbation theory to correctly reproduce the critical behavior of
the system near its second order phase transition.

Within the language of chiral perturbation theory the neglected
effects of thermal quark fluctuations may be described by an effective
temperature dependence of the parameter $F_0(T)$. We notice that the
temperature at which these corrections become important equals
approximately one third of the constituent quark mass $M_q(T)$ or the
sigma mass $m_\sigma(T)$, respectively, in perfect agreement with
fig.~\ref{Thresh}. As suggested by this figure the onset of the
effects from thermal fluctuations of heavy particles with a
$T$--dependent mass $m_H(T)$ is rather sudden for $T\gta m_H(T)/3$.
These considerations also apply to our two flavor quark meson model.
Within full QCD we expect temperature dependent initial values at
$k_\Phi$.

The dominant contribution to the temperature dependence of the initial
values presumably arises from the influence of the mesons containing
strange quarks as well as the strange quark itself.  Here the quantity
$\ol{m}^2_{k_\Phi}$ seems to be the most important one.  (The
temperature dependence of higher couplings like $\lambda(T)$ is not
very relevant if the IR attractive behavior remains valid, i.e.~if
$Z_{\Phi,k_\Phi}$ remains small for the range of temperatures
considered. We neglect a possible $T$--dependence of the current quark
mass $\hat{m}$.) In particular, for three flavors the potential
$U_{k_\Phi}$ contains a term
\begin{equation}
  \label{LLL12}
  -\frac{1}{2}\ol{\nu}_{k_\Phi}
  \left(\det\Phi+\det\Phi^\dagger\right)=
  -\ol{\nu}_{k_\Phi}\vph_s\Phi_{uu}\Phi_{dd}+\ldots
\end{equation}
which reflects the axial $U_A(1)$ anomaly. It yields a contribution to
the effective mass term proportional to the expectation value
$\VEV{\Phi_{ss}}\equiv\vph_s$, i.e.
\begin{equation}
  \label{LLL13}
  \Delta\ol{m}^2_{k_\Phi}=
  -\frac{1}{2}\ol{\nu}_{k_\Phi}\vph_s\; .
\end{equation}
Both, $\ol{\nu}_{k_\Phi}$ and $\vph_s$, depend on $T$.  We expect
these corrections to become relevant only for temperatures exceeding
$m_K(T)/3$ or $M_s(T)/3$. We note that the temperature dependent kaon
and strange quark masses, $m_K(T)$ and $M_s(T)$, respectively, may be
somewhat different from their zero temperature values but we do not
expect them to be much smaller. A typical value for these scales is
around $500\MeV$. Correspondingly, the thermal fluctuations neglected
in our model should become important for $T\gta170\MeV$. It is even
conceivable that a discontinuity appears in $\vph_s(T)$ for
sufficiently high $T$ (say $T\simeq170\MeV$). This would be reflected
by a discontinuity in the initial values of the $O(4)$--model leading
to a first order transition within this model.  Obviously, these
questions should be addressed in the framework of the three flavor
$SU_L(3)\times SU_R(3)$ quark meson model. Work in this direction is
in progress.

We note that the temperature dependence of $\ol{\nu}(T)\vph_s(T)$ is
closely related to the question of an effective high temperature
restoration of the axial $U_A(1)$ symmetry~\cite{PW84-1,Shu94-1}.  The
$\eta^\prime$ mass term is directly proportional to this combination,
$m_{\eta^\prime}^2(T)-m_\pi^2(T)\simeq\frac{3}{2}\ol{\nu}(T)
\vph_s(T)$ \cite{JW96-2}. Approximate $U_A(1)$ restoration would occur
if $\vph_s(T)$ or $\ol{\nu}(T)$ would decrease sizeable for large $T$.
For realistic QCD this question should be addressed by a three flavor
study. Within two flavor QCD the combination $\ol{\nu}_k\vph_s$ is
replaced by an effective anomalous mass term $\ol{\nu}_k^{(2)}$. The
temperature dependence of $\ol{\nu}^{(2)}(T)$ could be studied by
introducing quarks and the axial anomaly in the two flavor matrix
model of ref.~\cite{BW97-1}.  We add that this question has also been
studied within full two flavor QCD in lattice
simulations~\cite{BKLO96-1,MILC97-2,KLS97-1}. So far there does not
seem to be much evidence for a restoration of the $U_A(1)$ symmetry
near $T_c$ but no final conclusion can be drawn yet.

To summarize, we have found that the effective two flavor quark meson
model presumably gives a good description of the temperature effects
in two flavor QCD for a temperature range $T\lta170\MeV$. Its
reliability should be best for low temperature where our results agree
with chiral perturbation theory.  However, the range of validity is
considerably extended as compared to chiral perturbation theory and
includes, in particular, the critical temperature of the second order
phase transition in the chiral limit.  We have explicitly connected
the universal critical behavior for small $\abs{T-T_c}$ and small
current quark masses with the renormalized couplings at $T=0$ and
realistic quark masses. The main quantitative uncertainties from
neglected fluctuations presumably concern the values of $f_\pi^{(0)}$
and $T_c$ which, in turn, influence the non--universal amplitudes $B$
and $D$ in the critical region. We believe that our overall picture is
rather solid. Where applicable our results compare well with numerical
simulations of full two flavor QCD.

\newpage

\vspace{1cm}\noindent
{\LARGE\bf Appendices}

\appendix{The quark mass term}
\label{Source}
In this appendix we determine the source $\jmath={\rm
  diag}(j_u,j_d,\ldots)$ as a function of the average current quark
mass $\hat{m}$. In this context it is important to note that the
source $\jmath$ does not depend on the IR cutoff scale $k$.  Since
$\jmath$ is determined by the properties of the quark meson model at
the compositeness scale $k_{\Phi}$ and also enters directly the value
of the pion mass, which is determined at $k=0$, this relation provides
a bridge between the short and long distance properties of the quark
meson model.  This will allow us to compute the chiral condensate
$\VEV{\ol{\psi}\psi}$ or the parameter $B_0$ of chiral perturbation
theory \cite{GL82-1}. (We expect, however, sizeable corrections when
going from two to three flavors.  They arise because of the relevance
of strange quark physics at scales near $k_{\Phi}$.). In a more
general context we need the proportionality coefficient $a_q$ between
the source $\jmath_q$ and the current quark mass $m_q$,
$q=u,d,\ldots$, taken at the renormalization scale\footnote{We will
  occasionally use the notation $\hat{m}(\mu)$, $m_q(\mu)$ or
  $\VEV{\ol{\psi}\psi}(\mu)\equiv\VEV{\ol{\psi}\psi}_{k=0}(\mu)$ (not
  to be confused with
  $\VEV{\ol{\psi}\psi}_k\equiv\VEV{\ol{\psi}\psi}_k(\mu=k_\Phi)$) in
  order to indicate the renormalization scale $\mu$. If no argument is
  given $\mu=k_\Phi$ is assumed.} $\mu=k_{\Phi}$,
\begin{equation}
  \label{PPF00} 
  \jmath_q=\frac{Z_{\psi,k_{\Phi}}}{\ol{h}_{k_{\Phi}}} a_q m_q \; .
\end{equation}  
For a computation of the coefficient $a_q$ we need to look into the
details of the introduction of composite meson fields in QCD
\cite{EW94-1,JW96-4}. Let us assume that at the scale $k_{\Phi}$ a
part of the QCD average action for quarks $\Gamma_{k_\Phi}[\psi]$
factorizes in the quark bilinear
\begin{equation}
  \label{PPF01}
  \chi_{ab}(q)=-\int\frac{d^4p}{(2\pi)^4}\tilde{g}(p,q)
  \ol{\psi}_{Lb}(p) \psi_{Ra}(p+q)
\end{equation}
such that
\begin{equation}
  \label{Fk}
  \Gamma_{k_\Phi}[\psi]=-F_{k_\Phi}[\chi]+
  \Gamma_{k_\Phi}^{\prime}[\psi]\; .
\end{equation}
We can then introduce meson fields by inserting the identity
\begin{equation}
  \label{PPF03}
  N \int D\Phi \exp\left(-F_{k_\Phi}[\chi+\Phi]\right)=1
\end{equation}
into the path integral which formally defines $\Gamma_{k_\Phi}[\psi]$.
(Here $N$ is a field independent normalization factor.) This
effectively replaces in (\ref{Fk}) the term $-F_{k_\Phi}[\chi]$
by\footnote{The summation over internal indices as well as the
  integration over momenta has been suppressed. For complex
  $\chi_{ab}(q)$ similar terms have to be supplemented in the
  expansion. See ref.~\cite{EW94-1,JW96-4} for a more detailed
  description. In our Euclidean conventions one has
  $\chi^{\dagger}_{ab}\sim +\tilde{g}^* \ol{\psi}_{Rb} \psi_{La}$.}
\begin{equation}
  \label{PPF04}
  -F_{k_\Phi}[\chi]+F_{k_\Phi}[\chi+\Phi]=\ds{
    \frac{\partial{F_{k_\Phi}[\chi]}}
    {\partial\chi_{ab}(q)}\Phi_{ab}(q)
    +\frac{1}{2}\frac{\partial^2F_{k_\Phi}[\chi]}
    {\partial\chi_{ab}(q)\partial\chi_{cd}(q^\prime)}
  \Phi_{ab}(q)\Phi_{cd}(q^{\prime})+\ldots\; .}
\end{equation}
The original multi--quark interaction $-F_{k_\Phi}[\chi]$ is canceled
by the lowest order term in the Taylor expansion in $\Phi$.  Instead,
we have substituted mesonic self--interactions $F_{k_\Phi}[\Phi]$ and
interactions between mesons and quarks corresponding to the terms in
the expansion which contain powers of $\chi$ and $\Phi$. In
particular, we may specialize to the case where the derivative terms
in $F_{k_\Phi}$ are small and consider a local form $F_{k_\Phi}=\int
d^4x f_{k_\Phi}(\chi)$. A quark mass term is linear in $\chi$ and
translates into a source term for $\Phi$
\begin{equation}
  \label{source} 
  \ds{-\frac{Z_{\psi,k_\Phi}}{\tilde{g}}\mbox{Tr}
    (\chi^{\dagger}m+m^{\dagger}\chi)}\longrightarrow
  \ds{-\frac{Z_{\psi,k_\Phi}}{\tilde{g}}\mbox{Tr}
    (\Phi^{\dagger}m+m^{\dagger}\Phi)}
  =-\ds{\frac{1}{2}\mbox{Tr}(\Phi^{\dagger}\jmath+\jmath^{\dagger}\Phi)
    }
\end{equation}
where $m={\rm diag}(m_u,m_d,\ldots)$.  A factorizing four fermion
interaction yields
\begin{equation}
  \label{PPF06}
  \ol{m}_{k_\Phi}^2\mbox{Tr}\chi^{\dagger}\chi \longrightarrow
  \ol{m}_{k_\Phi}^2\mbox{Tr}\Phi^{\dagger}\Phi
  +\ol{m}_{k_\Phi}^2\mbox{Tr}(\Phi^{\dagger}\chi+\chi^{\dagger}\Phi)\; .
\end{equation}
The second term corresponds to the Yukawa interaction with
$\ol{h}_{k_\Phi}=\ol{m}_{k_\Phi}^2\tilde{g}$. We can therefore extract
$a_q$ from eq.\ (\ref{source}) as
\begin{equation}
  \label{aq}
  a_q=2 \ol{m}_{k_\Phi}^2\; .
\end{equation}
We note that only the terms linear and quadratic in $\chi$ influence
the value of $a_q$. We could either restrict the composite fields from
the beginning to the ones contained in the $O(4)$--symmetric linear
$\sigma$-model or work with all the fields contained in a complex $2
\times 2$ matrix $\Phi$.  In the latter case the anomaly term would
contribute to both the masses and the Yukawa coupling. The net result
is the same with $\ol{m}_{k_\Phi}^2$ denoting the relevant mass term
for the $O(4)$ vector. For our conventions with $\ol{h}_{k_\Phi}=1$ we
have to normalize with $\tilde{g}=\ol{m}_{k_\Phi}^{\ -2}$.  Finally an
eight fermion interaction becomes
\begin{equation}
  \label{eightfer}
  \ds{\frac{1}{2}\ol{\lambda}_{k_\Phi}\, \mbox{Tr}(\chi^{\dagger}\chi)^2}
  \longrightarrow \ds{\frac{1}{2}\, \ol{\lambda}_{k_\Phi}(\mbox{Tr}
    \Phi^{\dagger}\Phi)^2+\, \ol{\lambda}_{k_\Phi} 
    \mbox{Tr}\Phi^{\dagger}\Phi
    \,  \mbox{Tr}(\chi^{\dagger}\Phi+\Phi^{\dagger}\chi)+\ldots
    }\; .
\end{equation}
We see here the appearance of terms quadratic in the quarks involving
higher powers of $\Phi$.

There is an alternative, equivalent way of understanding the relation
between $\jmath$ and $m_q$: The quark masses in the picture with
mesons must be equal at the scale $k_{\Phi}$ to the current quark mass
$m_q(k_{\Phi})$. Let us consider an $O(4)$--symmetric fermionic
interaction $\ol{m}_{k_\Phi}^2 \mbox{Tr}
\chi^{\dagger}\chi+\frac{1}{2}\ol{\lambda}_{k_\Phi}(\mbox{Tr}
\chi^{\dagger}\chi)^2$ which leads to a meson potential
\begin{equation}
  \label{PPF07}
  U_{k_\Phi}=\ol{m}_{k_\Phi}^2\Tr\Phi^\dagger\Phi+
  \frac{1}{2}\ol{\lambda}_{k_\Phi}
  \left(\Tr\Phi^\dagger\Phi\right)^2\; .
\end{equation}
In the mesonic picture the quarks acquire masses through the Yukawa
coupling to $\Phi$
\begin{equation}
  \label{source2}
  M_{k}=\ds{\frac{\ol{h}_{k}}{Z_{\psi,k}}
    \left(1+\frac{\ol{\lambda}_{k}}
      {\ol{m}_{k}^2}\mbox{Tr}\langle\Phi^{\dagger}\rangle_k
      \VEV{\Phi}_k\right) \VEV{\Phi}_k }
\end{equation}
where the second term arises from the higher order coupling in
(\ref{eightfer}). Here $\VEV{\Phi}_k=\mbox{diag}
(\varphi_u,\varphi_d,\ldots)$ is the expectation value at the coarse
graining scale $k$ in the presence of the source term and
$M_k=\mbox{diag}(M_u,M_d,\ldots)$.  It is sufficient to specify the
dependence of $U_{k_\Phi}$ on real diagonal fields $\Phi_{qq}$. Then
the $\varphi_q$ are determined from the condition
\begin{equation}
  \label{PPF08}
  \ds{\frac{\partial U_{k_\Phi}}{\partial \Phi_{qq}}(\varphi_q)}
  =2\left(\ol{m}_{k_\Phi}^2+
  \ol{\lambda}_{k_\Phi}\sum_{q^{\prime}}\varphi_{q^{\prime}}^2
  \right)
  \varphi_q=\jmath_q\; .
\end{equation}
Identifying $M_{k=k_{\Phi}}$ in (\ref{source2}) with $m(k_\Phi)$ one
has
\begin{equation}
  \label{PPF09}
  a_q\left(1+\ds{\frac{\ol{\lambda}_{k_\Phi}}{\ol{m}_{k_\Phi}^2}
    \sum_{q^{\prime}}
    \varphi_{q^{\prime}}^2}\right)=\ds{\frac{\jmath_q}{\varphi_q}}=
  2\ol{m}_{k_\Phi}^2
  +2\ol{\lambda}_{k_\Phi}
  \sum_{q^{\prime}}\varphi_{q^{\prime}}^2
\end{equation}
and we recover (\ref{aq}) or, in our normalization with
$Z_{\psi,k_{\Phi}}=1$, $\ol{h}_{k_{\Phi}}=1$,
\begin{equation}
  \label{PPF10}
  \jmath=2 \ol{m}_{k_\Phi}^2 \hat{m}\; .
\end{equation}
It is remarkable that higher order terms (e.g. $\sim
\ol{\lambda}_{k_\Phi}$) do not influence the relation between $\jmath$
and $\hat{m}$.  Only the quadratic term $\ol{m}_{k_\Phi}^2$ enters
which is in our scenario the only relevant coupling. This feature is
an important ingredient for the predictive power of the model as far
as the absolute size of the current quark mass is concerned. An
appearance of higher order couplings in $a_q$ would make it very hard
to compute this quantity.  We emphasize that within our formalism
there is no difference of principle between the current quark mass and
the constituent quark mass.  Whereas the current quark mass
$m_q(k_\Phi)$ at the normalization scale $\mu=k_\Phi$ corresponds to
$M_{q,k}$ at the compositeness scale $k_{\Phi}$ the constituent quark
mass is $M_{q,k=0}$. As $k$ is lowered from $k_{\Phi}$ to zero one
observes that the quark mass increases, similarly to the running
current quark mass.  Once chiral symmetry breaking sets in at the
scale $k_{\chi SB}$ there is a large increase in the quark masses,
especially for $M_u$ and $M_d$.

The formalism of composite fields also provides the link
\cite{EW94-1} to the chiral condensate $\VEV{\ol{\psi}\psi}$ since
the expectation value $\VEV{\Phi}$ is related to the expectation
value of a composite quark-antiquark operator. For 
$\ol{\lambda}=0$ one has \cite{JW96-4}
\begin{equation}
  \label{PPF11}
  \VEV{\Phi}_k+\langle\Phi^{\dagger}\rangle_k=
  -\ds{\frac{1}{\ol{m}_{k_\Phi}^2}}
  \VEV{\ol{\psi}\psi}_k(k_{\Phi})+m_q(k_{\Phi})+m_q^{\dagger}(k_{\Phi})
\end{equation}
with $\VEV{\ol{\psi}\psi}_k(k_\Phi)$ a suitably regularized operator
normalized at $\mu=k_{\Phi}$.

\appendix{Threshold functions}
\label{ThresholdFunctions}

In this appendix we list the various definitions of dimensionless
threshold functions appearing in the flow equations and the
expressions for the anomalous dimensions for $T=0$. They involve the
inverse scalar average propagator $P(q)$ defined in (\ref{BBB01}) and
the corresponding fermionic function $P_F$ which can be chosen
as~\cite{JW96-1} 
\begin{equation}
  \label{JJJ000}
  P_F(q)=P(q)\equiv q^2\left(1+r_F(q)\right)^2\; .
\end{equation}
We abbreviate
\begin{equation}
  \label{LLL23}
  x= q^2\; ,\;\; P(x)\equiv P(q)\; ,\;\;
  \dot{P}(x)\equiv\frac{\prl}{\prl x}P(x)\; ,\;\;
  \widehat{\frac{\prl}{\prl t}}\dot{P}\equiv
  \frac{\prl}{\prl x}\widehat{\frac{\prl}{\prl t}}P\; ,
\end{equation}
etc., and use the formal definition
\begin{equation}
  \label{LLL24}
  \ds{\widehat{\frac{\prl}{\prl t}}}
  \equiv \ds{
    \frac{1}{Z_{\Phi,k}}\frac{\prl R_k}{\prl t}
    \frac{\prl}{\prl P} }
  + \ds{
    \frac{2}{Z_{\psi,k}} \frac{P_F}{1+r_F}
    \frac{\prl \left[Z_{\psi,k} r_F\right]}{\prl t}
    \frac{\prl}{\prl P_F} }\; .
\end{equation}
The bosonic threshold functions read
\begin{equation}
  \label{LLL20}
  \begin{array}{rcl}
    \ds{l_n^d (w;\eta_\Phi)} &=& \ds{
      l_n^d (w) - \eta_\Phi\hat{l}_n^d (w)
      }\nnn
    &=& \ds{
      \frac{n+\delta_{n,0}}{2} 
      k^{2n-d} \int_0^\infty d x\, x^{\frac{d}{2}-1}
      \left(\frac{1}{Z_{\Phi,k}} \frac{\prl R_k}{\prl t}\right)
        \left( P+w k^2\right)^{-(n+1)} }\nnn
    \ds{ l_{n_1,n_2}^d(w_1,w_2;\eta_\Phi)} &=& \ds{
      \l_{n_1,n_2}^d(w_1,w_2)
      -\eta_\Phi \hat{l}_{n_1,n_2}^d(w_1,w_2)}\nnn
    &=& \ds{
      -\hal k^{2(n_1+n_2)-d}
      \int_0^\infty dx\, x^{\frac{d}{2}-1}
      \widehat{\frac{\prl}{\prl t}} \left\{\left(
      P+w_1k^2\right)^{-n_1}
      \left( P+w_2k^2\right)^{-n_2} \right\}}
  \end{array}
\end{equation}
where $n,n_1,n_2\ge0$ is assumed. For $n\neq0$ the
functions $l_n^d$ may also be written as
\begin{equation}
  \label{LLL21}
  l_n^d (w;\eta_\Phi) =
  -\hal k^{2n-d} \int_0^{\infty} dx x^{\frac{d}{2}-1}
  \widehat{\frac{\prl}{\prl t}} \left( P+w k^2\right)^{-n}\; .
\end{equation}
The fermionic integrals $l_n^{(F)d} (w;\eta_\psi)=l_n^{(F)d} (w)-
\eta_\psi\check{l}_n^{(F)d} (w)$ are defined analogously as
\begin{equation}
  \label{JJJ001}
  \begin{array}{rcl}
    \ds{l_n^{(F)d} (w;\eta_\psi)} &=& \ds{
      \left(n+\delta_{n,0}\right)
      k^{2n-d} \int_0^\infty d x\, x^{\frac{d}{2}-1}
      \frac{1}{Z_{\psi,k}}\frac{P_F}{1+r_F}
      \frac{\prl\left[ Z_{\psi,k}r_F\right]}{\prl t}
        \left(P+w k^2\right)^{-(n+1)} }\; .
  \end{array}
\end{equation}
Furthermore one has
\begin{equation}
  \label{LLL22}
  \begin{array}{rcl}
    \ds{l_{n_1,n_2}^{(FB)d}(w_1,w_2;\eta_\psi,\eta_\Phi)} &=& \ds{
      l_{n_1,n_2}^{(FB)d}(w_1,w_2)
      -\eta_\psi \check{l}_{n_1,n_2}^{(FB)d}(w_1,w_2)
      -\eta_\Phi \hat{l}_{n_1,n_2}^{(FB)d}(w_1,w_2) 
      }\nnn
    && \ds{ \hspace{-2cm}
      = -\hal k^{2(n_1+n_2)-d}
      \int_0^\infty dx\, x^{\frac{d}{2}-1}
      \widehat{\frac{\prl}{\prl t}}\left\{
      \frac{1}{[P_F(x)+k^2w_1]^{n_1} [P(x)+k^2w_2]^{n_2} } \right\}
      }\nnn
    \ds{m_{n_1,n_2}^d (w_1,w_2;\eta_\Phi)} &\equiv& \ds{
      m_{n_1,n_2}^d (w_1,w_2) - \eta_\Phi 
      \hat{m}_{n_1,n_2}^d (w_1,w_2) 
      }\nnn
    && \ds{\hspace{-2cm} 
      = -\hal k^{2(n_1+n_2-1)-d}
      \int_0^\infty dx\, x^{\frac{d}{2}}
      \widehat{\frac{\prl}{\prl t}} \left\{
      \frac{\dot{P} (x)}
      {[P(x)+k^2 w_1]^{n_1} }
      \frac{\dot{P} (x)}
      {[P(x)+k^2 w_2]^{n_2}} \right\} 
      }\nnn
    \ds{m_4^{(F)d} (w;\eta_\psi)} &=& \ds{
      m_4^{(F)d} (w)-\eta_\psi \check{m}_4^{(F)d} (w) 
      }\nnn
    &=& \ds{ 
      -\hal k^{4-d}
      \int_0^\infty dx\, x^{\frac{d}{2}+1}
      \widehat{\frac{\prl}{\prl t}} \left(
      \frac{\prl}{\prl x}
      \frac{1+r_F(x)}{P_F(x)+k^2w}\right)^2
      \label{m4Fd} 
      }\nnn
    \ds{m_{n_1,n_2}^{(FB)d}(w_1,w_2;\eta_\psi,\eta_\Phi)} &=& \ds{
      m_{n_1,n_2}^{(FB)d}(w_1,w_2)
      -\eta_\psi \check{m}_{n_1,n_2}^{(FB)d}(w_1,w_2)
      -\eta_\Phi \hat{m}_{n_1,n_2}^{(FB)d}(w_1,w_2) 
      }\nnn
    && \ds{ \hspace{-2cm}
      = -\hal k^{2(n_1+n_2-1)-d}
      \int_0^\infty dx\, x^{\frac{d}{2}}
      \widehat{\frac{\prl}{\prl t}}\left\{
      \frac{1+r_F(x)}{[P_F(x)+k^2w_1]^{n_1}}
      \frac{\dot{P}(x)}{[P(x)+k^2w_2]^{n_2}} \right\} \; . 
      }
  \end{array}
\end{equation}
The dependence of the threshold functions on the anomalous dimensions
arises from the $t$--derivative acting on $Z_{\Phi,k}$ and
$Z_{\psi,k}$ within $R_k$ and $Z_{\psi,k}r_{F}$, respectively.  We
furthermore use the abbreviations
\begin{equation}
  \label{LLL25}
  \begin{array}{rcl}
  \ds{l_n^d(\eta_\Phi)\equiv l_n^d(0;\eta_\Phi)} &,& \ds{
    l_{n}^{(F)d}(\eta_\psi)\equiv l_{n}^{(F)d}(0;\eta_\psi)}\nnn
  \ds{l_n^d(w)\equiv l_n^d(w;0)} &,& \ds{
  l_n^d\equiv l_n^d(0;0)}
  \end{array}
\end{equation}
etc.~and note that in four dimensions the integrals
\begin{equation}
  \label{LLL26}
  l_2^4(0,0)=l_2^{(F)4}(0,0)=
  l_{1,1}^{(FB)4}(0,0)=m_4^{(F)4}(0)=m_{1,2}^{(FB)4}(0,0)=1
\end{equation}
are independent of the particular choice of the infrared cutoff.

\appendix{Temperature dependent threshold functions}
\label{AnomalousDimensions}

Non--vanishing temperature manifests itself in the flow equations
(\ref{AAA68}), (\ref{AAA91})---(\ref{AAA69}) only through a change to
$T$--dependent threshold functions. In this appendix we will define
these functions and discuss some subtleties regarding the definition
of the anomalous dimensions and the Yukawa coupling for $T\neq0$. The
corresponding $T=0$ threshold functions can be found in
appendix~\ref{ThresholdFunctions} where also some of our notation is
fixed.

The flow equation (\ref{AAA68}) for the effective average potential
involves a bosonic and a fermionic threshold function whose
generalization to finite temperature is straightforward
\begin{equation}
  \label{JJJ00}
  \begin{array}{rcl}
    \ds{l_n^d(w,\tilde{T};\eta_\Phi)} &=& \ds{
      \frac{(n+\delta_{n,0})}{2}
      \frac{v_{d-1}}{v_d}k^{2n-d+1}\tilde{T}
      \sum_{l\in\ZZZ}\int_0^\infty
      d x\,x^{\frac{d-3}{2}}
      Z_{\Phi,k}^{-1}
      \frac{\prl_t R_k(y)}
      {\left[ P(y)+k^2 w\right]^{n+1}}\; ,
      }\nnn
    \ds{l_n^{(F)d}(w,\tilde{T};\eta_\Phi)} &=& \ds{
      \left(n+\delta_{n,0}\right)
      \frac{v_{d-1}}{v_d}k^{2n-d+1}}\nnn
    &\times& \ds{
      \tilde{T}\sum_{l\in\ZZZ}\int_0^\infty
      d x\,x^{\frac{d-3}{2}}
      Z_{\psi,k}^{-1}\frac{P_F(y_F)}{1+r_F(y_F)}
      \frac{\prl_t\left[ Z_{\psi,k}r_F(y_F)\right]}
      {\left[P_F(y_F)+w k^2\right]^{n+1}}
      }
  \end{array}
\end{equation}
where $\tilde{T}=T/k$ and
\begin{equation}
  \label{JJJ01}
  \begin{array}{rcl}
    \ds{y} &=& \ds{x+(2 l\pi T)^2}\nnn
    \ds{y_F} &=& \ds{x+(2 l+1)^2\pi^2 T^2}
  \end{array}
\end{equation}

The computation of the anomalous dimensions $\eta_\Phi$, $\eta_\psi$
and the flow equation for the Yukawa coupling $h$ at non--vanishing
temperature requires some care. The anomalous dimensions determine the
IR cutoff scale dependence of $Z_{\Phi,k}$ and $Z_{\psi,k}$ according
to $\eta_\Phi=-\prl_t\ln Z_{\Phi,k}$, $\eta_\psi=-\prl_t\ln
Z_{\psi,k}$ with $t=\ln k/k_\Phi$.  It is important to realize that
for a computation of the scale dependence of the effective
three--dimensional $Z_{\Phi,k}$ and $Z_{\psi,k}$ for $T\neq0$ momentum
dependent wave function renormalization constants of the
four--dimensional theory are required.  This is a consequence of the
fact that in the three--dimensional model each of the infinite number
of different Matsubara modes of a four--dimensional bosonic or
fermionic field $\phi(Q)$ corresponds to a different value of
$Q_0=2\pi lT$ or $Q_0=(2l+1)\pi T$, respectively, with
$Q^2=Q_0^2+\vec{Q}^{\,2}$ and $l\in\ZZ$.  We will therefore allow for
momentum dependent wave function renormalizations, i.e.~for a kinetic
part of $\Gamma_k$ of the form
\begin{equation}
  \label{PPP00}
  \Gamma^{\rm kin}_k=
  \int\frac{d^d q}{(2\pi)^d}\left\{
  Z_{\Phi,k}(q^2)q^2\Tr\left(\Phi^\dagger(q)\Phi(q)\right)+
  Z_{\psi,k}(q^2)\ol{\psi}(q)\gamma^\mu q_\mu\psi(q)\right\}
\end{equation}
in momentum space.

In the $O(4)$--model the evolution equation for $Z_{\Phi,k}(Q)$ may
then be obtained by considering a background field configuration with
a small momentum dependence,
\begin{equation}
  \label{PPP01}
  \Phi_j(x)=\vph\delta_{j 1}+
  \left(\delta\vph e^{-i Q x}+
    \delta\vph^* e^{i Q x}\right)\delta_{j 2}\; ;\;\;\;
    j=1,\ldots,4
\end{equation}
supplemented by
\begin{equation}
  \label{PPP02}
  \psi_a=\ol{\psi}_a=0\; ;\;\;\; a=1,2\; .
\end{equation}
Expanding around this configuration at the minimum of the effective
average potential $U_k$ we observe that $\delta\vph$ corresponds to an
excitation in the Goldstone boson direction. The exact inverse
two--point function $\Gamma_k^{(2)}$ turns out to be block--diagonal
with respect to scalar and fermion indices for this configuration. It
therefore decays into corresponding matrices $\Gamma_{S k}^{(2)}$ and
$\Gamma_{F k}^{(2)}$ acting in the scalar and fermion subspaces,
respectively.  The scale dependence of the scalar wave function
renormalization for non--vanishing $T$ is obtained from (\ref{ERGE})
and (\ref{PPP00}) for the configuration (\ref{PPP01}) as
\begin{equation}
  \label{PPP03}
  \begin{array}{rcl}
    \ds{\frac{\prl}{\prl t}Z_{\Phi,k}(Q^2)} &=& \ds{
      \frac{1}{\vec{Q}^2}\Bigg(\lim_{\delta\vph\delta\vph^*\ra0}
      \frac{\delta}{\delta(\delta\vph\delta\vph^*)}
      \Bigg\{\frac{1}{2}\Tr\left[\left(
      \Gamma_{S k}^{(2)}+R_k\right)^{-1}
      \frac{\prl}{\prl t}R_k\right]}\nnn
    &-& \ds{
      \Tr\left[\left(\Gamma_{F k}^{(2)}+R_{F k}\right)^{-1}
      \frac{\prl}{\prl t}R_{F k}\right]\Bigg\}-
      (\vec{Q}\ra0)\Bigg) }\; .
  \end{array}
\end{equation}
In the three--dimensional theory there is now a different scalar wave
function renormalization $Z_{\Phi,k}(Q_0,\vec{Q})$ for each Matsubara
mode $Q_0$. As in the four--dimensional model for $T=0$ we neglect the
momentum dependence of the wave function renormalization constants and
evaluate $Z_{\Phi,k}$ for $\vec{Q}=0$ for each Matsubara mode. We will
furthermore simplify the truncation of the effective average action by
choosing the Matsubara zero--mode wave function renormalization
constant for all Matsubara modes, i.e., approximate
\begin{equation}
  \label{PPP03a}
  Z_{\Phi,k}(\tilde{T})=Z_{\Phi,k}(Q_0^2=0,\vec{Q}^2=0)\; .
\end{equation}
This is justified by the rapid decoupling of all massive Matsubara
modes within a small range of $k$ for fixed $T$ as discussed in
section~\ref{FiniteTemperatureFormalism}. This results in the
expression (\ref{AAA69}) for $\eta_\Phi$ but now with temperature
dependent threshold functions ($\tilde{T}=T/k$)
\begin{equation}
  \label{PPP04a}
  \begin{array}{rcl}
    \ds{m_{n_1,n_2}^{d}(w_1,w_2,\tilde{T};\eta_\Phi)}
    &=& \ds{
      m_{n_1,n_2}^{d}(w_1,w_2,\tilde{T})-\eta_\Phi 
      \hat{m}_{n_1,n_2}^{d}(w_1,w_2,\tilde{T}) 
      }\nnn
    && \ds{\hspace{-1cm}=
      -\hal k^{2(n_1+n_2-1)-d+1}
      \frac{d v_{d-1}}{(d-1)v_d}
      }\nnn
    && \ds{\hspace{-1cm}\times
      \tilde{T}\sum_{l\in\ZZZ}
      \int_0^\infty dx\, x^{\frac{d-1}{2}} 
      \widehat{\frac{\prl}{\prl t}} \left\{
      \frac{\dot{P}(y)}
      {[P(y)+k^2 w_1]^{n_1} }
      \frac{\dot{P} (y)}
      {[P(y)+k^2 w_2]^{n_2}} \right\} }\nnn
    \ds{m_{4}^{(F)d}(w,\tilde{T};\eta_\psi)}
    &=& \ds{
      m_4^{(F)d}(w,\tilde{T})-
      \eta_\psi\check{m}_4^{(F)d}(w,\tilde{T}) 
      }\nnn
    && \ds{ \hspace{-1cm}=
      -\frac{1}{2} k^{5-d}\frac{d v_{d-1}}{(d-1)v_d}
      \tilde{T}\sum_{l\in\ZZZ}
      \int_0^\infty dx\, x^{\frac{d-1}{2}}y_F
      \widehat{\frac{\prl}{\prl t}} \left(
      \frac{\prl}{\prl x}
      \frac{1+r_F(y_F)}{P_F(y_F)+k^2w}\right)^2}\; .
  \end{array}
\end{equation}
For further technical details we refer the reader to
ref.~\cite{JW96-1}.

The fermion anomalous dimension and the flow equation for the Yukawa
coupling can be obtained by considering a
field configuration
\begin{equation}
  \label{PPP05}
  \begin{array}{rcl}
    \ds{\Phi_j(x)} &=& \ds{\vph\delta_{j 1}\; ;\;\;\;
      j=1,\ldots,4}\nnn
    \ds{\psi_a(x)} &=& \ds{
      \psi_a e^{-i Q x}}\nnn
    \ds{\ol{\psi}_a(x)} &=& \ds{
      \ol{\psi}_a e^{i Q x}\; ;\;\;\;
      a=1,2}\; .
  \end{array}
\end{equation}
The derivation follows similar lines as for the scalar anomalous
dimension discussed above.  For computational details we refer the
reader to \cite{JW96-1}. An important difference as compared to
$Z_{\Phi,k}(Q)$ relates to the fact that there are no fermionic zero
modes. It would therefore be inconsistent to define
$Z_{\psi,k}(\tilde{T})$ or $h_k(\tilde{T})$ at $Q_0=0$ if $Q$ denotes
the external fermion momentum. Yet, we will again resort to the
approximation of using the same wave function renormalization constant
and Yukawa coupling for all fermionic Matsubara modes. For the same
reason as for $Z_{\Phi,k}(\tilde{T})$ we will use for this purpose the
mode with the lowest $T$--dependent mass, i.e.~define
\begin{equation}
  \label{PPP06}
  \begin{array}{rcl}
  \ds{Z_{\psi,k}(\tilde{T})} &=& \ds{
    Z_{\psi,k}(Q_0^2=\pi^2 T^2,\vec{Q}^{\,2}=0)}\nnn
  \ds{h_k(\tilde{T})} &=& \ds{
    h_k(Q_0^2=\pi^2 T^2,\vec{Q}^{\,2}=0)}\; ,
  \end{array}
\end{equation}
where we have neglected a possible dependence of $h_k$ on the external
scalar momentum of the Yukawa vertex.  This yields the expressions
(\ref{AAA70}) and (\ref{AAA69}) for the flow of $h^2$ and $\eta_\psi$,
respectively, but now with the $T$--dependent threshold functions
\begin{equation}
  \label{PPP08}
  \begin{array}{rcl}
    \ds{m_{1,2}^{(F B)d}(w_1,w_2,\tilde{T};\eta_\psi,\eta_\Phi)}
    &=& \ds{
      m_{1,2}^{(F B)d}(w_1,w_2,\tilde{T})
      }\nnn
    &-& \ds{
      \eta_\Phi\hat{m}_{1,2}^{(F B)d}(w_1,w_2,\tilde{T})-
      \eta_\psi\check{m}_{1,2}^{(F B)d}(w_1,w_2,\tilde{T})
      }\nnn
    && \ds{ \hspace{-4cm}
      = -\frac{1}{2} k^{2(n_1+n_2)-d-1}
      \frac{d v_{d-1}}{(d-1)v_d}
      \tilde{T}\sum_{l\in\ZZZ}
      \int_0^\infty dx\, x^{\frac{d-1}{2}}
      }\nnn
    && \ds{\hspace{-4cm}\times
      \widehat{\frac{\prl}{\prl t}}\left\{
      \frac{1+r_F(y_F)}{[P_F(y_F)+k^2w_1]^{n_1}}
      \frac{\dot{P}(y)}{[P(y)+k^2w_2]^{n_2}} \right\}
      }\nnn
    \ds{l_{n_1,n_2}^{(FB)d}(w_1,w_2,\tilde{T};\eta_\psi,\eta_\Phi)} &=& \ds{
      l_{n_1,n_2}^{(FB)d}(w_1,w_2,\tilde{T})}\nnn
    &-& \ds{
      \eta_\psi \check{l}_{n_1,n_2}^{(FB)d}(w_1,w_2,\tilde{T})
      -\eta_\Phi \hat{l}_{n_1,n_2}^{(FB)d}(w_1,w_2,\tilde{T}) 
      }\nnn
    && \ds{ \hspace{-4cm}
      = -\hal k^{2(n_1+n_2)-d+1}
      \frac{v_{d-1}}{v_d}\tilde{T} 
      \times\sum_{l\in\ZZZ}
      \int_0^\infty dx\, x^{\frac{d-3}{2}}
      }\nnn
    && \ds{ \hspace{-4cm}
      \widehat{\frac{\prl}{\prl t}}\left\{
      \frac{1}{[P_F(y_F)+k^2w_1]^{n_1} [P(y)+k^2w_2]^{n_2} } \right\}
      }\; .
  \end{array}
\end{equation}

\end{document}